\documentclass[aps,pra,nobalancelastpage,superscriptaddress,twocolumn,reprint]{revtex4-2}
\usepackage{graphicx}
\usepackage{amsmath}
\usepackage{braket}
\usepackage{centernot}

\usepackage{standalone, gensymb, tensor, amsmath,amsfonts,amssymb,amsthm,bbm,bm, braket}
\usepackage[T1]{fontenc}
\usepackage{hyperref}
\usepackage[inline]{enumitem}
\usepackage{scalerel, physics}
\usepackage{wasysym}
\usepackage{comment}
\usepackage{enumerate}
\usepackage{graphicx}
\usepackage{mathtools}
\usepackage{soul}
\usepackage{xparse}
\usepackage{dsfont}
\usepackage{ifthen}
\usepackage{tensor}
\usepackage{tikz}
\usepackage{pgfplots}
\usepackage{tikz-network}
\usepackage{simplewick} 
\usepackage{ mathrsfs }

\usetikzlibrary{patterns,decorations.pathreplacing}
\theoremstyle{break}        

\theoremstyle{break}

\usepackage{color}
\definecolor{alessandrored}{RGB}{250, 153, 140}
\definecolor{OliveGreen}{RGB}{85,107,47}
\definecolor{NavyBlue}{RGB}{0,0,128}
\definecolor{alessandrogreen}{RGB}{17, 127, 35}
\definecolor{jonasgreen}{RGB}{81, 160, 37}
\usepackage{tensor}
\usepackage{xifthen}
\usepackage{xargs}

\definecolor{blue1}{RGB}{0, 0, 255}
\definecolor{blue2}{RGB}{0, 150, 255}
\definecolor{blue3}{RGB}{0, 71, 171}
\definecolor{blue4}{RGB}{100, 149, 237}
\definecolor{blue5}{RGB}{93, 63, 211}

\definecolor{red1}{RGB}{238, 75, 43}
\definecolor{red2}{RGB}{233, 116, 81}
\definecolor{red3}{RGB}{222, 49, 99}
\definecolor{red4}{RGB}{250, 160, 160}
\definecolor{red5}{RGB}{236, 88, 0}

\definecolor{green1}{RGB}{80, 180, 152}
\definecolor{green2}{RGB}{166,218,149}

\definecolor{orange1}{RGB}{255, 116, 23}

\definecolor{grey1}{RGB}{98,98,98}
\definecolor{grey2}{RGB}{211,211,211}
\definecolor{grey3}{RGB}{192,192,192}
\definecolor{grey4}{RGB}{169,169,169}
\definecolor{yellow1}{RGB}{253, 218, 13}
\definecolor{purple1}{RGB}{3, 38, 241}
\definecolor{green3}{RGB}{147, 197, 114}
\definecolor{orange2}{RGB}{255, 170, 51}

\newcommandx{\fineq}[5][1=-.8ex,2=1,3=1,5=0]{
	\begin{tikzpicture}[baseline={([yshift=#1]current  bounding  box.center)}, scale = #2, every node/.style={scale = #3},rotate around={#5:(0,0)},every node/.style={transform shape}]
		#4
	\end{tikzpicture}
}

\definecolor{bertinired}{RGB}{232,102,102}
\definecolor{bertiniblue}{RGB}{101,147,245}
\definecolor{bertinigreyblue}{RGB}{101,147,245}
\definecolor{bertinigreyred}{RGB}{232,102,102}
\definecolor{bertinivioletc}{RGB}{45,130,60}
\definecolor{bertinigreen}{RGB}{166,218,149}
\definecolor{bertiniorange}{RGB}{255, 116, 23}
\definecolor{OliveGreen}{RGB}{85,107,47}
\definecolor{NavyBlue}{RGB}{0,0,128}
\definecolor{bertiniviolet}{RGB}{210,145,178}
\definecolor{bertinigrey1}{RGB}{98,98,98}
\definecolor{bertinigrey2}{RGB}{211,211,211}
\definecolor{bertinigrey3}{RGB}{192,192,192}
\definecolor{bertinigrey4}{RGB}{169,169,169}

\newcommandx{\tikzdiagup}{
	\tikz {\draw[thick] (0,0)--(0.15,0.15); \draw (0,0) rectangle (0.15,0.15);}
}

\newcommandx{\gatecross}[1][1=0.5]{
	\pgfmathparse{#1/2.0}
	\let\x\pgfmathresult
	\draw[thick] (-\x,-\x) -- (\x,\x);
	\draw[thick] (\x,-\x) -- (-\x,\x);
}
\newcommandx{\gatesqu}[2][1=0.25,2=]{
	\pgfmathparse{#1/2.0}
	\let\x\pgfmathresult
	\ifthenelse{\equal{#2}{}}{
		\draw[thick, fill=white, rounded corners=2pt] (-\x,\x) rectangle (\x,-\x);
	}{
		\draw[thick, fill=#2, rounded corners=2pt] (-\x,\x) rectangle (\x,-\x);
	}
}

\newcommandx{\gatemark}[2][1=0.075,2=tr]{
	\pgfmathparse{#1}
	\let\l\pgfmathresult
	\ifthenelse{\equal{#2}{topleft}}{
		\draw[thick] (0,\l) -- ++(-\l,0) --++ (0,-\l);
	}{}
	\ifthenelse{\equal{#2}{topright}}{
		\draw[thick] (0,\l) -- ++(\l,0) --++ (0,-\l);
	}{}
	
	\ifthenelse{\equal{#2}{bottomleft}}{
		\draw[thick] (0,-\l) -- ++(-\l,0) --++ (0,\l);
	}{}
	\ifthenelse{\equal{#2}{bottomright}}{
		\draw[thick] (0,-\l) -- ++(\l,0) --++ (0,\l);
	}{}
	
}

\newcommandx{\squaregate}[3][1=0,2=0,3=white]
{
	\begin{scope}[shift={(#1,#2)},rounded corners= 2pt]
		\draw[thick,fill=#3] (-.13,-.13) rectangle (.15,.15);
	\end{scope}
}

\newcommandx{\roundgate}[6][1=0,2=0,3=1,4=topright,5=white,6=-1]{
	\pgfmathparse{#3}
	\let\l\pgfmathresult
	\begin{scope}[shift={(#1,#2)}]
		\gatecross[\l]
			\pgfmathparse{\l/2.0}
		\let\s\pgfmathresult
		\gatesqu[\s][#5]
		\pgfmathparse{\l*0.15}
		\let\m\pgfmathresult
	\ifthenelse{\equal{#6}{-1}}{		\gatemark[\m][#4]
	}{	\node at ({0},{0}) {\scalebox{1.3}{$#6$}};}
\end{scope}
}

\newcommandx{\wcirc}[2]{\begin{scope}
		\draw[fill=white] (#1,#2) circle (0.15);	\end{scope}} 
\newcommandx{\wcircc}[2]{\begin{scope}
		\draw[fill=white] (#1,#2) circle (0.13);	\end{scope}} 
\newcommandx{\wsqr}[2]{\begin{scope}
		\draw[fill=white,shift={(#1,#2)}] (-.13,.13) rectangle (.13,-.13);	\end{scope}} 
\newcommandx{\wsqrr}[2]{\begin{scope}
		\draw[fill=white,shift={(#1,#2)}] (-.11,.11) rectangle (.11,-.11);	\end{scope}}
\newcommandx{\bcirc}[2]{\begin{scope}
		\draw[fill=black] (#1,#2) circle (0.15);	\end{scope}} 
\newcommandx{\thetastate}[4][1=0,2=0,3=1,4=]{
	\pgfmathparse{#3/2}
	\let\l\pgfmathresult
	\pgfmathparse{\l*0.15}
	\let\m\pgfmathresult
	\begin{scope}[shift={(#1,#2)}]
		\draw[thick] (0,0)--(\l,\l);
		\draw[thick] (0,0)--(-\l,\l);
		\ifthenelse{\equal{#4}{}}{
			\draw[fill=white] (0,0) circle (0.15);
		}{
			\draw[thick, fill=#4] (0,0) circle (0.15);
		}
	\end{scope}
}

\newcommandx{\thetastateflipped}[4][1=0,2=0,3=1,4=]{
	\pgfmathparse{#3/2}
	\let\l\pgfmathresult
	\pgfmathparse{\l*0.15}
	\let\m\pgfmathresult
	\begin{scope}[shift={(#1,#2)}]
		\draw[thick] (0,0)--(\l,-\l);
		\draw[thick] (0,0)--(-\l,-\l);
		\ifthenelse{\equal{#4}{}}{
			\draw[fill=white] (0,0) circle (0.15);
		}{
			\draw[thick, fill=#4] (0,0) circle (0.15);
		}
	\end{scope}
}

\newcommandx{\vertgate}[5][1=0,2=0,3=4,4=bertiniorange,5=topright]
{
	\begin{scope}[shift={(#1,#2)}]
		\ifthenelse{\equal{#3}{1}}{
			\roundgate[0][0][1][#5][#4]
		}{
			\foreach \n[evaluate=\n as \y using {2*\n-2}] in {1,...,#3}{
				\roundgate[0][\y][1][#5][#4]
			}
		}
	\end{scope}
}

\newcommandx{\tsfmatV}[8][1=0,2=0,3=l,4=4,5=tr,6=init,7=bertiniorange,8=topright]{
	\begin{scope}[shift={(#1,#2)}]
		\ifthenelse{\equal{#3}{l}}{
			\pgfmathsetmacro{\flag}{0}
		}{
			\pgfmathsetmacro{\flag}{1}
		}
		
		\foreach \y[evaluate=\y as \x using {mod(\y+\flag,2)}] in {1,...,#4}{
			\roundgate[\x][\y][1][#8][#7]
		}
		\ifthenelse{\equal{#5}{tr}}{
			\foreach \y[evaluate=\y as \x using {mod(\y+\flag,2)}] in {#4}{
				\draw [fill=white] (\x-0.5,\y+0.5) circle (0.15);
				\draw [fill=white] (\x+0.5,\y+0.5) circle (0.15);
			}
		}{}
		\ifthenelse{\equal{#6}{init}}{
			\thetastate[\flag][0][1][#7]
		}{}
	\end{scope}
}

\newcommandx{\leftriangle}[5][1=0,2=0,3=4,4=bertiniorange,5=topright]{
	\begin{scope}[shift={(#1,#2)}]
		\pgfmathsetmacro{\t}{#3}
		\pgfmathsetmacro{\steps}{ceil(\t/2)}
		\foreach \i[evaluate=\i as \x using -\t+2*\i-1, evaluate=\i as \ylim using \t-2*\i+2] in {1,...,\steps}{
			\foreach \y[evaluate=\y as \thisx using {\x+\y-1}] in {1,...,\ylim}{
				\roundgate[\thisx][\y][1][#5][#4][2]
			}
		}
	\end{scope}
}

\newcommandx{\rightriangle}[5][1=0,2=0,3=4,4=bertiniorange,5=topright]{
	\begin{scope}[shift={(#1,#2)}]
		\pgfmathsetmacro{\t}{#3}
		\pgfmathsetmacro{\steps}{ceil(\t/2)}
		\foreach \i[evaluate=\i as \x using -\t+2*\i-1, evaluate=\i as \ylim using \t-2*\i+2] in {1,...,\steps}{
			\foreach \y[evaluate=\y as \thisx using {-\x-\y+1}] in {1,...,\ylim}{
				\roundgate[\thisx][\y][1][#5][#4][-1]
			}
		}
	\end{scope}
}

\newcommandx{\eigenVL}[8][1=0,2=0,3=l,4=5,5=tr,6=init,7=bertiniorange,8=topright]{
	\begin{scope}[shift={(#1,#2)}]
		\pgfmathsetmacro{\t}{#4}
		\leftriangle[0][0][\t][#7][#8]
		
		\ifthenelse{\equal{#6}{init}}{
			\drawinitstate[0][0][l][\t][#7]
		}{}
		
		\ifthenelse{\equal{#5}{tr}}{
			\draw[fill=white] \foreach \x in {0,...,\t} {(\x-0.5-\t,0.5+\x) circle (0.15)};
			\ifthenelse{\equal{#3}{r}}{
				\draw[fill=white] (0.5,\t+0.5) circle (0.15);
			}{}
		}{}
		\ifthenelse{\equal{#5}{parttr}}{
			\draw[fill=white] \foreach \x in {0,...,\t} {(\x-0.5-\t,0.5+\x) circle (0.15)};
		}{}
	\end{scope}
}

\newcommandx{\eigenVR}[8][1=0,2=0,3=l,4=5,5=tr,6=init,7=bertiniorange,8=topright]{
	\begin{scope}[shift={(#1,#2)}]
		\pgfmathsetmacro{\t}{#4}
		\rightriangle[0][0][\t][#7][#8]
		
		\ifthenelse{\equal{#6}{init}}{
			\drawinitstate[0][0][r][\t][#7]
		}{}
		
		\ifthenelse{\equal{#5}{tr}}{
			\draw[fill=white] \foreach \x in {0,...,\t}{(-\x+0.5+\t,0.5+\x) circle (0.15)};
			\ifthenelse{\equal{#3}{l}}{
				\draw[fill=white] (-0.5,\t+0.5) circle (0.15);
			}{}
		}{}
	\end{scope}
}

\newcommandx{\tra}[2][1]{\underset{#1}{\text{tr}}\left[#2\right]}

\newcommandx{\tsfmatDgate}[7][1=0,2=0,3=l,4=4,5=tr,6=bertiniorange,7=topright]
{
	\begin{scope}[shift={(#1,#2)}]
		\ifthenelse{\equal{#3}{l}}{
			\pgfmathsetmacro{\flag}{-1}
		}{
			\pgfmathsetmacro{\flag}{1}
		}
		\pgfmathsetmacro{\t}{#4}
		\foreach \i[evaluate=\i as \x using {\flag*\i}, evaluate=\i as \y using \i] in {1,...,\t}{
			\roundgate[\x][\y][1][#7][#6]
		}
		
		\ifthenelse{\equal{#5}{tr}}{
			\foreach \i[evaluate=\i as \x using {\flag*\i}, evaluate=\i as \y using \i] in {\t}{
				\draw [fill=white] (\x-0.5,\y+0.5) circle (0.15);
				\draw [fill=white] (\x+0.5,\y+0.5) circle (0.15);
			}  
		}{}
	\end{scope}
	
}

\newcommandx{\tsfmatD}[8][1=0,2=0,3=l,4=4,5=tr,6=init,7=bertiniorange,8=topright]{
	\begin{scope}[shift={(#1,#2)}]
		\ifthenelse{\equal{#6}{init}}{
			\thetastate[0][0][1][#7]
		}{}
		
		\ifthenelse{\equal{#3}{l}}{
			\pgfmathsetmacro{\flag}{-1}
		}{
			\pgfmathsetmacro{\flag}{1}
		}
		
		\pgfmathsetmacro{\t}{#4}
		\foreach \i[evaluate=\i as \x using {\flag*\i}, evaluate=\i as \y using \i] in {1,...,\t}{
			\roundgate[\x][\y][1][#8][#7]
		}
		
		\ifthenelse{\equal{#5}{tr}}{
			\foreach \i[evaluate=\i as \x using {\flag*\i}, evaluate=\i as \y using \i] in {\t}{
				\draw [fill=white] (\x-0.5,\y+0.5) circle (0.15);
				\draw [fill=white] (\x+0.5,\y+0.5) circle (0.15);
			}  
		}
		\ifthenelse{\equal{#5}{parttr}}{
			\foreach \i[evaluate=\i as \x using {\flag*\i}, evaluate=\i as \y using \i] in {\t}{
				\draw [fill=white] (\x+0.5*\flag,\y+0.5) circle (0.15);
			}  
		}
		{}
	\end{scope}
}

\newcommandx{\drawinitstate}[5][1=0,2=0,3=l,4=4,5=bertiniorange]{
	\pgfmathsetmacro{\t}{#4}
	\begin{scope}[shift={(#1,#2)}]
		\pgfmathsetmacro{\steps}{ceil((\t-1)/2)}
		\ifthenelse{\equal{#3}{l}}{
			\foreach \i[evaluate=\i as \x using -\t+2*\i] in {0,...,\steps}{
				\thetastate[\x][0][1][#5]
			}
		}{
			\foreach \i[evaluate=\i as \x using -\t+2*\i] in {0,...,\steps}{      
				\thetastate[-\x][0][1][#5]
			}
		}
	\end{scope}
}

\newcommandx{\drawinitstateflipped}[5][1=0,2=0,3=l,4=4,5=bertiniorange]{
	\pgfmathsetmacro{\t}{#4}
	\begin{scope}[shift={(#1,#2)}]
		\pgfmathsetmacro{\steps}{ceil((\t-1)/2)}
		\ifthenelse{\equal{#3}{l}}{
			\foreach \i[evaluate=\i as \x using -\t+2*\i] in {0,...,\steps}{
				\thetastateflipped[\x][0][1][#5]
			}
		}{
			\foreach \i[evaluate=\i as \x using -\t+2*\i] in {0,...,\steps}{      
				\thetastateflipped[-\x][0][1][#5]
			}
		}
	\end{scope}
}

\newcommandx{\eigenDL}[6][1=0,2=0,3=l,4=4,5=bertiniorange,6=topright]{
	\begin{scope}[shift={(#1,#2)}]
		\pgfmathsetmacro{\t}{#4}
		\ifthenelse{\equal{#3}{l}}{
			\eigenVL[0][0][l][\t][tr][init][#5][#6]
			\pgfmathsetmacro{\t}{#4-1}
			\rightriangle[1][0][\t][#5][#6]
			\drawinitstate[1][0][r][\t][#5]
		}{
			\begin{scope}[shift={(-0.5,0.5)}]
				\foreach \i[evaluate=\i as \x using \i, evaluate=\i as \y using \i] in {0,...,\t}{      
					\draw (\x,\y)--++(0.5,0);
					\draw[fill=white] (\x,\y) circle (0.15);
				}
			\end{scope}
		}
	\end{scope}
}

\newcommandx{\eigenDR}[6][1=0,2=0,3=l,4=4,5=bertiniorange,6=topright]{
	\begin{scope}[shift={(#1,#2)}]
		\pgfmathsetmacro{\t}{#4}
		\ifthenelse{\equal{#3}{r}}{
			\eigenVR[0][0][r][\t][tr][init][#5][#6]
			\pgfmathsetmacro{\t}{#4-1}
			\leftriangle[-1][0][\t][#5][#6]
			\drawinitstate[-1][0][l][\t][#5]
		}{
			\begin{scope}[shift={(0.5,0.5)}]
				\foreach \i[evaluate=\i as \x using \i, evaluate=\i as \y using \t-\i] in {0,...,\t}{      
					\draw (\x,\y)--++(0.5,0);
					\draw[fill=white] (\x+0.5,\y) circle (0.15);
				}
			\end{scope}
		}
	\end{scope}
}

\newcommandx{\idonpurity}[2][1=0,2=0]
{
	\begin{scope}[shift={(#1,#2)}]
		\draw[thick] (-0.5,0)--++(-0.1,0.1)--++(0,0.2)--++(0.1,-0.1);
		\draw[thick] (-0.5,0.4)--++(-0.1,0.1)--++(0,0.2)--++(0.1,-0.1);
		\draw[thick] (0.5,0)--++(0.1,0.1)--++(0,0.2)--++(-0.1,-0.1);
		\draw[thick] (0.5,0.4)--++(0.1,0.1)--++(0,0.2)--++(-0.1,-0.1);
	\end{scope}
}

\newcommandx{\swaponpurity}[2][1=0,2=0]
{
	\begin{scope}[shift={(#1,#2)}]
		\draw[thick] (-0.5,0)--++(-0.2,0.2)--++(0,0.6)--++(0.2,-0.2);
		\draw[thick] (-0.5,0.2)--++(-0.075,0.075)--++(0,0.2)--++(0.075,-0.075);
		\draw[thick] (+0.5,0)--++(+0.2,0.2)--++(0,0.6)--++(-0.2,-0.2);
		\draw[thick] (+0.5,0.2)--++(+0.075,0.075)--++(0,0.2)--++(-0.075,-0.075);
	\end{scope}
}

\newcommandx{\hook}[4][1=0,2=0,3=t,4=l]{
	\begin{scope}[shift={(#1,#2)}]
		\ifthenelse{\equal{#3}{t}}{
			\ifthenelse{\equal{#4}{l}}{\draw[thick] (0.5,-0.5) arc (45:-90:0.15);}{\draw[thick] (0.5,-0.5) arc (45:270:0.15);}
		}{\ifthenelse{\equal{#4}{l}}{\draw[ thick] (0.5,-0.5) arc (-45:90:0.15);}{\draw[ thick] (0.5,-0.5) arc (315:90:0.15);}
		}
	\end{scope}
}

\newcommandx{\hhook}[4][1=0,2=0,3=t,4=l]{
	\begin{scope}[shift={(#1,#2)}]
		\ifthenelse{\equal{#3}{t}}{
			\ifthenelse{\equal{#4}{l}}{\draw[thick] (0.5,-0.5) arc (-45:175:0.15);}{\draw[thick] (0.5,-0.5) arc (225:0:0.15);}
		}{\ifthenelse{\equal{#4}{l}}{\draw[ thick] (0.5,-0.5) arc (-45:180:-0.15);}{\draw[ thick] (0.5,-0.5) arc (45:-180:0.15);}
		}
	\end{scope}
}

\newcommandx{\Pproj}[3][3=$P_\Lambda$]{
\begin{scope}[shift={(#1-.5,#2-1)}]
\draw[thick,fill=white] (0,0)rectangle (1,2);
\draw[thick] (0,1.5)--(-.5,1.5);
\draw[thick] (1,1.5)--(1.5,1.5);
\draw[thick] (0,.5)--(-.5,.5);
\draw[thick] (1,.5)--(1.5,.5);
\node[scale=2] at (.5,1) {#3};
\end{scope}}

\definecolor{FcolU}{rgb}{0.71,0.78,0.91}
\definecolor{colLines}{rgb}{0.31,0.31,0.31}
\definecolor{colVMPSLines}{rgb}{0.11,0.11,0.11}
\definecolor{IcolUc}{rgb}{0.71,0.41,0.42}
\definecolor{IcolU}{rgb}{0.71,0.8,0.76}
\definecolor{IcolVMPSc}{rgb}{0.73,0.69,0.7}
\definecolor{IcolVMPS}{rgb}{0.81,0.77,0.78}
\definecolor{colObs}{rgb}{1.,1.,1.}

\newcommandx{\eightlegs}[2][1=0,2=0]{
	\begin{scope}[shift={(#1,#2)}]
		\foreach \x in {1,...,8}{
			\draw (\x, 0)--++(0,0.25);
			\draw[fill] (\x,0) circle (0.05);
		}
		\foreach \x in {1,3}{
			\pgfmathsetmacro\result{2*\x-1} 
			\node () at (\result,-0.5) {$i_{\x}$};
			\pgfmathsetmacro\result{2*\x}
			\node () at (\result,-0.5) {$j_{\x}$};	
		}
		\foreach \x in {2,4}{
			\pgfmathsetmacro\result{2*\x} 
			\node () at (\result,-0.5) {$i_{\x}$};
			\pgfmathsetmacro\result{2*\x-1}
			\node () at (\result,-0.5) {$j_{\x}$};	
		}
	\end{scope}
}

\newcommandx{\MPSinitialstate}[5][1=0,2=0,3=bertiniorange,4=topright,5=-1]{
\begin{scope}[shift={(#1,#2)},rounded corners=1.5pt]
	\draw[black,thick,fill=#3] 
	(-0.25,.25)--++(.5,0)--++(0,-.3)--++(-.5,0)--cycle;
	\draw[thick] (-.25,.25)--++(-.25,.25);
	\draw[thick] (.25,.25)--++(.25,.25);
	\draw[very thick] (-1,.-.05)--++(2,0);
\ifthenelse{\equal{#5}{-1}}{
	\ifthenelse{\equal{#4}{topright}}{\draw[thick,rounded corners=0.3]
	(-.1,.15)--++(.2,0)--++(0,-0.1); }{}
	\ifthenelse{\equal{#4}{topleft}}{\draw[thick,rounded corners=0.3]
	(.1,.15)--++(-.2,0)--++(0,-0.1); }{}
	\ifthenelse{\equal{#4}{bottomleft}}{\draw[thick,rounded corners=0.3]
	(.1,-.15)--++(-.2,0)--++(0,0.1); }{}	\ifthenelse{\equal{#4}{bottomright}}{\draw[thick,rounded corners=0.3]
	(-.1,-.15)--++(.2,0)--++(0,0.1); }{}}{			\node at ({0},{0.085}) {\scalebox{1.}{{$#5$}}};}
\end{scope}
}

\newcommandx{\Cmatrix}[6][1=0,2=0,3=2,4=bertiniorange,5=,6=topright]{
	\pgfmathsetmacro\result{#3-1} 
	\begin{scope}[shift={(#1,#2)}]
		\foreach \i in {0,...,\result}
		{\foreach \j in {0,...,\i}
			{\roundgate[\i+\j][\i-\j][1][#6][#4]}
		}
		\ifthenelse{\equal{#5}{init}}{
			\foreach \i in {0,...,#3}
			{
				\MPSinitialstate[-1+2*\i][-1][#4]
			}
		}{}
	\end{scope}
}

\renewcommand{\bcirc}{\fineq[-0.5ex][0.7][1]{\cstate[0][0][][black]}}

\newcommandx{\cstate}[4][1=0,2=0,3= ,4=white]{
	\begin{scope}[shift={(0,0)}]
				\draw[fill=#4,thick] (#1,#2) circle (0.13);
				\node[scale=1.1] at (#1,#2) {$#3$};
\end{scope}
}
\newcommandx{\sqrstate}[4][1=0,2=0,3= ,4=white]{
	\begin{scope}[shift={(#1,#2)}]
		\draw[thick,fill=#4] (-0.13,-0.13) rectangle (0.13,0.13) ;
		\node[scale=1.1] at (0,0) {$#3$};
	\end{scope}
}
\newcommandx{\pairproduct}[2][1=0,2=0]
{\begin{scope}[shift={(#1 ,#2)}]
\draw[thick] (-.5,.5) arc(-135:-45:1/1.414);
\sqrstate[0][.5-.1414][][black]	
\end{scope}
}
\newcommandx{\bellpair}[2][1=0,2=0]
{\begin{scope}[shift={(#1 ,#2)}]
		\draw[thick] (-.5,.5) arc(-135:-45:1/1.414);
	\end{scope}
}
\newcommandx{\charge}[3][1=0,2=0,3=black]
{
	\ifthenelse{\equal{#3}{blue}}{\def \chargecolor{bertinigreyblue}
	}{
	\ifthenelse{\equal{#3}{red}}{\def \chargecolor{bertinigreyred}}{\def \chargecolor {#3}}}
\begin{scope}[shift={(#1 ,#2)}]
	\draw[ fill=\chargecolor] circle (0.08);        
\end{scope}
}
\newcommandx{\trianglediag}[7][1=0,2=0,3=1,4=bertiniorange,5= ,6=-1,7=topright]
{\begin{scope}[shift={(#1 ,#2)}]
	\foreach \i in {0,...,#3}
	{	\foreach \j in {0,...,\i}
		{	\roundgate[-\j+2*\i][\j][1][topright][#4][#6]
		}
	}
	\foreach \i in {-1,...,#3}
	{\ifthenelse{\equal{#5}{bellpair}}{\bellpair[\i*2+1][-1]}{\ifthenelse{\equal{#5}{pairproduct}}{\pairproduct[\i*2+1][-1]}{\MPSinitialstate[\i*2+1][-1][#4][#7][#6]}}}
\end{scope}
}
\newcommandx{\projectorleg}[4][1=0,2=0,3=R,4=left]
{
	\begin{scope}[shift={(#1 ,#2)}]
		{\ifthenelse{\equal{#3}{R}}{		\draw[thick] (-.25,-.25)--++(.5,.5);}{\ifthenelse{\equal{#3}{L}}{		\draw[thick] (-.25,.25)--++(.5,-.5);}{\draw[thick] (-.25,0)--++(.5,0);}}
		}
	\draw[thick, fill=white] circle (0.13);
	\ifthenelse{\equal{#4}{right}}{
		\draw[thick] (.0,.07)--++(.07,0)--++(0,-.07);
		\node[scale=0.5] at (0,-.02) {$\alpha$};
	}{}
	\ifthenelse{\equal{#4}{left}}{	
	\draw[thick] (0,.07)--++(-.07,0)--++(0,-.07);
	\node[scale=0.5] at (0,-.03) {$\beta$};
	}{}
	\end{scope}
}
\usetikzlibrary{patterns,decorations.pathreplacing}

\newcommand{\be}{\begin{equation}}
\newcommand{\ee}{\end{equation}}
\newcommand{\ea}{\end{aligned}}
\newcommand{\bea}{\begin{equation}\begin{aligned}}
\newcommand{\eea}{\end{aligned}\end{equation}}

\newcommand{\pif}[1]{\frac{\pi}{#1}}

\pgfplotsset{
    colormap={springpastels}{ rgb255=(253, 127, 111) rgb255=(126, 176, 213) rgb255=(178, 224, 97) rgb255=(189, 126, 190) rgb255=(255, 181, 90)  rgb255=(190, 185, 219) rgb255=(253, 204, 229) rgb255=(139, 211, 199)}
    }

\definecolorseries{foo}{hsb}{step}[hsb]{0,1,1}[hsb]{.618,0,0}

\newtheorem{property}{Property}

\newcommand{\Mcal}[0]{\mathcal{M}}

\let\oc\undefined
\newcommand{\oc}[0]{\ocircle}
\newcommand{\wt}[1]{\widetilde{#1}}
\newcommand{\kb}[2]{\ketbra*{#1}{#2}}

\definecolor{Zfcolor}{RGB}{194, 243, 255}
\definecolor{Zfccolor}{RGB}{255, 155, 155}
\definecolor{ZXfcolor}{RGB}{67,147,195}
\definecolor{ZXfccolor}{RGB}{255, 45, 55}
\newcommand{\Zfcolor}[0]{Zfcolor}
\newcommand{\ZXfcolor}[0]{ZXfcolor}
\newcommand{\Zfccolor}[0]{Zfccolor}
\newcommand{\ZXfccolor}[0]{ZXfccolor}

\definecolor{Rfcolor}{RGB}{44,162,95}

\newcommand{\Rscale}[0]{1.5}

\newcommandx{\rgate}[2][1=0,2=0]
{
\begin{scope}[shift={(#1 ,#2)}]
    \roundgate[0][-1][1][topright][\Afcolor]
    \roundgate[0][1][1][topright][\Afcolor]
    \roundgate[1][0][1][topright][\ZXfcolor][1]
    \roundgate[-1][0][1][topright][\ZXfcolor][1]
\end{scope}
}
\newcommandx{\layer}[4][1=0,2=0,3=0,4=bertiniorange]
{
\begin{scope}[shift={(#1 ,#2)}]
    \pgfmathparse{#3}
	\foreach \i in {0,...,\pgfmathresult}
	{	
        \roundgate[2*\i][0][1][topright][#4]
	}
\end{scope}
}
\newcommandx{\transfer}[7][1=0,2=0,3=0,4=n,5=bertiniorange,6=bertiniorange,7=1]
{
\begin{scope}[shift={(#1 ,#2)}]
    \pgfmathparse{#3}
    \ifthenelse{\equal{#4}{n}}
    {
    \roundgate[0][0][1][topright][#5][n]
    \roundgate[2*\pgfmathresult+2][0][1][topright][#5][n]
    \roundgate[1][1][1][topright][#6][n]
    }
    {
    \roundgate[0][0][1][#4][#5][1]
    \roundgate[2*\pgfmathresult+2][0][1][#4][#5][1]
    \roundgate[1][1][1][#4][#6][1]
    }
    \ifthenelse{\equal{#7}{1}}
    {
    \cstate[-.5][.5]
    \cstate[-.5][-.5]
    
    \ifthenelse{\equal{#4}{topright}}{
        \cstate[2*\pgfmathresult+2+0.9][.5]
        \cstate[2*\pgfmathresult+2+0.9][-.5]
    }{
        \sqrstate[2*\pgfmathresult+2+.5][.5]
        \sqrstate[2*\pgfmathresult+2+.5][-.5]
    }
    }{}
	\foreach \i in {1,...,\pgfmathresult}
	{	
        \ifthenelse{\equal{#4}{n}}
        {
        \roundgate[2*\i][0][1][topright][#6][n]
        \roundgate[2*\i+1][1][1][topright][#6][n]
        }
        {
        \roundgate[2*\i][0][1][#4][#6][1]
        \roundgate[2*\i+1][1][1][#4][#6][1]
        }
	}
\end{scope}
}
\newcommandx{\transferT}[7][1=0,2=0,3=0,4=n,5=bertiniorange,6=bertiniorange,7=1]
{
\begin{scope}[shift={(#1 ,#2)}]
    \pgfmathparse{#3}
    \ifthenelse{\equal{#4}{n}}
    {
    \roundgate[0][1][1][topright][#5][n]
    \roundgate[2*\pgfmathresult+2][1][1][topright][#5][n]
    \roundgate[1][0][1][topright][#6][n]
    }
    {
    \roundgate[0][1][1][#4][#5]
    \roundgate[2*\pgfmathresult+2][1][1][#4][#5]
    \roundgate[1][0][1][#4][#6]
    }
    \ifthenelse{\equal{#7}{1}}
    {
    \cstate[-.5][1.5]
    \cstate[-.5][.5]
    \sqrstate[2*\pgfmathresult+2+.5][1.5]
    \sqrstate[2*\pgfmathresult+2+.5][.5]
    }{}
	\foreach \i in {1,...,\pgfmathresult}
	{	
        \ifthenelse{\equal{#4}{n}}
        {
        \roundgate[2*\i][1][1][topright][#6][n]
        \roundgate[2*\i+1][0][1][topright][#6][n]
        }
        {
        \roundgate[2*\i][1][1][#4][#6]
        \roundgate[2*\i+1][0][1][#4][#6]
        }
	}
\end{scope}
}

\newcommandx{\rectangletwotone}[7][1=0,2=0,3=1,4=2,5=1,6=2, 7=]
{\begin{scope}[shift={(#1,#2)}]
    \pgfmathparse{#4 - 1}
    \foreach \i in {0,...,\pgfmathresult}
        {\pgfmathparse{#5 - 1}
        \foreach \j in {0,...,\pgfmathresult}
            {	
                \pgfmathparse{2*\i + 1}
                \pgfmathparse{\pgfmathresult/#6 - floor(\pgfmathresult/#6) < 0.001}
                \ifthenelse{\equal{\pgfmathresult}{0}}
                {\roundgate[2*\i][2*\j][#3][topright][\ZXfcolor][1]}
                {\roundgate[2*\i][2*\j][#3][topright][\Zfcolor][1]}

                \pgfmathparse{2*\i + 2}
                \pgfmathparse{\pgfmathresult/#6 - floor(\pgfmathresult/#6) < 0.001}
                \ifthenelse{\equal{\pgfmathresult}{0}}
                {\roundgate[2*\i+1][2*\j+1][#3][topright][\ZXfcolor][1]}
                {\roundgate[2*\i+1][2*\j+1][#3][topright][\Zfcolor][1]}
            }
        }
    \pgfmathparse{#4 - 1}
    \foreach \i in {0,...,\pgfmathresult}{
        \ifthenelse{\equal{#7}{bellpair}}
        {\bellpair[2*\i+1][-1]}
        {\MPSinitialstate[2*\i+1][-1][\ZXfcolor][1][]}
    }
\end{scope}
}
\newcommandx{\trianglediagtwotone}[7][1=0,2=0,3=1,4=2,5=1,6=,7=]
{\begin{scope}[shift={(#1,#2)}]
    \pgfmathparse{#4}
    \foreach \i in {0,...,\pgfmathresult}
    	{\foreach \j in {0,...,\i}
    		{	
                \pgfmathparse{1-\j+2*\i}
                \pgfmathparse{\pgfmathresult/#5 - floor(\pgfmathresult/#5) < 0.001}
                \ifthenelse{\equal{\pgfmathresult}{0}}
                {
                    \ifthenelse{\equal{#7}{topright}}{\roundgate[-\j+2*\i][\j][#3][topright][\ZXfcolor]}{\roundgate[-\j+2*\i][\j][#3][topright][\ZXfcolor][#7]}
                }
                {
                    \ifthenelse{\equal{#7}{topright}}{\roundgate[-\j+2*\i][\j][#3][topright][\Zfcolor]}{\roundgate[-\j+2*\i][\j][#3][topright][\Zfcolor][#7]}
                }
    		}
    	}
    \pgfmathparse{#4}
    \foreach \i in {-1,...,\pgfmathresult}{
        \ifthenelse{\equal{#6}{bellpair}}
        {\bellpair[2*\i+1][-1]}
        {
            \ifthenelse{\equal{#7}{topright}}{\MPSinitialstate[2*\i+1][-1][\ZXfcolor]}{\MPSinitialstate[2*\i+1][-1][\ZXfcolor][][#7]}
        }
    }
\end{scope}
}

\newcommandx{\trianglediagtwotonebounded}[9][1=0,2=0,3=1,4=2,5=1,6=,7=,8=0,9=100]
{\begin{scope}[shift={(#1,#2)}]
    \pgfmathparse{#4}
    \foreach \i in {0,...,\pgfmathresult}
    	{\foreach \j in {0,...,\i}
    		{	
                \pgfmathparse{1-\j+2*\i > #8}
                \ifthenelse{\equal{\pgfmathresult}{1}}
                {
                \pgfmathparse{1-\j+2*\i < #9}
                \ifthenelse{\equal{\pgfmathresult}{1}}
                {
                    \pgfmathparse{1-\j+2*\i}
                    \pgfmathparse{\pgfmathresult/#5 - floor(\pgfmathresult/#5) < 0.001}
                    \ifthenelse{\equal{\pgfmathresult}{0}}
                    {
                        \ifthenelse{\equal{#7}{topright}}{\roundgate[-\j+2*\i][\j][#3][topright][\ZXfcolor]}{\roundgate[-\j+2*\i][\j][#3][topright][\ZXfcolor][#7]}
                    }
                    {
                        \ifthenelse{\equal{#7}{topright}}{\roundgate[-\j+2*\i][\j][#3][topright][\Zfcolor]}{\roundgate[-\j+2*\i][\j][#3][topright][\Zfcolor][#7]}
                    }
                }{}
                }{}
    		}
    	}
    \pgfmathparse{#4}
    \foreach \i in {-1,...,\pgfmathresult}{
        \pgfmathparse{\i*2+1 > #8-1}
        \ifthenelse{\equal{\pgfmathresult}{1}}{
        \pgfmathparse{\i*2+1 < #9-1}
        \ifthenelse{\equal{\pgfmathresult}{1}}{
            \ifthenelse{\equal{#6}{bellpair}}
            {\bellpair[2*\i+1][-1]}
            {
                \ifthenelse{\equal{#7}{topright}}{\MPSinitialstate[2*\i+1][-1][\ZXfcolor]}{\MPSinitialstate[2*\i+1][-1][\ZXfcolor][][#7]}
            }
        }{}
        }{}
    }
\end{scope}
}

\begin{document}

\title{Asymptotically Solvable Quantum Circuits}

\author{Samuel H. Pickering}
\affiliation{School of Physics and Astronomy, University of Birmingham, Edgbaston, Birmingham, B15 2TT, UK}

\author{Bruno Bertini}
\affiliation{School of Physics and Astronomy, University of Birmingham, Edgbaston, Birmingham, B15 2TT, UK}
	
\begin{abstract}
The discovery of chaotic quantum circuits with (partially) solvable dynamics has played a key role in our understanding of non-equilibrium quantum matter and, at the same time, has helped the development of concrete platforms for quantum computation. It was shown that solvability does not prevent the generation of chaotic dynamics, however, it imposes non-trivial constraints on the generated correlations. A natural question is then whether it is possible to gain insight into the generic case despite the latter being very hard to access. To address this question here we introduce a family of `asymptotically solvable' quantum circuits where the solvability constraints only affect correlations on length scales beyond a tuneable threshold. This means that their dynamics are only solvable for long enough times: for times shorter than the threshold they are generic. We show this by computing both their dynamical correlations on the equilibrium (infinite temperature) state and their thermalisation dynamics following quantum quenches from compatible (asymptotically solvable) non-equilibrium initial states. The class of systems we introduce is generically ergodic but contains a non-interacting point, which we use to provide exact analytical results, complementing those of numerical experiments, on the non-solvable early time regime. 
\end{abstract}

\maketitle

\section{Introduction}

Coherent quantum many-body systems out of equilibrium are currently among the most actively researched topics in physics. They are ubiquitous in Nature --- occurring at scales ranging from astronomical to subatomic --- and understanding their dynamics can both fill long-standing gaps in our understanding of physics and yield to the discovery of new emergent phenomena. Specifically, some of the key questions include understanding the general principles underlying the onset of ergodicity and the emergence of classical behaviours at late times~\cite{eisert2015quantum, calabrese2016introduction, serbyn2021quantum, bastianello2022introduction}, determining the general laws controlling the spreading and scrambling of quantum information~\cite{calabrese2005evolution, hayden2007black, sekino2008fast, nahum2017quantum, shenker2014multiple, kitaev2015talk, hosur2016chaos, bertini2022growth}, classifying new phases occurring out of equilibrium~\cite{sacha2018time, khemani2019brief}, and characterising the effects of quantum noise or monitoring through measurements~\cite{skinner2019measurement, li2019measurement}.

Addressing these questions was essentially inconceivable until a few years ago as there was no efficient way to access non-equilibrium quantum many-body dynamics in the presence of interactions. The last decade has brought a sharp change in this state of affairs with the discovery of interacting many-body systems with (partially) accessible dynamics~\cite{potter2022entanglement, fisher2023random, bertini2025exactly, bertini2026non}. The latter take the form of  \emph{quantum circuits}, i.e., many-body systems in discrete space-time where the interactions are implemented by discrete unitary operations in a way that closely resembles the architecture of quantum processors (see, e.g., Ref.~\cite{arute2019quantum}). The solvable instances of quantum circuits have been obtained by either averaging under random unitary operations \`a la random matrix theory~\cite{potter2022entanglement, fisher2023random} or imposing constraints (or symmetries)~\cite{bertini2025exactly}.  Remarkably, one can impose these constraints and still observe ergodicity~\cite{bertini2019exact, yu2024hierarchical}, as well as other complex dynamical behaviours~\cite{klobas2021exact, bertini2022exact, klobas2026inprep}. For instance, the so called \emph{dual-unitary circuits}~\cite{bertini2019exact}, which exhibit a duality under the exchange of space and time, are generically ergodic.

Solvable quantum circuits played a key role in characterising the universal features of the dynamics, e.g., entanglement growth~\cite{nahum2017quantum, bertini2019entanglement, gopalakrishnan2019unitary, piroli2020exact, zhou2022maximal}, operator spreading~\cite{nahum2018operator, vonKeyserlingk2018operator, claeys2020maximum}, and (deep) thermalisation~\cite{piroli2020exact, ho2022exact, ippoliti2023dynamical}, of quantum many-body systems 
as well as yielding insight into the spectrum~\cite{chan2018solution, chan2018spectral, bertini2018exact, friedman2019spectral, bertini2021random} and eigenstates~\cite{fritzsch2021eigenstate, fritzsch2025eigenstate}. At the same time they also provided valuable benchmarks for real-world quantum computers~\cite{mi2021information, chertkov2022holographic, fischer2024dynamical} (see also Ref.~\cite{bertini2026mitigated}). To attain solvability, however, one has to introduce simplifications that inevitably limit the generality of the observed phenomenology. For instance, in the case of random circuits one attains solvability by computing averages, which are informative on single instances only when fluctuations are small~\cite{zhou2019emergent}. On the other hand, imposing constraints one can compute exact behaviours of single instances but correlations are restricted~\cite{bertini2019entanglement}. Therefore, it is important to understand what happens as the constraints are gradually lifted.

So far, both these routes have proven difficult to follow: random circuit fluctuations could be rigorously controlled only in certain asymptotic limits~\cite{zhou2019emergent} and while there has been success remixing the basic building blocks of dual-unitary circuits \cite{Kasim_2023, claeys_2024_hadamard, claeys2024from, rampp2025solvable} no systematic way to lift the solvability constraints has so far been found. A remarkable attempt is the one of Ref.~\cite{yu2024hierarchical}, which proposed a family of `hierarchical' generalisations of dual-unitary circuits where each member of the hierarchy encounters a constraint on the `space evolution' that is weaker than the previous one (dual-unitary circuits are at the bottom of the hierarchy as the most constrained ones). Whereas the first step away from dual-unitary circuits continues to be exactly solvable~\cite{foligno2024quantum, bertini2024exact} the subsequent ones are not efficiently treatable by any known means. 

Here we propose an alternative, we argue more effective, route to lift the space-time duality constraint: we introduce a family of \emph{asymptotically solvable} quantum circuits where the `unitarity'~\footnote{Or better, isometric property, see the discussion in Sec.~\ref{sec:solvabilityspaceevolution}.} of the space evolution is imposed only after a tuneable number of steps. We show that this implies that the circuits are only solvable for long enough time scales while their behaviour for shorter time scales is not affected by solvability constraints and departs substantially from that of solvable cases. 

In the rest of this introduction we present a more detailed discussion of the setting considered, recall the main ideas of solvability via space-time duality, and summarise our main results.

\subsection{Solvability Via Space-Time Duality}
\label{sec:solvabilityspaceevolution}

We consider local quantum circuits, i.e., a collection of $2L$ quantum variables with $d$ internal states --- \emph{qudits} --- that are arranged at the vertices of a one-dimensional chain and are evolved by discrete applications of a unitary operator implementing local interactions. In particular, we focus on cases where the evolution operator can be written as 
\be
\mathbb{U} =\mathbb{U}_o\mathbb{U}_e,\,\, \mathbb{U}_e=\bigotimes_{x=0}^{L-1} U_{x,x+1/2},\,\,\mathbb{U}_o=\bigotimes_{x=1}^{L} U_{x-1/2,x}\,,
\label{eq:floquetoperator}
\ee
where we labelled the sites by half integers and denoted by $A_{x,y}$ the operator acting non-trivially only on the qudits at $x$ and $y$. The unitary operator $U$ (which can in principle be different at different spatial points) is typically referred to as the \emph{local gate} and determines the dynamical properties of the system. Quantum dynamical systems whose evolution operator takes the form in Eq.~\eqref{eq:floquetoperator} are known as `brickwork quantum circuits'. We stress that, although we focus on this setting for the sake of simplicity, the concept of solvability via space-time duality can also be defined for quantum circuits in higher dimensions or with more general forms of local interactions~\cite{bertini2019exact, prosen2021many, jonay2021triunitary, suzuki2022computational, claeys2024from, rampp2025geometric, rampp2025solvable, bertini2025exactly, milbradt2023ternary, mestyan2024multi}.

The analysis of these systems is greatly aided by adopting Penrose's notation, whereby one represents the local gates as boxes with four legs protruding to represent the incoming and outgoing state of the qudits it acts on 
\be
U = \fineq[-0.8ex][0.75][1]{
    \roundgate[0][0][1][topright][orange1]
},\qquad U^\dag = \fineq[-0.8ex][0.75][1]{
    \roundgate[0][0][1][bottomright][green1]
}, 
\ee
and represents matrix multiplication by joining the appropriate legs~\footnote{Here we use the convention that matrix multiplication goes from bottom to top, i.e. $AB$ is represented with the symbol for $A$ \emph{above} the one for $B$.}. In fact, to study operator dynamics is more convenient to introduce a graphical notation for the `doubled' local gate $U \otimes U^*$, which can be used to implement Heisenberg evolution 
\be
\label{eq:foldedgate}
U \otimes U^* = \fineq[-0.8ex][0.75][1]{
\roundgate[0.25][0.15][1][topright][green1]
\roundgate[0][0][1][topright][orange1]
} = \fineq[-0.8ex][0.75][1]{
   \roundgate[0][0][1][topright][\ZXfcolor][1]
}, 
\ee
together with a symbol to represent contraction of indices pertaining to the two the two copies, i.e.,  
\be
\label{eq:bulletstate}
\ket*{\ocircle}=\sum_{i=1}^d \ket*{i}\otimes\ket*{i} = \fineq[-0.8ex][0.6][1]{
    \draw(0, 0)--++(0, .5);
    \cstate[0][0]
    }, \quad\,\,\, \bra*{\ocircle}=\sum_{i=1}^d \bra*{i}\otimes\bra*{i} = \fineq[-0.8ex][0.6][1]{
    \draw(0, 0)--++(0, .5);
    \cstate[0][.5]
    }. 
\ee
In this way, we can conveniently represent the unitarity of the local gate by means of the following graphical identities
\be
\label{eq:unitarity}
\fineq[-0.8ex][0.7][1]{
  \roundgate[0][0][1][topright][\ZXfcolor][1]
     \cstate[-0.5][0.5]
    \cstate[0.5][0.5]
}=\fineq[-0.8ex][0.7][1]{
    \draw(-.5, -.5)--++(0,1.0);
    \draw(.5,  -.5)--++(0,1.0);
    \cstate[-0.5][0.5]
    \cstate[0.5][0.5]
},\qquad \fineq[-0.8ex][0.7][1]{
  \roundgate[0][0][1][topright][\ZXfcolor][1]
     \cstate[-0.5][-0.5]
    \cstate[0.5][-0.5]
}=\fineq[-0.8ex][0.7][1]{
    \draw(-.5, -.5)--++(0,1);
    \draw(.5,  -.5)--++(0,1);
    \cstate[-0.5][-0.5]
    \cstate[0.5][-0.5]
}.
\ee
We refer the reader to Ref.~\cite{bertini2025exactly} for an extended discussion of this graphical notation. 

\begin{figure}
\includegraphics[width=1.0\columnwidth]{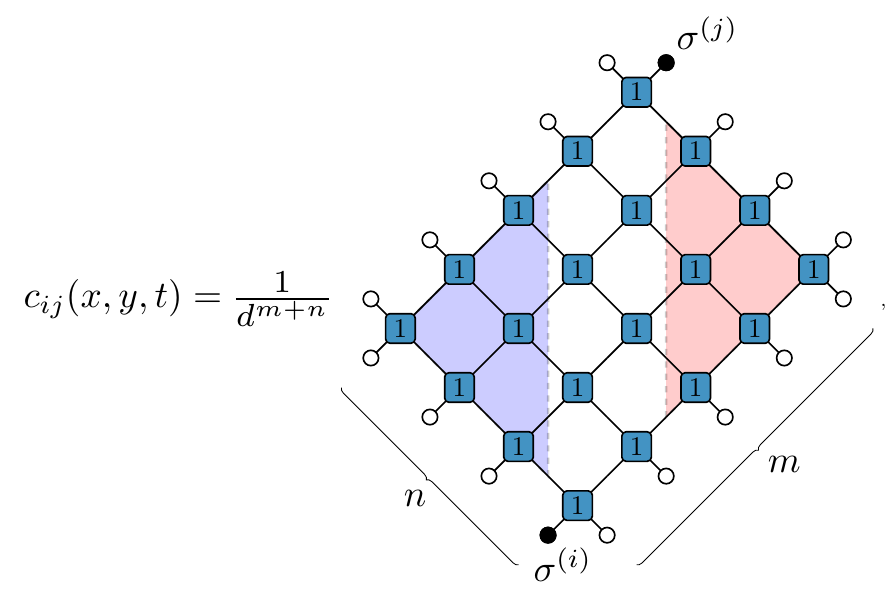}
\caption{Example of the tensor network resulting from a dynamical correlation function. Shaded in blue and red are respectively the left and right fixed points of the space evolution (influence matrices). These objects encode the effects of the surrounding system upon the local region containing the two operators.}
\label{fig:correlationcircuit}
\end{figure}

As hinted at before, a convenient way to identify solvable systems in this setting is to exchange the roles of space and time, i.e., consider the evolution in the spatial direction rather than the temporal one, see, e.g., Ref.~\cite{bertini2025exactly}. To illustrate this point explicitly we follow Ref.~\cite{bertini2019exact} and consider the example of dynamical correlation functions on the infinite temperature state, which is generically the only invariant state in a brickwork circuit. The latter can be expressed as  
\be
\label{eq:twopointcorrelation}
c_{ij}(x,y,t)={\frac{\tr[\sigma^{(i)}_x \mathbb{U}^{-t}\sigma^{(j)}_y \mathbb{U}^t ]}{\tr[\mathds{1}_{2L}]}}, 
\ee
where $\mathds{1}_{x}$ represents the identity operator in $\mathbb C^{d^x}$ and $\{\sigma^{(i)}\}_{i=0}^{d^2-1}$ (${\sigma^{(0)}=\mathds{1}_1}$) denotes the set of generalised Gell-Mann matrices: a basis of ${\rm End}(\mathbb C^{d})$ that is orthonormal under the Hilbert-Schmidt scalar product, i.e.
\be
\braket*{\sigma^{(i)}}{\sigma^{(j)}}\equiv\frac{1}{d}\tr[\sigma^{(i)}\sigma^{(j)\dagger}]=\delta_{i,j}\,. 
\ee
Using the unitarity conditions in Eq.~\eqref{eq:unitarity} one can easily show that the correlations $c_{ij}(x,y,t)$ are restricted to a causal light cone as depicted in Fig~\ref{fig:correlationcircuit}. For typical circuits, the computational difficulty in calculating the correlation function in Eq.~\eqref{eq:twopointcorrelation} grows exponentially in the minimum of the two light cone separations $\min(t\pm |x-y|)$. This increasing complexity is due to the increasingly complicated effect the environment has upon the subsystem containing the operators. The shaded regions in Fig.~\ref{fig:correlationcircuit} indicate the tensors, often referred to as influence matrices~\cite{lerose2020influence}, responsible for encoding this effect. Solvable circuits can be identified in the cases where these tensors take simple forms.

\begin{widetext}
\begin{figure*}
\centering
\includegraphics{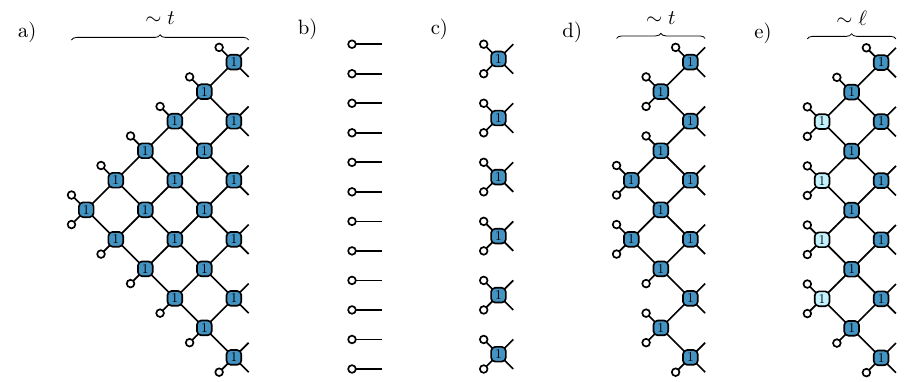}
\caption{The types of fixed points/influence matrices seen in solvable and non-solvable circuits. a) the fixed points seen in generic gates, with no simplification the growing width with time implies there is no efficient representation of the state. b) the infinite temperature product state that comes from dual-unitarity (DU). c) the pair-correlated state the results from DU2. d) the fixed point resulting from DU3 that still has a width that scales with time. e) the proposal in this paper, to insert inhomogeneities into an otherwise generic circuit of type (a) that truncate the fixed point to always have finite width.}
\label{fig:fixedpoint}
\end{figure*}
\end{widetext}

The influence matrices can be seen as the evolution of the infinite temperature state in space (i.e. evolving with the gates acting from left to right or right to left). We can bring some simple physical intuition to this process to help understand when the influence matrices will be simple objects. To begin with, one should consider the objects appearing at either the left or the right corner of the diagram. Focussing on the former for definiteness we have  
\be
\label{eq:cornerdiagram}
\begin{aligned}
\fineq[-0.8ex][0.7][1]{
  \roundgate[0][0][1][topright][\ZXfcolor][1]
    \cstate[-0.5][-0.5]
    \cstate[-0.5][0.5]
}
=
\fineq[-0.8ex][0.7][1]{
    \draw(-.5, -.5)--++(1.0, 0);
    \draw(-.5,  .5)--++(1.0, 0);
    \cstate[-0.5][-0.5]
    \cstate[-0.5][0.5]
}
+
\sum_{i,j=1}^{d^2-1} A_{ij}
\fineq[-0.8ex][0.7][1]{
    \draw(-.5, -.5)--++(1.0, 0);
    \draw(-.5,  .5)--++(1.0, 0);
	\node[scale=1.3] at (-1.1,-.5)  {$\sigma^{(j)}$};
	\node[scale=1.3] at (-1.1,.5)  {$\sigma^{(i)}$};
    \cstate[-0.5][-0.5][][black]
    \cstate[-0.5][0.5][][black]
},
\end{aligned}
\ee
where we introduced 
\be
A_{ij} \equiv 
\frac{1}{d^2} \fineq[-0.8ex][0.7][1]{
  \roundgate[0][0][1][topright][\ZXfcolor][1]
    \cstate[-0.5][-0.5]
    \cstate[-0.5][0.5]
    \cstate[0.5][-0.5][][]
    \cstate[0.5][0.5][][]
	\node[scale=1.5] at (1.45,-.5)  {$(\sigma^{(j)})^*$};
	\node[scale=1.5] at (1.45,.5)  {$(\sigma^{(i)})^*$};
},
\ee
and
\be
\sum_{i,j=1}^d (\sigma^{(k)})_{ij} \ket*{i}\otimes\ket*{j} = \fineq[-0.2ex][0.6][1]{
    \draw(0, 0)--++(0, .5);
    \node[scale=1.5] at (0.5,-0.25)  {$\sigma^{(k)}$};
    \cstate[0][0][][black]
    }\!.
\ee
Physically, Eq.~\eqref{eq:cornerdiagram} describes one step of spatial evolution of the infinite temperature state. The evolution then continues by repeated applications of rectangular matrices of the form 
\be
\fineq[-0.8ex][0.6][1]{
\foreach \j in {0,...,3}{
\roundgate[0][2*\j][1][topright][\ZXfcolor][1]      
}
\draw [decorate, decoration = {brace}]   (-.65,.5)--++(0,5);
\node[scale=1.3] at (-1.1,3) {$2m$};
\cstate[-0.5][-0.5]
\cstate[-0.5][6.5]
}\,,
\ee
where $m=1,2,\ldots$ denotes the step. Since spatial evolution is generically not unitary, the infinite temperature state is not invariant. Instead, it produces a complicated superposition of operators that proliferate as $m$ increases. These operator strings build correlations between the different sites of the influence matrix that represent the same qubit at different times. As such they represent the environment acting in a non-Markovian manner on the subsystems containing $\sigma^{(i)}$ and $\sigma^{(j)}$. By this we mean that it is a bath capable of storing information from previous time steps --- i.e.\ it has a memory. This memory requires the temporal state representing the environment (cf.\ Figs.~\ref{fig:correlationcircuit} and~\ref{fig:fixedpoint}) to have a growing Schmidt rank. Therefore, it is the proliferation of non-trivial operator strings that a solvable circuit must restrict.

The simplest restriction on the non-trivial operators strings is to ask the infinite temperature state to be invariant also under space evolution, i.e., set to zero all the coefficients $A_{ij}$ in Eq.~\eqref{eq:cornerdiagram}. This request is equivalent to asking the spatial evolution to be unitary, like the temporal one, and the local gates fulfilling this have been dubbed `dual-unitary' (DU) gates~\cite{bertini2019exact}. 
For DU gates one finds that all two-point correlations vanish inside the light cone and are only non-zero on the light cone edge, where they can be easily determined in terms of a finite-dimensional quantum channel~\cite{bertini2019exact}. More generally, the DU property implies that several aspects of non-equilibrium dynamics can be characterised exactly~\cite{bertini2019entanglement, gopalakrishnan2019unitary, piroli2020exact, claeys2020maximum, bertini2020scrambling,kos2023scrambling, foligno2022growth, giudice2021temporal, foligno2023temporal, bertini2020operator, bertini2020operator2, reid2021entanglement, ho2022exact, claeys2022emergentquantum, breach2025solvable}, see also the recent review~\cite{bertini2025exactly}. 

To generate richer patterns of correlations --- and hence  go beyond the example of DU gates --- one has to consider cases where the infinite temperature state produces extra operator strings. 
Ref.~\cite{yu2024hierarchical} suggested that a way to do this is to impose `hierarchical conditions', known as DU$n$, that effectively restrict the lifetimes of the unwanted operators. Their general form reads as 
\be
\label{eq:dun_intro}
\begin{aligned}
\fineq[-0.8ex][0.7][1]{
    \roundgate[1][1][1][topright][\ZXfcolor][1]
    \roundgate[2][2][1][topright][\ZXfcolor][1]
    \roundgate[4][4][1][topright][\ZXfcolor][1]
    \draw [thick, black, dotted] (2.5,2.5) -- (3.5,3.5);
    \cstate[0.5][0.5]
    \cstate[0.5][1.5]
    \cstate[1.5][2.5]
    \cstate[3.5][4.5]
    \draw [decorate, decoration = {brace}]   (4.65,3.5) -- (1.5,.35);
    \node[scale=1.5] at (3.5,1.75) {$n$};
}
=
\fineq[-0.8ex][0.7][1]{
    \roundgate[1][2][1][topright][\ZXfcolor][1]
    \roundgate[3][4][1][topright][\ZXfcolor][1]
    \draw [thick, black, dotted] (1.5,2.5) -- (2.5,3.5);
    \draw(.5, .5)--++(1.0, 0);
    \cstate[.5][0.5]
    \cstate[.5][1.5]
    \cstate[.5][2.5]
    \cstate[2.5][4.5]
    \draw [decorate, decoration = {brace}]   (3.65,3.5) -- (1.5,1.35);
    \node[scale=1.5] at (3.35,2.25) {$n-1$};
}.
\end{aligned}
\ee
The physical interpretation of these conditions is that the amplitude of the non-identity operators that are produced from the infinite temperature state vanish after $n-1$ spatial steps provided they do not interact with other operator strings. This then allows for certain gates to be removed from the influence matrix as if they were DU. For $n>2$, however, the strings produced at one step can interact with those produced at later steps and avoid annihilation: this results in a growth of complexity (cf.\ Fig.~\ref{fig:fixedpoint}) that can be quantified by the scaling of temporal entanglement~\cite{banuls2009matrix, lerose2020influence, foligno2023temporal}. Applying a minimal cut bound~\cite{casini2016spread} to the temporal state after maximal simplification via Eq.~\eqref{eq:dun_intro}, cf.~Fig.~\ref{fig:fixedpoint}, we get the following estimate for the scaling of the temporal entanglement entropy 
\be
S_{TE} \propto \left(\frac{n-2}{n}\right) t, \qquad n\geq 2. 
\ee
The above scaling suggests that, while the ${{\rm DU}n=2}$ condition leads to restricted two-point correlations~\cite{yu2024hierarchical} and treatable dynamics~\cite{foligno2024quantum}, ${{\rm DU}n>2}$ is not solvable.

In summary, a solvable circuit where the environment has a tuneable amount of memory, i.e.\ a tuneable bond dimension, is missing in the literature (although cases with finite bond dimension have been found in certain integrable permutation circuits~\cite{klobas2021exact,klobas2021exact2, klobas2026inprep}).

\subsection{Main results}
\label{sec:mainres}

\begin{figure}
\includegraphics[width=1.0\columnwidth]{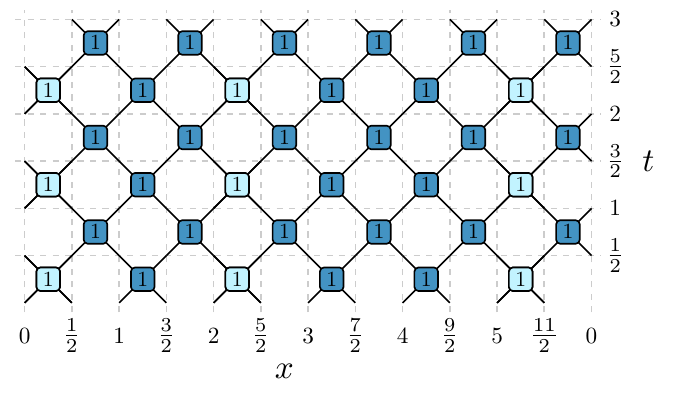}
\caption{Example of the structure of the circuit. Pictured here is $(\mathbb{U}\otimes\mathbb{U}^*)^t$ the folded evolution operator applied $t=3$ times. The evolution operator acts on $2L=12$ qubits positioned on the half integers. The light blue gates represent positions at which $s_x=0$. Here we placed them at $x=0, 2, 5$ but our arguments hold for arbitrary positions of these inhomogeneities.  
}
\label{fig:circuitexample}
\end{figure}

In this paper we propose an alternative mechanism to contrast the proliferation of non-identity operators under spatial evolution. Instead of annihilating non-trivial operators only after they evolve freely for a number of steps, we control their proliferation in such a way that they can be always fully annihilated by the application of a different gate. Namely, for $n=0,1,2,\dots$ we propose local relations of the form 
\be
\label{eq:inhom_condition}
\begin{aligned}
&\fineq[-0.8ex][0.7][1]{
    \roundgate[0][0][1][topright][\Zfcolor][1]
    \roundgate[1][1][1][topright][\ZXfcolor][1]
    \roundgate[3][3][1][topright][\ZXfcolor][1]
    \roundgate[4][4][1][topright][\Zfcolor][1]
    \cstate[-0.5][-0.5]
    \cstate[-0.5][0.5]
    \cstate[0.5][1.5]
    \cstate[2.5][3.5]
    \cstate[3.5][4.5]
       \draw [thick, black, dotted] (1.5,1.5) -- (2.5,2.5);
     \draw [decorate, decoration = {brace}]   (3.65,2.5) -- (1.5,1.35-1);
    \node[scale=1.5] at (3,1.25) {$n$};
}
=
\fineq[-0.8ex][0.7][1]{
    \roundgate[0][4][1][topright][\Zfcolor][1]
    \draw(-.5, -.5)--++(1.0, 0);
    \draw(-.5, .5)--++(1.0, 0);
    \draw(-.5, 2.5)--++(1.0, 0);
    \cstate[-.5][-0.5]
    \cstate[-.5][0.5]
    \cstate[-.5][2.5]
    \cstate[-.5][3.5]
    \cstate[-.5][4.5]
    \node[scale=1.5] at (-.5,1.75) {$\vdots$};
    \draw [decorate, decoration = {brace}]   (.65,2.65)--++(0,-2.35);
\node[scale=1.5] at (1.1,1.5) {$n$};
}.
\end{aligned}
\ee
The conditions in Eq.~\eqref{eq:inhom_condition} produce a generalisation of the DU condition that is not part of the hierarchy of Ref.~\cite{yu2024hierarchical} and, to be applicable, require a inhomogeneous circuit as the one depicted in Fig.~\ref{fig:circuitexample}.  The position of the inhomogeneities does not matter as long as their density is finite in the thermodynamic limit. Note that the inclusion of the $n=0$ case implies that the light blue gates are DU2 which we will adopt as their moniker. 

Before proceeding it is useful to make two observations: (i) One should not associate the $n$ in Eq.~\eqref{eq:inhom_condition} with that in Eq.~\eqref{eq:dun_intro} as in the former all gates but the final one are removed on the left hand side. For this reason, and what we discuss in Sec.~\ref{sec:asymptotic_to_du2}, the conditions for $n>0$ can be interpreted as asymptotic DU2 conditions. (ii) If the light blue gates are chosen to be invertible when acting in space, one can remove them from Eq.~\eqref{eq:inhom_condition} to obtain an equation where all the gates are removed for identities. This would then imply dual-unitarity of both the light and dark blue gates. Therefore, in order for Eq.~\eqref{eq:inhom_condition} to be non-trivial one should consider light blue gates that are DU2 but not DU.

We show that Eq.~\eqref{eq:inhom_condition} can be solved by a family of gates $U_s$ that are dependent on a continuous parameter $s$ for which we define
\be
\label{eq:identification}
\fineq[-0.8ex][0.7][1]{
  \roundgate[0][0][1][topright][\ZXfcolor][1]
}= U_{s\neq0} \otimes U_{s\neq0}^*, \qquad \fineq[-0.8ex][0.7][1]{
    \roundgate[0][0][1][topright][\Zfcolor][1]
}= U_{s=0} \otimes U_{s=0}^*, 
\ee
where for circuits on qubits ($d=2$) the gate $U_s$ can be expressed as
\be
\label{eq:asysolvable} 
\begin{aligned}
&U_s = (w_-\otimes w_+) e^{i\frac{\pi}{4} (Z\otimes Z + s Y\otimes Y)},\\
&w_\pm = e^{ih_\pm Z} e^{i\frac{\pi}{4} X}\!\!,
\end{aligned}
\ee
with $X,Y$ and $Z$ Pauli matrices and $s \in[0,1]$.

This family generally entails non-integrable interactions and interpolates between DU2 and DU gates by varying $s$ from 0 to $1$. Interestingly, the gates in this family maximise the average amount of entanglement produced when they are applied to product states, i.e., have the maximal `entangling power'~\cite{zanardi2000entangling, aravinda_2021} among two-qubit gates. In Appendix~\ref{app:gateformulations}, we show that the time evolution operator generated by the gates in Eq.~\eqref{eq:asysolvable} (with arbitrary values of $s_x\in[0,1]$ and $h_\pm=h_x$ at each position $x=0,1/2, \ldots$) can be expressed as an inhomogeneous kicked Ising model (KIM), see Refs.~\cite{prosen2002general, prosen2007chaos}. Namely, it can be written as 
\be
\mathbb{U} = {\rm T}\!\exp[i \int_0^1 {\rm d}\tau H(\tau)],
\ee
in terms of the following time-periodic Hamiltonian 
\begin{align}
    &H(t) = H_I(t) + \sum_{m=-\infty}^{\infty} \delta(t-m/2) H_K\,, \label{eq:timedepham1} \\
    &H_I(t) = \pif{4}\sum_{x} ((s_x\!-\!1) \Theta_{x+t}+1)\, Z_{x-\frac{1}{2}} Z_{x} +\! \sum_{x} h_x Z_{x}\,, \notag \\
    &H_K = \pif{4} \sum_{x} X_{x}\,, \notag
\end{align}
where $s_x\in[0,1]$, we used the shorthand notation 
\be
\sum_x f(x) \equiv \sum_{x\in \{0,1/2,1,\ldots, L-1/2\}} f(x),
\ee
and introduced the periodic step function
\bea
    \Theta_y &= 
    \begin{cases}
    0, & \lfloor 2y \rfloor\,\, {\rm even} \\
    1, & \lfloor 2y \rfloor\,\, {\rm odd}
    \end{cases}.
\eea
Note that, in contrast with the regular KIM, the Hamiltonian in Eq.~\eqref{eq:timedepham1} has modulated Ising terms. When one chooses $h_x=0$ this model is may be mapped to free fermions with a Jordan-Wigner transformation (cf. Appendix~\ref{app:diagonalisation}).

The family in Eq.~\eqref{eq:asysolvable} can be generalised to arbitrary $d$ by introducing the $d \times d$ `clock' $Z_d$ and `shift' $X_d$ operators fulfilling 
\be
X_d Z_d = \omega   Z_d X_d,
\ee
with $\omega=e^{i 2\pi/d}$, and reducing to the Pauli $Z$ and $X$ matrices for $d=2$ (cf.\ App.~\ref{sec:clockalgebra}). In terms of these we can set
\be
\label{eq:asysolvablelarged}
\hspace{-1.5cm}
\begin{aligned}
&U_s =    (w_-\!\otimes\! w_+)  e^{\frac{i\pi}{2d} s \sum_{j=1}^{d} X_d^j \otimes X_d^j} \left(\sum_{j=1}^d  \ketbra{j}{j}\otimes Z_d^j\right)\!\!,\\
&w_\pm = e^{i \sum_{j=1}^{d-1} h_{j,\pm} Z^j_q} H_q ,
\end{aligned}\hspace{-1cm}
\ee
where $H_q$ is the generalised Hadamard matrix (cf.\ App.~\ref{sec:clockalgebra}). Making the identification in Eq.~\eqref{eq:identification} these continue to solve Eq.~\eqref{eq:inhom_condition} while they reduce to those in Eq.~\eqref{eq:asysolvablelarged} for $d=2$ modulo global phases. Note that for $d>4$ this family does not contain anymore a dual-unitary point (cf.\ App.~\ref{sec:proofgend}).

\begin{figure}
\includegraphics[width=1.0\columnwidth]{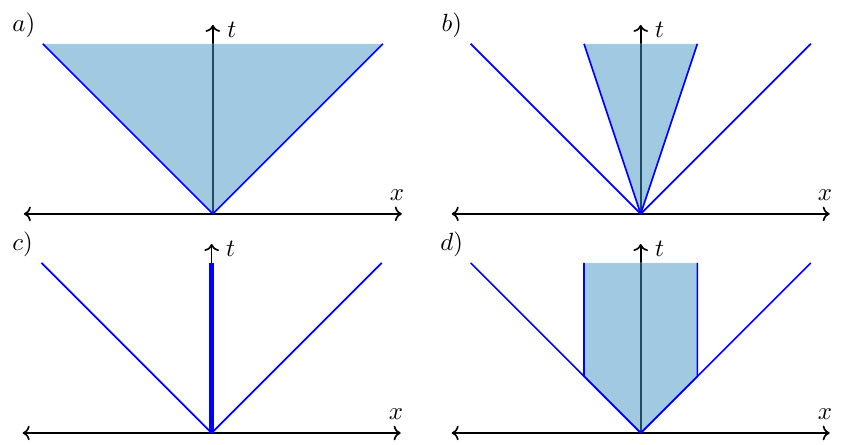}
\caption{Schematic of the dynamical correlations $c_{ij}(x,0,t)$ seen in the different types of circuits. a) Generic non-solvable dynamics are only constrained by the unitarity of the gates giving a simple lightcone support to the correlations. b) Partially restricted yet non-solvable dynamics such as those seen in DU3 circuits where the correlations exist on the $|x|=t$ line and within a steeper interior lightcone. c) An example of a solvable circuit where the correlations are restricted to individual lines, in this case $|x|=0, t$ as seen in DU2 circuits. d) The support of correlations seen in the asymptotically solvable circuits seen in this paper. Correlations spread in a generic manner up until the inhomogeneities of the circuit restrict them. The only correlations outside of these regions exist along the $|x|=t$ lines.}
\label{fig:correlationsschematic}
\end{figure}

Moving on to physical properties, we show exactly that in inhomogeneous circuits built of gates fulfilling Eq.~\eqref{eq:inhom_condition} dynamical correlations are supported both on the light cone edge and within a region of a configurable width, giving them what we will refer to as a ``dagger shape'' (cf.~Fig~\ref{fig:correlationsschematic} (d)) and derive conditions for their ergodicity. Moreover, we identify a family of compatible initial states for which we can characterise the thermalisation dynamics of local observables and the growth of entanglement between local subsystems for times larger than $\ell^*$, the maximal distance between two DU2 (light blue) gates in Fig.~\ref{fig:circuitexample}. Importantly, we show that for large times the slope of entanglement growth divided by $\log d$ --- the ``entanglement velocity'' --- becomes independent of the R\'enyi index and is determined by the DU2 gates. We also show that these circuits have a DU2-like entanglement membrane, see Refs.~\cite{jonay2018coarse, zhou2019emergent, zhou2020entanglement}. Due to these properties we dub the circuits made of the gates in Eq.~\eqref{eq:inhom_condition} ``asymptotically solvable'' circuits. 

Despite the emergence of solvability at long times, for $\ell^* \gg t$, these circuits can substantially depart from the DU2 behaviour and show generic physics. This is true for both dynamical correlations, which are nontrivial in the full causal light cone, and for quench dynamics. For instance, the entanglement velocity does generically depend on the R\'enyi index at short times, which is a sign of absence of dual-unitary physics. Interestingly, however, our numerical results suggest that the entanglement velocity in the regime $\ell^*\gg t \gg 1$ is bounded from below by that of the circuit where all gates are DU2.

To shed more light on these early-time behaviours we consider the non-interacting limit ($h_x=0$) of the Hamiltonian in Eq.~\eqref{eq:timedepham1}. We show that in this case the time evolution operator can be written as the product of three mutually commuting terms, i.e.,   
\be
\mathbb{U} = e^{i\pif{2} \sum_{x} X_{x}} \mathbb U_{1} \mathbb U_{2}, 
\label{eq:floquet_free}
\ee
with
\be 
\begin{aligned}
&\mathbb U_{1} = e^{i\pif{4}\sum_{x=1}^{L} Y_{x}Y_{x+\frac{1}{2}}} e^{i\pif{4}\sum_{x=1}^{L} Z_{x-\frac{1}{2}}Z_{x}},\\
&\mathbb U_{2} = e^{i\pif{4} \sum_{x=1}^{L}s_x Y_{x-\frac{1}{2}}Y_{x}} e^{i \pif{4} \sum_{x=1}^{L} s_{x+\frac{1}{2}} Z_{x}Z_{x+\frac{1}{2}}}\!.
\end{aligned}
\ee
The first of these operators is dual unitary (its spectrum coincides with that of the self-dual kicked Ising model~\cite{akila2016particle, bertini2018exact} at the non-interacting point), whilst $\mathbb U_2$ produces correlations spanning the full causal light cone. Of course, this clearly changes if at some position $x$ we have $s_x=0$, in which case $\mathbb U_2$ cannot propagate information across $x$. If there is more than one point with $s_x=0$ then $\mathbb U_2$ acts non-trivially only within the subsystems enclosed between each two of these points.

The key observation is that $\mathbb U_1$ and $\mathbb U_2$ act on disjoint Hilbert subspaces (cf.\ Appendix~\ref{app:diagonalisation}), which means that they \emph{propagate correlations independently}. 
This explains the entanglement growth being bounded from below by the DU2 circuit: the modes of $\mathbb U_1$ always move at the maximal speed and their contribution is independent of $s_x$, while the modes of $\mathbb U_2$ are completely static at the DU2 point $s=0$ (they get reflected by the impurities) and then contribute positively when $s\neq0$. In fact, this free-fermionic picture also provides a heuristic interpretation for the diagrammatic relations in Eq.~\eqref{eq:inhom_condition}. Remarkably, the results described continue to hold also beyond the non-interacting point, where no decomposition of the form Eq.~\eqref{eq:floquet_free} exists. Instead, one must return to the circuit formulation where the structure of the dynamics results from the local relations Eq.~\eqref{eq:asysolvable} that hold irrespective of the longitudinal fields $h_x$.

The rest of the paper is structured as follows. In Sec.~\ref{sec:system}, we study the local relations. Namely, we show that the local gates in Eq.~\eqref{eq:identification}, with $U_s$ in Eq.~\eqref{eq:asysolvablelarged}, fulfil Eq.~\eqref{eq:inhom_condition} and discuss the relation between the latter and the DU2 condition of Ref.~\cite{yu2024hierarchical}. In Sec.~\ref{sec:correlations} we consider the behaviour of dynamical correlations in these circuits and discuss their ergodicity properties. In Sec.~\ref{sec:quenchdynamics} we discuss the quench dynamics of asymptotically solvable circuits by introducing an appropriate class of asymptotically solvable states. We show that the physical picture emerging for dynamical correlations also occurs for post-quench correlations and entanglement dynamics. We also discuss the $\ell^*\gg t$ regime where the circuit undergoes generic dynamics and is thus difficult to describe.  Finally, in Sec.~\ref{sec:conclusions} we present our conclusions and a discussion of future work.  Some further, technical details, and proofs are reported in the appendixes.

\section{Local Relations}
\label{sec:system}

Let us begin considering the properties of the local gates in Eqs.~\eqref{eq:asysolvable} and~\eqref{eq:asysolvablelarged}: first we show that they do indeed fulfil the diagrammatic relation in Eq.~\eqref{eq:inhom_condition}, then we discuss the relation between Eq.~\eqref{eq:inhom_condition} and the DU2 condition.  

\subsection{Informal proof of Eq.~\eqref{eq:inhom_condition}}
\label{sec:proofd2}

To help give a simple understanding of where the conditions Eq.~\eqref{eq:asysolvable} fit into our previous discussion we give a summary of their proof in ${d=2}$. The actual proof in the general case follows similar lines and is reported in Appendix~\ref{sec:proofgend}. Introducing the diagrammatic notation 
\be
\fineq[-0.8ex][0.75][1]{
\roundgate[0][0][1][topright][white][$Z$]}=e^{i\frac{\pi}{4} Z\otimes Z}, \quad 
\fineq[-0.8ex][0.75][1]{
\roundgate[0][0][1][bottomright][grey2][$Z$]}= e^{-i\frac{\pi}{4} Z\otimes Z},
\ee
we have 
\be
\fineq[-0.8ex][0.75][1]{
\roundgate[0][0][1][topright][white][$Z$]
\roundgate[-1][0][1][topright][grey2][$Z$]
} = \fineq[-0.8ex][0.7][1]{
    \draw(-.25, -.5)--++(0,1.0);
    \draw(.25,  -.5)--++(0,1.0);
} + \fineq[-0.8ex][0.7][1]{
    \draw(-.25, -.5)--++(0,1.0);
    \draw(.25,  -.5)--++(0,1.0);
    \cstate[-0.25][0][][]
     \cstate[0.25][0][][]
    \node[scale=1.3] at (-.65,0)  {$Z$};
    \node[scale=1.3] at (.65,0)  {$Z$};
} = \fineq[-0.8ex][0.7][1]{
    \draw(-.5, -.25)--++(1.0, 0);
    \draw(-.5,  .25)--++(1.0, 0);
} + \fineq[-0.8ex][0.7][1]{
    \draw(-.5, -.25)--++(1.0, 0);
    \draw(-.5,  .25)--++(1.0, 0);
       \cstate[0][0.25][][]
     \cstate[0][-0.25][][]
    \node[scale=1.3] at (0,-.65)  {$Z$};
    \node[scale=1.3] at (0,.65)  {$Z$};
}.
\ee
The first equality may be easily computed using 
\be
e^{i\frac{\pi}{4}Z\otimes Z} = \frac{1}{\sqrt{2}}(\mathbbm{1}\otimes\mathbbm{1} + iZ\otimes Z) \,.
\ee 
The second equality can be verified by observing that the matrix elements coincide. Analogously one can show that for the off-diagonal elements of the operator basis $\alpha = X, Y$
\be
\fineq[-2.2ex][0.75][1]{
\roundgate[0][0][1][topright][white][$Z$]
\roundgate[-1][0][1][topright][grey2][$Z$]
    \cstate[-0.5][.5][][]
    \node[scale=1.3] at (-.5,.85)  {$\alpha$};
} = 0, \qquad
\fineq[0.8ex][0.75][1]{
\roundgate[0][0][1][topright][white][$Z$]
\roundgate[-1][0][1][topright][grey2][$Z$]
    \cstate[-0.5][-.5][][]
    \node[scale=1.3] at (-.2,-.75)  {$\alpha$};
} = 0
\label{eq:operator_deaths}
\ee
Multiplying the top left and right legs respectively by $w_+^\dag$ and $w_+$ and folding the gray gate below the white the second relation above is represented as
\be
\begin{aligned}
\fineq[-0.8ex][0.7][1]{
    \roundgate[0][0][1][topright][\Zfcolor][1]
    \cstate[-0.5][-0.5][][]
    \cstate[-0.5][0.5]
	\node[scale=1.4] at (-1.1,-.5)  {$\alpha$};
}
= {0},
\end{aligned}
\label{eq:operator_deaths_folded}
\ee
Now, to understand the effect of adding $e^{i\pif{4}s Y\otimes Y}$ we define
\bea
\fineq[-0.8ex][0.75][1]{
\roundgate[0][0][1][topright][red5]}&=e^{i\pif{4}(Z\otimes Z + s Y\otimes Y)}, \\
\fineq[-0.8ex][0.75][1]{
\roundgate[0][0][1][bottomright][green2]}&= (e^{i\pif{4}(Z\otimes Z + s Y\otimes Y)})^\dag,
\eea
and using  
\be
e^{i \frac{\pi s}{4} Y\otimes Y} =  \cos(\frac{\pi s}{4}) \mathbbm{1}\otimes\mathbbm{1} + i \sin(\frac{\pi s}{4}) Y \otimes Y,  
\ee
we find 
\begin{align}
\fineq[-1ex][0.7][1]{
    \roundgate[0][0][1][topright][red5]
    \roundgate[-1][0][1][topleft][green2]
}
&= \cos^2(\frac{\pi s}{4}) 
\fineq[-.8ex][0.7][1]{
    \roundgate[-1][0][1][topright][grey2][$Z$]
    \roundgate[0][0][1][topright][white][$Z$]
}
+ \sin^2(\frac{\pi s}{4}) 
\fineq[-2.2ex][0.7][1]{
    \roundgate[-1][0][1][topright][grey2][$Z$]
    \roundgate[0][0][1][topright][white][$Z$]
    \draw[thick](.5,  .5)--++(0.1, 0.1);
    \draw[thick](-1.5,  .5)--++(-0.1, 0.1);
    \cstate[.4][0.4][][]
    \cstate[-1.4][0.4][][]
    \node[scale=1.3] at (-1.5,.85)  {$Y$};
    \node[scale=1.3] at (.5,.85)  {$Y$};
} \notag\\
&= 
\fineq[-.4ex][0.7][1]{
    \draw(-.5, -.25)--++(1.0, 0);
    \draw(-.5,  .25)--++(1.0, 0);
} 
+\cos(\frac{\pi s}{2}) 
\fineq[-.4ex][.7][1]{
    \draw(-.5, -.25)--++(1.0, 0);
    \draw(-.5,  .25)--++(1.0, 0);
    \cstate[0][0.25][][]
    \cstate[0][-0.25][][]
    \node[scale=1.3] at (0,.7)  {$Z$};
    \node[scale=1.3] at (0,-.65)  {$Z$};
}.
\end{align}
where the some terms have cancelled due to Eq.~\eqref{eq:operator_deaths}.
Multiplying the top left and right legs respectively by $w_+^\dag$ and $w_+$ and folding the green gate below the orange the above relation is represented as
\be
\label{eq:properties_pauli}
\fineq[-0.8ex][0.7][1]{
  \roundgate[0][0][1][topright][\ZXfcolor][1]
    \cstate[-0.5][-0.5]
    \cstate[-0.5][0.5]
}
=
\fineq[-0.8ex][0.7][1]{
    \draw(-.5, -.5)--++(1.0, 0);
    \draw(-.5,  .5)--++(1.0, 0);
    \cstate[-0.5][-0.5]
    \cstate[-0.5][0.5]
}+ 
\cos(\frac{\pi s}{2})
\fineq[-0.8ex][0.7][1]{
    \draw(-.5, -.5)--++(1.0, 0);
    \draw(-.5,  .5)--++(1.0, 0);
	\node[scale=1.3] at (-1.1,-.5)  {$Z$};
	\node[scale=1.3] at (-1.1,.5)  {$\widetilde{Y}$};
    \cstate[-0.5][-0.5][][black]
    \cstate[-0.5][0.5][][black]
},
\ee
where we set 
\bea
\widetilde{Y} &= w_+ Z w_+^\dag = \cos(2h_+) Y + \sin(2h_+) X \\
\widetilde{X} &= w_+ X w_+^\dag = \cos(2h_+) X - \sin(2h_+) Y 
\ea
\label{eq:tildeops}
\ee
For generic $s$ these gates will not satisfy the conditions shown in Eq.~\eqref{eq:operator_deaths_folded}. However, instead of setting the amplitude of the operator strings to zero, they do the next best thing and confine the operators to a subspace. By this we mean that the following overlap vanishes
\bea
\hspace{-0.3cm}
\fineq[-0.8ex][0.7][1]{
  \roundgate[0][0][1][topright][\ZXfcolor][1]
    \cstate[-0.5][-0.5][][]
    \cstate[-0.5][0.5]
    \cstate[0.5][0.5][][]
    \node[scale=1.5] at (-0.2,-.7)  {$\alpha$};
    \node[scale=1.3] at (0.9,.5)  {$Z$};
}
&= \tr_1 [U^\dag_s (\mathbbm{1}\otimes Z) U_s (\alpha\otimes\mathbbm{1})] \\
&= - \tr_1 [e^{-i\pif{4}Z\otimes Z}(\mathbbm{1}\otimes Y) e^{i\pif{4}Z\otimes Z} (\alpha\otimes\mathbbm{1})]\\
&= 0
\eea
where again $\alpha = X, Y$ and we indicate taking a partial trace over the left spin with $\tr_1$. The first equality uses the fact that $w_\pm^\dag Z w_\pm = -Y$ and that $[e^{i\pif{4} Y\otimes Y}, \mathbbm{1}\otimes Y] = 0$. The last equality holds as a consequence of Eq.~\eqref{eq:operator_deaths}. This restriction means that no matter how many darker gates are applied in a chain, the uppermost component of the operator string will not leave the $\alpha = X, Y$ subspace, i.e. 
\be
\begin{aligned}
\fineq[-0.8ex][0.6][1]{
  \roundgate[0][0][1][topright][\ZXfcolor][1]
    \roundgate[1][1][1][topright][\ZXfcolor][1]
    \roundgate[3][3][1][topright][\ZXfcolor][1]
    \cstate[4-0.5][4-0.5][][]
    \cstate[-0.5][-0.5][][]
    \cstate[-0.5][0.5]
    \cstate[0.5][1.5]
    \cstate[2.5][3.5]
    \draw [thick, black, dotted] (1.5,1.5) -- (2.5,2.5);
    \node[scale=1.5] at (-1,-.75)  {$\alpha$};
    \node[scale=1.5] at (4,3.5)  {$\beta$};
         \draw [decorate, decoration = {brace}]   (3.65,2.5) -- (.5,.35-1);
    \node[scale=1.5] at (2.5,.75) {$n$};
}
= {0},
\qquad
\begin{split}
    \alpha &\in \{X, Y\}, \\
    \beta &\in \{\mathbbm{1},\, Z\}.
\end{split}
\label{eq:properties3}
\end{aligned}
\ee
An immediate consequence of this is that the following holds
\be
\begin{aligned}
\fineq[-0.8ex][0.6][1]{
  \roundgate[0][0][1][topright][\ZXfcolor][1][1]
    \roundgate[1][1][1][topright][\ZXfcolor][1]
    \roundgate[3][3][1][topright][\ZXfcolor][1]
    \roundgate[4][4][1][topright][\Zfcolor][1]
    \cstate[-0.5][-0.5][][]
    \cstate[-0.5][0.5]
    \cstate[0.5][1.5]
    \cstate[2.5][3.5]
    \cstate[3.5][4.5]
    \draw [thick, black, dotted] (1.5,1.5) -- (2.5,2.5);
    \node[scale=1.5] at (-1,-.75)  {$\alpha$};
         \draw [decorate, decoration = {brace}]   (3.65,2.5) -- (.5,.35-1);
    \node[scale=1.5] at (2.5,.75) {$n$};
}
= {0},
\qquad
\alpha \in \{ X, Y\}.
\label{eq:properties3}
\end{aligned}
\ee
Considering now the l.h.s.\ of Eq.~\eqref{eq:inhom_condition}, repeatedly applying Eq.~\eqref{eq:properties_pauli} (the first time with $s=0$, then for $s\neq0$) and Eq.~\eqref{eq:properties3} we readily find the r.h.s., thereby concluding the proof.

\subsection{Corollaries of the relations in Eq.~\eqref{eq:inhom_condition}}
\label{sec:graphical_identity_subeqs}
The local relations in Eq.~\eqref{eq:inhom_condition} appear to only allow simplifications of the non-DU2 gates between two impurities. However, if we take them and contract on both sides with the folded identity on the bottom leg
\be
\begin{aligned}
&\fineq[-0.8ex][0.7][1]{
    \roundgate[0][0][1][topright][\Zfcolor][1]
    \roundgate[1][1][1][topright][\ZXfcolor][1]
    \roundgate[3][3][1][topright][\ZXfcolor][1]
    \roundgate[4][4][1][topright][\Zfcolor][1]
    \cstate[-0.5][-0.5]
    \cstate[.5][-0.5]
    \cstate[-0.5][0.5]
    \cstate[0.5][1.5]
    \cstate[2.5][3.5]
    \cstate[3.5][4.5]
       \draw [thick, black, dotted] (1.5,1.5) -- (2.5,2.5);
     \draw [decorate, decoration = {brace}]   (3.65,2.5) -- (1.5,1.35-1);
    \node[scale=1.5] at (3,1.25) {$n$};
}
=
\fineq[-0.8ex][0.7][1]{
    \roundgate[0][4][1][topright][\Zfcolor][1]
    \draw(-.5, -.5)--++(1.0, 0);
    \draw(-.5, .5)--++(1.0, 0);
    \draw(-.5, 2.5)--++(1.0, 0);
    \cstate[.5][-0.5]
    \cstate[-.5][-0.5]
    \cstate[-.5][0.5]
    \cstate[-.5][2.5]
    \cstate[-.5][3.5]
    \cstate[-.5][4.5]
    \node[scale=1.5] at (-.5,1.75) {$\vdots$};
    \draw [decorate, decoration = {brace}]   (.65,2.65)--++(0,-2.35);
\node[scale=1.5] at (1.1,1.5) {$n$};
}
\end{aligned}
\ee
we may then use the unitarity condition (Eq.~\eqref{eq:unitarity}) in order to remove the bottom left gate such that we are left with
\be
\label{eq:inhom_condition_subeq}
\begin{aligned}
&\fineq[-0.8ex][0.7][1]{
    \roundgate[1][1][1][topright][\ZXfcolor][1]
    \roundgate[3][3][1][topright][\ZXfcolor][1]
    \roundgate[4][4][1][topright][\Zfcolor][1]
    \cstate[0.5][0.5]
    \cstate[0.5][1.5]
    \cstate[2.5][3.5]
    \cstate[3.5][4.5]
       \draw [thick, black, dotted] (1.5,1.5) -- (2.5,2.5);
     \draw [decorate, decoration = {brace}]   (3.65,2.5) -- (1.5,1.35-1);
    \node[scale=1.5] at (3,1.25) {$n$};
}
=
\fineq[-0.8ex][0.7][1]{
    \roundgate[0][4][1][topright][\Zfcolor][1]
    \draw(-.5, .5)--++(1.0, 0);
    \draw(-.5, 2.5)--++(1.0, 0);
    \cstate[-.5][0.5]
    \cstate[-.5][2.5]
    \cstate[-.5][3.5]
    \cstate[-.5][4.5]
    \node[scale=1.5] at (-.5,1.75) {$\vdots$};
    \draw [decorate, decoration = {brace}]   (.65,2.65)--++(0,-2.35);
\node[scale=1.5] at (1.1,1.5) {$n$};
}.
\end{aligned}
\ee
This then shows us that the non-DU2 gates cancel telescopically into the DU2 impurity gates even when they are not sandwiched between two impurities.

\subsection{Relation between Eq.~\eqref{eq:inhom_condition} and DU2}
\label{sec:asymptotic_to_du2}

Considering the bulk of a circuit where the columns of light blue gates ($U_{s=0}$) are arranged in a regular spatial sub-lattice, i.e.\ every $2\ell$ sites, one can recognise `coarse grained gates' acting on $4\ell$ qudits and fulfilling the DU2 condition. Namely, centring ourselves about a column of light blue gates, we can consider 
\be
\label{eq:RGstep}
\fineq[-0.8ex][0.7 * \Rscale][1]{
    \roundgate[0][0][1][topright][yellow1][1]
}
\equiv
\fineq[-0.4ex][0.7][1]{
    \roundgate[0][2][1][topright][\Zfcolor][1]
    \roundgate[0][0][1][topright][\Zfcolor][1]
    \roundgate[0][-2][1][topright][\Zfcolor][1]
    
    \roundgate[-1][1][1][topright][\ZXfcolor][1]
    \roundgate[1][1][1][topright][\ZXfcolor][1]
    \roundgate[-2][0][1][topright][\ZXfcolor][1]
    \roundgate[2][0][1][topright][\ZXfcolor][1]
    \roundgate[-1][-1][1][topright][\ZXfcolor][1]
    \roundgate[1][-1][1][topright][\ZXfcolor][1]
    
    \draw [decorate, decoration = {brace}]   (2.65,-.65) -- (.65,-2.65);
    \node[scale=1.5] at (2,-2) {$2\ell$};
},
\ee
and, with the use of Eq.~\eqref{eq:inhom_condition}, show that they fulfil the DU2 condition 
\be
\label{eq:du2_renorm}
\begin{aligned}
\fineq[-0.8ex][0.6*\Rscale][1]{
    \roundgate[1][1][1][topright][yellow1][1]
    \roundgate[2][2][1][topright][yellow1][1]
    \cstate[0.5][0.5]
    \cstate[0.5][1.5]
    \cstate[1.5][2.5]
}
=
\fineq[-0.8ex][0.6*\Rscale][1]{
    \roundgate[1][2][1][topright][yellow1][1]
    \draw(.5, .5)--++(1.0, 0);
    \cstate[.5][0.5]
    \cstate[.5][1.5]
    \cstate[.5][2.5]
}.
\end{aligned}
\ee
This property holds for arbitrary $\ell$ and regardless of the choice of free parameters used in the other gates. 

Therefore, there exists a simple renormalisation group flow --- or tensor blocking procedure --- 
\be
\fineq[-0.8ex][.7][1]{
		\roundgate[0][0][1][topright][\ZXfcolor][1]
} \longmapsto
\fineq[-0.4ex][0.7][1]{
    \roundgate[0][2][1][topright][\ZXfcolor][1]
  \roundgate[0][0][1][topright][\ZXfcolor][1]
    \roundgate[0][-2][1][topright][\ZXfcolor][1]
    
    \roundgate[-1][1][1][topright][\ZXfcolor][1]
    \roundgate[1][1][1][topright][\ZXfcolor][1]
    \roundgate[-2][0][1][topright][\ZXfcolor][1]
    \roundgate[2][0][1][topright][\ZXfcolor][1]
    \roundgate[-1][-1][1][topright][\ZXfcolor][1]
    \roundgate[1][-1][1][topright][\ZXfcolor][1]
    
    \draw [decorate, decoration = {brace}]   (2.65,-.65) -- (.65,-2.65);
    \node[scale=1.5] at (2,-2) {$\ell$};
}, 
\ee
under which infinite, $2\ell$-site translationally invariant asymptotically solvable circuits approach DU2 circuits in one step (and remain DU2 thereafter as the latter is a fixed point). This property, however, does not mean that the dynamics of asymptotically solvable circuits immediately reduces to that of DU2. In essence, this is because non-equilibrium dynamics involves non-trivial time boundaries. An example of this will be the dynamics of entanglement between the halves of the system: if the cut is not directly centred of a light blue gate then the entanglement growth cannot be calculated by the application of Eq.~\eqref{eq:du2_renorm} alone and as we will show only becomes asymptotically equivalent to the DU2 result only in the $t\gg\ell$ limit. 

It is also worth emphasising that the above argument cannot be applied in the absence of strict translation symmetry, i.e., when the separation between blue columns varies throughout the circuit. In this case one still observes the eventual emergence of DU2 physics, but the time-scale for it to occur is different at different spatial points and is difficult to be characterised by simple coarse graining.

\section{Correlation Functions}
\label{sec:correlations}


In this section we demonstrate that for asymptotically solvable circuits the correlations inside the causal light cone take the dagger shape where they are exactly confined to a vertical strip with the only leakage allowed along the $|x-y|=t$ line. This allows all the correlation functions to be described by the repeated application of a fixed channel.

\begin{figure}
\includegraphics[width=1.0\columnwidth]{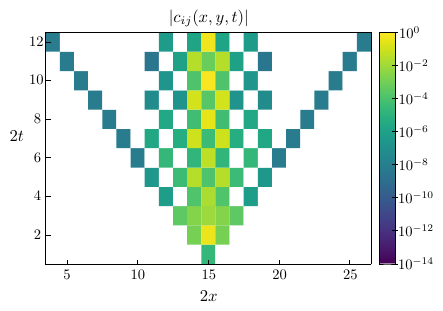}
\caption{A correlation function between two identical random traceless Hermitian operators supported on 4 sites. The position of the second operator is fixed with its leftmost spin at $2y=15$. All gates use a small homogeneous longitudinal field $h=\pi/800$. Gates with $s_j=0$ act on the $13$th and $19$th bonds in the system, all other gates use $s_j = \pi/8$. As a result, inside the light cone we see non-trivial correlations at all times when the two operators $|x-y|\leq4$.}
\label{fig:correlations_numerics}
\end{figure}

Let us consider circuits built of the gates in Eq.~\eqref{eq:asysolvablelarged} with a position dependent parameter $s_x\in[0,1]$. The only requirement we make on the distribution $\{s_x\}$ is that the distance between two zeros, where we place a light blue gate, does not scale with the system size. Choosing without loss of generality $s_0=0$ we let $\ell$ be the position of the next zero such that $s_\ell = 0$. Considering $\ell=2$ for clarity, we can express the dynamical correlation between two operators in $[0,\ell]$ as
\begin{widetext}
\be
\label{eq:inhom_correlation_diagram}
\begin{aligned}
c_{ij}(x, y, t) &= \frac{1}{d^{2t+1}}
\fineq[-0.8ex][0.6][1]{
    \foreach \i in {0,...,3}
	{
        \foreach \j in {0,...,5}
    	{
            \pgfmathparse{\i+\j}
            \ifthenelse{\equal{\pgfmathresult}{2.0}}
            {\roundgate[\i+\j][-\i+\j][1][topright][\Zfcolor][1]}
            {
                \ifthenelse{\equal{\pgfmathresult}{6.0}}
                {\roundgate[\i+\j][-\i+\j][1][topright][\Zfcolor][1]}
                {\roundgate[\i+\j][-\i+\j][1][topright][\ZXfcolor][1]}
            }
        }
        \cstate[\i-0.5][-\i-0.5]
        \cstate[\i+5+0.5][-\i+5+0.5]
    }
    \foreach \j in {0,...,5}
    {
        \cstate[\j-0.5][\j+0.5]
        \cstate[\j+3+0.5][\j-3-0.5]
    }
    \cstate[3.5][-3.5][][black]
	\node[scale=1.5] at (4, -4)  {$\sigma^{(i)}$};
    \cstate[5.5][5.5][][black]
	\node[scale=1.5] at (6, 6)  {$\sigma^{(j)}$};
}
=\frac{1}{d^{2t+1}}
\fineq[-0.8ex][0.6][1]{
    \foreach \i in {0,...,3}
	{
        \foreach \j in {0,...,5}
    	{
            \pgfmathparse{\i+\j}
            \ifthenelse{\equal{\pgfmathresult}{2.0}}
            {\roundgate[\i+\j][-\i+\j][1][topright][\Zfcolor][1]}
            {
                \ifthenelse{\equal{\pgfmathresult}{6.0}}
                {\roundgate[\i+\j][-\i+\j][1][topright][\Zfcolor][1]}
                {
                    \pgfmathparse{\i+\j < 2.0}
                    \ifthenelse{\equal{\pgfmathresult}{1}}
                    {}
                    {
                    \pgfmathparse{\i+\j > 6.0}
                    \ifthenelse{\equal{\pgfmathresult}{1}}
                    {}
                    {\roundgate[\i+\j][-\i+\j][1][topright][\ZXfcolor][1]}
                    }
                }
            }
        }
    }
    \foreach \j in {0,...,3}
    {
        \cstate[\j+1.5][\j+2.5]
        \cstate[1.5][1.5-\j]
        \cstate[\j+3+0.5][\j-3-0.5]
        \cstate[6+0.5][\j+1-0.5]
    }
    \cstate[1.5][-2.5]
    \cstate[2.5][-3.5]
    \cstate[3.5][-3.5][][black]
	\node[scale=1.5] at (4, -4)  {$\sigma^{(i)}$};
    \cstate[5.5][5.5][][black]
    \cstate[6.5][4.5]
	\node[scale=1.5] at (6, 6)  {$\sigma^{(j)}$};
},
\end{aligned}
\ee
where we have applied our identities in Eq.~\eqref{eq:inhom_condition} and \eqref{eq:inhom_condition_subeq} to reduce all gates outside the zeros to being effectively DU gates. There is generally no further reduction that can take place. This means we must include the full region in-between the nearest zeros to the operators. This length scale then corresponds to the finite memory of the environment. One can also verify that if the two operators are between different pairs of zeros then the correlation is identically zero unless they are separated by a light cone distance $|x-y|=t$. Namely, denoting by $\{\ell_1,\ldots,\ell_{\mathcal N}\}$ the positions of the zeros in $\{s_x\}$, we have
\be
\!\!\!c_{ij}(x, y, t) = \delta_{y-x,\pm t}c_{ij}(x, x\pm t, t) +\!\! \sum_{m=1}^{\mathcal N}
\frac{\chi_{[\ell_{m},\ell_{m+1}]}(x)\chi_{[\ell_{m},\ell_{m+1}]}(y)}{d^{2(\ell_{m+1}-\ell_{m})}}
\mel*{\underbrace{\ocircle\cdots \ocircle\sigma^{(j)}}_{2y-\ell_m}\ocircle\cdots\ocircle}{T_{\ell_{m+1}-\ell_m}^t}{\underbrace{\ocircle\ocircle\sigma^{(i)}}_{2x-\ell_m}\ocircle\cdots\ocircle},
\ee
\end{widetext}
where the sign in the first term is $+$ if $x$ is integer and $-$ if is half odd integer, while in the second term we indicated by $\chi_{A}(x)$ the characteristic function of the interval $A$ and we introduced the channel 
\be
\label{eq:transfer_1rep}
\begin{aligned}
T_\ell = \frac{1}{d^2} \,
\fineq[-0.8ex][0.6][1]{
    \roundgate[0][0][1][topright][\Zfcolor][1]
    \roundgate[1][1][1][topright][\ZXfcolor][1]
    \roundgate[2][0][1][topright][\ZXfcolor][1]
    \roundgate[3][1][1][topright][\ZXfcolor][1]
    \roundgate[4][0][1][topright][\Zfcolor][1]
    \cstate[-0.5][-0.5]
    \cstate[-0.5][0.5]
    \cstate[4.5][-0.5]
    \cstate[4.5][0.5]

    \draw [decorate, decoration = {brace}]   (3.5,-.7) -- (0.5,-.7);
    \node[scale=1.5] at (2,-1.3) {$2\ell$};
}\,.
\end{aligned}
\ee
Similar statements can be made for larger operator strings as well (as long as the operator size is smaller than $2\ell$). Therefore, we can conclude that all non-zero correlation functions between two local operators on the interior of each other's light cone are determined by the channels in Eq.~\eqref{eq:transfer_1rep}. An example of such a correlation function is pictured in Fig~\ref{fig:correlations_numerics}. 

As pictured in Eq.~\eqref{eq:transfer_1rep}, the transfer matrix $T_\ell$ functions as a unitary evolution $\mathbb{U}\otimes\mathbb{U}^*$ that has been projected onto the sub-sector where the environment both starts and ends in the infinite temperature state, i.e., 
\be
    T_\ell=\frac{1}{d^2}(\bra*{\ocircle}\otimes \mathbbm{1}_{\ell-2}\otimes\bra*{\ocircle})\mathbb{U}\otimes\mathbb{U}^*(\ket*{\ocircle}\otimes \mathbbm{1}_{\ell-2}\otimes\ket*{\ocircle}).
\ee
A rephrasing of this is that the action of the rest of the system upon the subsystem is completely depolarising and Markovian. In the context of brickwork circuits, such channels have been considered when studying systems with explicitly DU environments~\cite{Kos_2021}. 

It is easily shown that the eigenvalues of this matrix lie on the unit disk (i.e.\ it is non-expanding). For any normalised $\ket*{v}$
\be
\begin{aligned}
|\!\mel{v}{T_\ell}{v}\! | &= |\!\mel{\ocircle v\ocircle}{\mathbb{U}\otimes\mathbb{U}^*}{\ocircle v\ocircle}\! | \\
&\leq
\left \| 
\fineq[-0.8ex][0.55][1]{
    \roundgate[1][1][1][topright][\ZXfcolor][1]
    \roundgate[3][1][1][topright][\ZXfcolor][1]
}
\right \| 
\!\cdot\!
\left \| 
\fineq[-0.8ex][0.55][1]{
    \roundgate[0][0][1][topright][\Zfcolor][1]
    \roundgate[2][0][1][topright][\ZXfcolor][1]
    \roundgate[4][0][1][topright][\Zfcolor][1]
}
\right \|\\
& \leq 1.
\end{aligned}
\ee
On top of this, as a consequence of unitarity, the transfer matrix has a trivial leading eigenvalue that is the infinite temperature state 
\be
T_\ell\ket*{\ocircle\dots\ocircle} = \ket*{\ocircle\dots\ocircle}.
\ee
Only using first principles, however, we are not immediately able to identify if there are any further eigenvectors with eigenvalue 1. This property is crucial as it decides the asymptotic behaviour of correlations. Representing the magnitude-ordered eigenvalues as $\lambda_i$ for $i=0,1,\dots$ we have two cases: (i) $|\lambda_1|<1$, i.e., the magnitude of the second ordered eigenvalue is less than 1, and (ii) $|\lambda_0| = |\lambda_1| = 1$, i.e., the magnitude of the leading and second ordered eigenvalues are equal to 1. Case (i) is the generic case corresponding to exponential decay of correlations over a timescale set by the gap $1-|\lambda_1|$. In Case (ii) some correlations do not decay and the system is either ergodic but not mixing (for $\lambda_1=e^{i\theta}\neq1$) or non-ergodic (for $\lambda_1=1$). Finally, we note that the conjugate matrix $T_\ell^*$ is related to the original $T_\ell$ by a unitary similarity transformation that swaps the forward and backward replicas. As a consequence the spectrum is symmetric under complex conjugation.

As discussed in Sec.~\ref{sec:mainres}, at the integrable point $d=2$ and $h_x=0$ one can identify subspaces of the Hilbert space which are exponentially large in the distance between consecutive zeros $\ell$, where the time evolution acts unitarily (cf.\ Appendix~\ref{app:diagonalisation}). This means that there are exponentially large (in $\ell$) subspaces where $T_\ell$ acts as a unitary matrix. In fact, for this local unitary dynamics to exist inside only between two specific zeros, say those at $x=0, \ell$, one need only that the longitudinal field is zero inside this region $h_x=0, x\in(0,\ell]$. The natural question is then whether there are any other regions than the non-interacting one at which the gap can close. In order to help answer this we state the following property.

\begin{property}
\label{prop:ergodic}
For $d=2$, the spectral gap of $T_\ell$ can only close if the longitudinal fields acting on the first two qubits $h_{\frac{1}{2}}, h_{1}$ and the last two qubits $h_{\ell-\frac{1}{2}}, h_{\ell}$ of the subsystem all belong to $\{0, \pi/4\}$.
\end{property}

The proof of the above property is given in Appendix~\ref{app:nonergodicity}. This is done by considering how local operators can avoid leaking out of the boundaries of the subsystem: this can only happen for special values of the longitudinal fields near those boundaries. An immediate consequence of Property~\ref{prop:ergodic} is that we do not expect to see any uni-modular eigenvalues appearing for uniform longitudinal fields outside of $h_x = 0,\pi/4$ and for random configurations of the longitudinal fields as the gap only closes on a set of measure zero. Importantly however, this tells us nothing about the size of the gap and on how it depends on $\ell$. It may appear concerning that $h_x=\pi/4$ emerges here as a possible non-ergodic point as this would seem to be the strongest the interactions can get in the system. However, in the following we confirm numerically that Property \ref{prop:ergodic} is not overly weak in this respect and that indeed there do exist special cases where the gap may close for $h_x=\pi/4$.

\begin{figure}
\includegraphics[width=1.0\columnwidth]{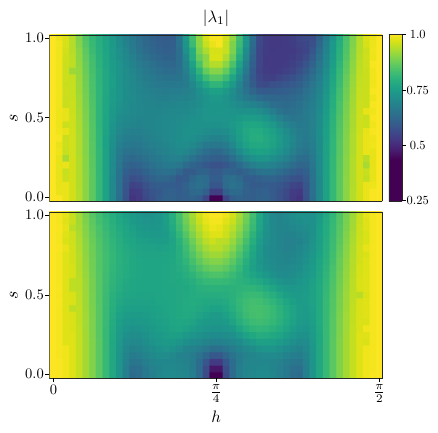}
\caption{The magnitude of the subleading eigenvalue $|\lambda_1|$ of the channel $T_\ell$ for $2\ell=12$} in the \textit{upper panel} and $2\ell=8$ in the \textit{lower panel} against $(s, h)$ parameters used for all gates in the subsystem except for the DU2 edge gates that use $h$ but have $s_x=0$.
\label{fig:subleading_numerics}
\end{figure}

We now consider a homogeneous channel where the values of $s_x$ and $h_x$ are constant in space except for the gates on the edge where $s_0=s_\ell=0$. For this channel we determine the magnitude of the second largest eigenvalue numerically: the results are displayed in Fig.~\ref{fig:subleading_numerics}. Immediately apparent is the emergence of conserved quantities at $h=\pi/4$ and $s=1$. At this point, all gates within the channel are Clifford gates. It is then the case that there exist operator strings that map to local strings with no amplitude leakage due to the fine tuned angles in the gates. The region of slow decay about this point then corresponds to a perturbed Clifford channel with small amounts of amplitude leakage of the previously conserved string. Away from this region, we also see the expected non-ergodic behaviour for the $h=0$ and $h=\pi/2$ lines. The dynamics at $h=\pi/2$ is equivalent to the dynamics at $h=0$ up to a single site rotation of the initial state. These regions have locally free fermionic dynamics and the conserved quantities here correspond to the fermions trapped between the $s_x=0$ points. We also note the overall change in the typical $|\lambda_1|$ values seen in the comparison between the upper and lower panels of Fig.~\ref{fig:subleading_numerics} that correspond to $2\ell = 12$ and $2\ell = 8$ respectively. For the ergodic channels, the increasing gap at larger $\ell$ can be rationalised by viewing the perturbative regions around the non-ergodic points as shrinking instead. This occurs as if we perturb each gate by a certain amount then the channel made of more gates will end up being perturbed by a greater amount, effectively leaving the full channel further away from the non-ergodic point.

It is important to note that this is not a discussion that generalises well to early times $t\sim\ell$ for two reasons. First, we have made no comment on the overall distribution of eigenvalues. While the gap may be growing with $\ell$ so too may the number of eigenvalues with similar magnitude to $\lambda_1$. The relative scaling of these two effects is then very important at early times. Secondly we have made no comment to the Jordan block structure of $T_\ell$. It is very likely that most operators will still decay quicker at short times for the smaller channels as a result of said channels having smaller Jordan blocks than the larger channels. This is expected physically as operators initially positioned in the bulk will only begin to decay once they have expanded to reach the depolarising boundaries of the channel $T_\ell$, something that will occur faster for smaller $\ell$.

As we know from Property~\ref{prop:ergodic} no conserved quantities can emerge away from $h=0,\pi/4$ but it is significant that we do not see non-ergodic behaviour for arbitrary $s$ values at $h=\pi/4$ and must also tune $s=1$. Even at the point $s=0$ where again all gates are Clifford and DU2 we do not see non-ergodic behaviour. It seems in this respect that Property~\ref{prop:ergodic} is weak and could stand to be improved by adding restrictions to the values of $s_x$ within the channel.

\section{Quench Dynamics}
\label{sec:quenchdynamics}

Let us now move on to consider the non-equilibrium dynamics of asymptotically solvable circuits. Our starting point is that, just like DU~\cite{piroli2020exact,foligno2025nonequilibrium} and DU2~\cite{yu2024hierarchical} circuits, asymptotically solvable circuits admit families of states that allow for the application of the solvability conditions when evaluating dynamical quantities. These are again found by considering two-site translation invariant matrix product states of the form 
\begin{equation}
    \!\!\!\!\ket*{\Psi(\Mcal)} \!\!=\!\! \sum_{\{i_j\}} \tr\!\left[\Mcal^{i_1,i_2}\dots\Mcal^{i_{2L-1},i_{2L}}\right] \!\!\ket*{i_1,\dots,i_{2L}},
\end{equation}
where $\{\Mcal^{i,j}\}$ are matrices acting on a $\chi$-dimensional auxiliary space. Graphically, we set
\begin{equation}
\begin{aligned}
\Mcal^{(i,j)}\otimes (\Mcal^{(k,l)})^* = \frac{1}{d}
\fineq[-0.8ex][0.8][1]{
    \MPSinitialstate[0][0][\ZXfcolor][][1]
	\node[scale=1] at (-0.9, 0.7)  {$(i,k)$};
	\node[scale=1] at (0.9, 0.7)  {$(j,l)$};
}
\end{aligned},
\end{equation}
where $(\cdot,\cdot)$ labels in the diagram denote an index on the front and back copy respectively where the thick lines represents two folded copies of the auxiliary space. We normalise the state such that in the thermodynamic limit
\begin{equation}
\label{eq:statenormalisation}
    \lim_{L\rightarrow\infty} \braket{\psi_0}{\psi_0} = 1.
\end{equation}
The above condition can be rewritten as 
\begin{equation}
    \lim_{L\rightarrow\infty} \tr\left(\mathcal T^L\right) = 1, 
\end{equation}
where we defined the MPS transfer matrix 
\begin{equation}
\begin{aligned}
\mathcal T\equiv \frac{1}{d}
\fineq[-0.8ex][0.8][1]{
    \MPSinitialstate[0][0][\ZXfcolor][][1]
	\cstate[0.5][0.5]
    \cstate[-0.5][0.5]
}.
\end{aligned}
\end{equation}
We then have that Eq.~\eqref{eq:statenormalisation} corresponds to the condition of $\mathcal T$ having a non-degenerate maximal eigenvalue $\lambda_0 =1$: we shall denote the corresponding left and right eigenvectors by $\bra{L}$ and $\ket{R}$ respectively.

We now define a solvable MPS to be one for which the following is true:
\begin{itemize}
\item[1)] The eigenvectors $\bra{L}$ and $\ket{R}$ fulfil  
\be
\label{eq:solvablestate_1}
(\bra{L})^\dag = \ket{R} = \sum_{i=1}^{\chi} \ket{i}\otimes\ket{i}.
\ee
Since these equations take the same form as Eq.~\eqref{eq:bulletstate} we represent them graphically as 
\be
(\bra{L})^\dag = \ket{R} = 
\fineq[-0.8ex][0.8][1]{
\draw[very thick] (-.5,0)--++(1,0);
\cstate[-0.5][0]
},
\ee
where the thicker line is to indicate that the sum is over the auxiliary space. 
\item[2)] The following conditions are satisfied
\begin{equation}
\begin{aligned}
&\fineq[-0.8ex][0.8][1]{
    \MPSinitialstate[0][0][\ZXfcolor][][1]
    \cstate[0.5][0.5][][]
    \node[scale=1.3] at (0.9, 0.7)  {$\beta$};
    \cstate[-0.5][0.5]
    \cstate[-1][0]
}
= 0\\
& \fineq[-0.8ex][0.8][1]{
    \MPSinitialstate[0][0][\ZXfcolor][][1]
    \cstate[0.5][0.5]
    \cstate[-0.5][0.5][][]
    \node[scale=1.3] at (-0.9, 0.7)  {$\beta$};
    \cstate[1][0]
}
= 0
\end{aligned}
\label{eq:solvablestate_2}
\end{equation}
where $\beta=\text{diag}\{\beta_1, \dots, \beta_d\} : \tr(\beta) = 0$ is an arbitrary traceless diagonal matrix.
\end{itemize}

Importantly, for $d=2$ low bond-dimension examples of the solvable states above include product states 
\be
\ket{\psi}  = \bigotimes_{x=1}^{2L} \ket*{\psi_x}, 
\ee
where each qubit is such that $\bra*{\psi_x}Z\ket*{\psi_x}=0$ (i.e. each qubit is polarised in the $xy$ plane), and all forms of entangled bell pair product states 
\be
\ket{\psi} = \bigotimes_{x=1}^{L} (\sum_{i,j=1}^{2} [v_x]_{i,j} \ket{ij}),
\ee
where $v_xv_x^\dag=\mathbbm{1}$. The conditions of solvability are such that the following rule holds
\begin{align}
\label{eq:solvability_with_mps}
\fineq[-0.8ex][0.6][1]{
    \MPSinitialstate[0][0][\ZXfcolor][][1]
    \roundgate[1][1][1][topright][\ZXfcolor][1]
    \roundgate[2][2][1][topright][\ZXfcolor][1]
    \roundgate[3][3][1][topright][\Zfcolor][1]
    \cstate[-0.5][0.5]
    \cstate[-1][0]
    \cstate[0.5][1.5]
    \cstate[1.5][2.5]
    \cstate[2.5][3.5]
}
=
\fineq[-0.8ex][0.6][1]{
    \roundgate[0][3][1][topright][\Zfcolor][1]
    \draw[very thick] (-.5,0)--++(1,0);
    \draw(-.5,.5)--++(1,0);
    \draw(-.5,1.5)--++(1,0);
    \cstate[-0.5][0.5]
    \cstate[-0.5][0]
    \cstate[-0.5][1.5]
    \cstate[-0.5][2.5]
    \cstate[-0.5][3.5]
}
\end{align}
for any number of intermediary gates. An analogous relation holds when flipping the diagram left to right. 

Using these conditions, one can simplify the tensor networks describing dynamical quantities after a quench from solvable states in the same way that could be done for the infinite temperature case. As was the true there, the core of all these simplifications is the ability to describe the evolution of a local subsystem without the need to consider its complement. The unique feature of asymptotically solvable circuits, compared to DU or DU2, is that the networks can only reduce up to the nearest $s_x=0$ points to the subsystem and, consequently, the real simplification only occurs for large times. 

To show this we consider the effective evolution of a subsystem $A$ defined as the region $x\in(0, \ell]$ where $s_{0}=s_{\ell}=0$. The reduced density matrix on this region may be written as
\begin{equation}
\begin{aligned}
\hspace{-.4cm}\ket*{\rho_A(t)} \!=\! \frac{1}{d^L}
\cdots
\fineq[-0.8ex][0.5][1]{
    \foreach \i in {0,...,3}{
        \foreach \j in {0,...,1}{	
            \roundgate[2*\i][2*\j][1][topright][\ZXfcolor][1]
            \roundgate[2*\i+1][2*\j+1][1][topright][\ZXfcolor][1]
        }
        \MPSinitialstate[2*\i+1][-1][\ZXfcolor][][1]
    }
    \foreach \j in {0,...,1}{	
        \roundgate[2*1][2*\j][1][topright][\Zfcolor][1]
        \roundgate[2*3][2*\j][1][topright][\Zfcolor][1]
    }
    
    \cstate[.5+2*0][3.5]
    \cstate[1.5+2*0][3.5]
    \cstate[.5+2*3][3.5]
    \cstate[1.5+2*3][3.5]
}
\!\!\cdots.
\end{aligned}
\end{equation}
We can apply unitarity and then use the existence of the fixed points Eq.~\eqref{eq:solvablestate_1} to simplify the diagram in the following manner
\begin{widetext}
\be
\ket*{\rho_A(t)} = \frac{1}{d^L}
\fineq[-0.8ex][0.5][1]{
    \foreach \i in {-2,...,5}{
        \MPSinitialstate[2*\i+1][-1][\ZXfcolor][][1]
    }
    \foreach \i in {0,...,4}{
        \pgfmathparse{min(3, 4-\i)}
        \foreach \j in {0,...,\pgfmathresult}{
            \roundgate[2*\i+\j][\j][1][topright][\ZXfcolor][1]
        }
    }
    \foreach \j in {0,...,1}{	
        \roundgate[2*1][2*\j][1][topright][\Zfcolor][1]
        \roundgate[2*3][2*\j][1][topright][\Zfcolor][1]
    }
    \foreach \i in {0,...,3}{
        \cstate[-1.5+\i][-.5+\i]
        \cstate[9.5-\i][-.5+\i]
    }
    \cstate[-3.5][-.5]
    \cstate[-2.5][-.5]
    \cstate[10.5][-.5]
    \cstate[11.5][-.5]

}
\hspace{-.6cm}= \frac{1}{d^{2t+\ell}\chi}\hspace{-.6cm}
\fineq[-0.8ex][0.5][1]{
    \foreach \i in {-1,...,4}{
        \MPSinitialstate[2*\i+1][-1][\ZXfcolor][][1]
    }
    \foreach \i in {0,...,4}{
        \pgfmathparse{min(3, 4-\i)}
        \foreach \j in {0,...,\pgfmathresult}{
            \roundgate[2*\i+\j][\j][1][topright][\ZXfcolor][1]
        }
    }
    \foreach \j in {0,...,1}{	
        \roundgate[2*1][2*\j][1][topright][\Zfcolor][1]
        \roundgate[2*3][2*\j][1][topright][\Zfcolor][1]
    }
    \foreach \i in {0,...,3}{
        \cstate[-1.5+\i][-.5+\i]
        \cstate[9.5-\i][-.5+\i]
    }
    \cstate[-2][-1]
    \cstate[10][-1]

}\hspace{-.6cm},
\ee
\end{widetext}
where the the thermodynamic limit is implicit in second equality and we used 
\be
\lim_{L\to\infty}\frac{1}{d^L}\left(\fineq[-0.8ex][0.8][1]{
    \MPSinitialstate[0][0][\ZXfcolor][][1]
	\cstate[0.5][0.5]
    \cstate[-0.5][0.5]
}\right)^L = \frac{1}{\chi}\fineq[-0.8ex][0.8][1]{
\draw[very thick] (.25,0)--++(.5,0);
\draw[very thick] (-.25,0)--++(-.5,0);
\cstate[-0.25][0]
\cstate[0.25][0]
}.
\ee
This is now in a position where the solvability conditions in Eq.~\eqref{eq:solvability_with_mps} can be applied. Using them we obtain 
\be
\begin{aligned}
\hspace{-.5cm}\ket*{\rho_A(t)} = \frac{1}{d^{2t+\ell}\chi}
\fineq[-0.8ex][0.5][1]{
    \transfer[0][0][4][topright][\Zfcolor][\ZXfcolor][1]
    \transfer[0][2][4][topright][\Zfcolor][\ZXfcolor][1]
    \foreach \i in {0,...,1}{
        \MPSinitialstate[2*\i+1][-1][\ZXfcolor][][1]
    }
    \cstate[0][-1]
    \cstate[4][-1]
}
= (T_\ell)^t \ket*{\rho_A(0)}\!,
\end{aligned}
\ee
where we have identified the column as the repeated application of the transfer matrix defined in Eq.~\eqref{eq:transfer_1rep}. Equivalently, we can say that the expectation value of any operator $\mathcal O_A$ with non-trivial support on $A$ evolves as 
\be
\label{eq:local_dynamics}
    \braket*{\mathcal O_A}{\rho_A(t)} = \bra*{\mathcal O_A} (T_\ell)^t \ket*{\rho_A(0)}.
\ee
This tells us that the analysis performed in Sec~\ref{sec:correlations} can be immediately applied to assess the asymptotic behaviour of the local state, i.e., whether it eventually reaches thermal equilibrium (at infinite temperature), relaxes to a generalised Gibbs ensemble, or fails to relax. Of course, for times $t$ and the support of the operator $\mathcal O$ both smaller than $\ell$ the local dynamics Eq.~\eqref{eq:local_dynamics} reduces to the standard form for an operator evolving in a circuit composed of gates using $s_x\neq0$. In the upcoming subsections Sec.~\ref{sec:asytime} and~\ref{sec:earlytime} we analyse the behaviour of quantum information the two different regimes.

\subsection{Asymptotic Times}
\label{sec:asytime}

To understand the behaviour of quantum information in the late time regime we begin by considering the entanglement between $A$, a continuous block of $2L_A$ qudits, and the rest of the system. To quantify the latter we will use the R\'enyi entropies of the reduced state $\rho_A(t)$, i.e., 
\be
S_A^{(\alpha)}(t) = \frac{1}{1-\alpha} \log \tr[\rho_A(t)^\alpha], \qquad \alpha \in \mathbb R. 
\ee
It is known (see, e.g., Ref.~\cite{foligno2024quantum}) that the behaviour of R\'enyi entropies in the limit of large subsystems (after the thermodynamic limit has been taken) is controlled by the boundaries between $A$ and the rest. Namely, specialising the treatment to $\alpha=n \in \mathbb Z$ we can express the R\'enyi entropies as
\begin{equation}
\hspace{-.35cm}    \lim_{L_A\rightarrow\infty}\lim_{L\rightarrow\infty} S_A^{(n)}(t) = 
    \frac{1}{1-n}\left[\log C_l(t) + \log C_r(t)\right],
\end{equation}
where we defined the diagrams $C_{l/r}$ as
\begin{widetext}
\be
\label{eq:entanglement_cone}
\begin{aligned}
C_{l/r}(t) = \frac{1}{\chi^{n}d^{2nt}}\hspace{-1cm}
\fineq[-0.8ex][0.55][1]{
    \trianglediagtwotone[0][0][1][6][5][mps][n]
    \roundgate[8][0][1.05][topright][\Zfcolor][n]
    \roundgate[8][2][1.05][topright][\Zfcolor][n]
    \roundgate[8][4][1.05][topright][\Zfcolor][n]
    \roundgate[0][0][1.05][topright][\Zfcolor][n]
    \roundgate[12][0][1.05][topright][\Zfcolor][n]
    \roundgate[9][1][1.05][topright][\ZXfcolor][n]
    \roundgate[9][3][1.05][topright][\ZXfcolor][n]
    \foreach \i in {0,...,7}{
        \cstate[-\i+5.5][-\i+6.5]	
        \sqrstate[\i+6.5][-\i+6.5]	
    }
    \cstate[-2][-1]	
    \sqrstate[14][-1]	
},
\end{aligned}
\ee
\end{widetext}
where the $l/r$ subscript denotes whether the diagram is centred on the left or right edge of the subsystem respectively, we introduced a diagrammatic representation for replicated gates 
\be
\label{eq:foldedgaten}
(U \otimes U^*)^{\otimes n} = \fineq[-0.8ex][0.75][1]{
   \roundgate[0][0][1][topright][\ZXfcolor][n]
},
\ee
the states representing relevant index contractions
\be
\begin{aligned}  
&\ket*{\ocircle}=\hspace{-.5cm}\sum_{i_1,\ldots, i_n=1}^d \ket*{i_1}\otimes\ket*{i_1}\otimes \cdots \otimes \ket*{i_n}\otimes\ket*{i_n} = \fineq[-0.8ex][0.6][1]{
    \draw(0, 0)--++(0, .5);
    \cstate[0][0]
    }, \\
&\ket*{\Box}=\hspace{-.5cm}\sum_{i_1,\ldots, i_n=1}^d \ket*{i_n}\otimes\ket*{i_1}\otimes \cdots \otimes \ket*{i_{n-1}}\otimes\ket*{i_n} = \fineq[-0.8ex][0.6][1]{
    \draw(0, 0)--++(0, .5);
    \sqrstate[0][0]
    }\,,
\end{aligned}
\ee
and their analogues in the auxiliary space. This structure is conditioned on unitarity alone. Now, we can apply our identities to simplify the diagram to within the nearest pair of zeros surrounding the centre
\be
\begin{aligned}
C_{l/r}(t)  = \frac{1}{\chi^n d^{2nt}} &\fineq[-0.8ex][0.5][1]{
    \trianglediagtwotonebounded[0][0][1][6][5][mps][n][4][10]
    \roundgate[8][0][1.05][topright][\Zfcolor][n]
    \roundgate[8][2][1.05][topright][\Zfcolor][n]
    \roundgate[8][4][1.05][topright][\Zfcolor][n]
    \foreach \i in {0,...,2}{
        \cstate[-\i+5.5][-\i+6.5]	
        \sqrstate[\i+6.5][-\i+6.5]	
    }
    \foreach \i in {0,...,5}{
        \cstate[3.5][\i-0.5]	
        \sqrstate[8.5][\i-0.5]	
    }
    \cstate[4][-1]	
    \sqrstate[8][-1]	
}
\end{aligned}
\ee
which may then be recognised as the repeated application of a single transfer matrix that connects the local initial state and the cut
\be
\begin{aligned}
C_{l/r}(t) = \frac{1}{\chi^n d^{n(2t+\ell_{l/r})}} &\fineq[-0.8ex][0.5][1]{
    \trianglediagtwotonebounded[0][0][1][6][5][mps][n][4][10]
    \roundgate[8][0][1.05][topright][\Zfcolor][n]
    \roundgate[8][2][1.05][topright][\Zfcolor][n]
    \roundgate[8][4][1.05][topright][\Zfcolor][n]
    \roundgate[8][6][1.05][topright][\Zfcolor][n]
    \roundgate[4][6][1.05][topright][\Zfcolor][n]
    \roundgate[7][7][1.05][topright][\ZXfcolor][n]
    \roundgate[5][7][1.05][topright][\ZXfcolor][n]
    
    \foreach \i in {0,...,7}{
        \cstate[3.5][\i-0.5]	
        \sqrstate[8.5][\i-0.5]	
    }
    \cstate[4.5][7.5]
    \cstate[5.5][7.5]
    \sqrstate[6.5][7.5]
    \sqrstate[7.5][7.5]
    \cstate[4][-1]	
    \sqrstate[8][-1]	
},
\label{eq:ent_cone_diagram}
\end{aligned}
\ee
where we have introduced the lengths $\ell_{l/r}$ that describe the length of interval between the zeros surrounding the left and right edges respectively. We define the multi-replica transfer matrices

\be
\begin{aligned}
T_{\ell}^{(n)} = \frac{1}{d^{2n}}
\fineq[-0.8ex][0.6][1]{
    \roundgate[0][0][1][topright][\Zfcolor][n]
    \roundgate[1][1][1][topright][\ZXfcolor][n]
    \roundgate[2][0][1][topright][\ZXfcolor][n]
    \roundgate[3][1][1][topright][\ZXfcolor][n]
    \roundgate[4][0][1][topright][\Zfcolor][n]
    \cstate[-0.5][-0.5]
    \cstate[-0.5][0.5]
    \sqrstate[4.5][-0.5]
    \sqrstate[4.5][0.5]
        \draw [decorate, decoration = {brace}]   (3.5,-.7) -- (0.5,-.7);
    \node[scale=1.5] at (2,-1.3) {$2\ell$};
}
\end{aligned},
\label{eq:transferN}
\ee
such that
\begin{equation}
    C_{l/r}(t) = 
    \bra*{\ocircle\dots \ocircle \Box\dots\Box} T_{\ell_{l/r}}^{(n)\, t} \ket*{\rho_{l/r}^{(n)}}. 
\end{equation}
where the additional normalisation has been absorbed into $\ket*{\rho_{l/r}^{(n)}}$ which, as shown in Eq.~\eqref{eq:ent_cone_diagram}, is the $n$-fold replicated MPS state contained within the region of size $\ell_{l/r}$ with the square and circle vectors applied at the edges. We therefore must study the spectra of $T_{\ell}^{(n)}$ to understand the behaviour of the entanglement.

In contrast to the 1-replica case where $T_\ell$ has the maximally mixed state $\ket*{\ocircle\dots\ocircle}$ as a leading eigenvector with eigenvalue $\lambda_0=1$, in the $n$-replica case this state is not even an eigenvector. By applying the solvability conditions in Eq.~\eqref{eq:inhom_condition}, it is elementary to verify that the following are eigenvectors
\begin{align}
    \ket*{R_\pm^{(n)}} &= \frac{1}{\sqrt{2A_\pm}}\left(
    \fineq[-0.8ex][0.5][1]{
    \draw [decorate, decoration = {brace}]   ( (0.5, 1.6) -- (3.5, 1.6);
    \node[scale=1.5] at (2,2.2) {$2\ell$};
        \roundgate[0][0][1][topright][\Zfcolor][n]
        \roundgate[1][1][1][topright][\ZXfcolor][n]
        \roundgate[2][0][1][topright][\ZXfcolor][n]
        \roundgate[3][1][1][topright][\ZXfcolor][n]
        \roundgate[1][-1][1][topright][\ZXfcolor][n]
        \roundgate[0][-2][1][topright][\Zfcolor][n]
        \cstate[-0.5][-0.5]
        \cstate[-0.5][0.5]
        \cstate[-0.5][-2.5]
        \cstate[-0.5][-1.5]
        \sqrstate[2.5][-0.5]
        \sqrstate[3.5][0.5]
        \sqrstate[0.5][-2.5]
        \sqrstate[1.5][-1.5]
    }
    \,\pm\,
    \fineq[-0.8ex][0.5][1]{
        \draw [decorate, decoration = {brace}]   ( (0.5, 1.6) -- (3.5, 1.6);
    \node[scale=1.5] at (2,2.2) {$2\ell$};
        \roundgate[1][1][1][topright][\ZXfcolor][n]
        \roundgate[2][0][1][topright][\ZXfcolor][n]
        \roundgate[3][1][1][topright][\ZXfcolor][n]
        \roundgate[4][0][1][topright][\Zfcolor][n]
        \roundgate[3][-1][1][topright][\ZXfcolor][n]
        \roundgate[4][-2][1][topright][\Zfcolor][n]
        \sqrstate[4.5][-0.5]
        \sqrstate[4.5][0.5]
        \sqrstate[4.5][-2.5]
        \sqrstate[4.5][-1.5]
        \cstate[1.5][-0.5]
        \cstate[.5][0.5]
        \cstate[3.5][-2.5]
        \cstate[2.5][-1.5]
    }
    \right) \notag \\ \notag\\
    \bra*{L_\pm^{(n)}} &= \frac{1}{\sqrt{2A_\pm}}\left(
    \fineq[-0.8ex][0.5][1]{
        \roundgate[0][0][1][topright][\Zfcolor][n]
        \roundgate[1][1][1][topright][\ZXfcolor][n]
        \roundgate[2][0][1][topright][\ZXfcolor][n]
        \roundgate[0][2][1][topright][\Zfcolor][n]
        \draw(3.5,0)--++(0,-.5);
        \cstate[-0.5][-0.5]
        \cstate[-0.5][0.5]
        \cstate[-0.5][1.5]
        \cstate[-0.5][2.5]
        \sqrstate[3.5][0]
        \sqrstate[2.5][0.5]
        \sqrstate[1.5][1.5]
        \sqrstate[0.5][2.5]
         \draw [decorate, decoration = {brace}]   (3.5,-.7) -- (0.5,-.7);
         \node[scale=1.5] at (2,-1.3) {$2\ell$};
    }
    \,\pm\,
    \fineq[-0.8ex][0.5][1]{
        \roundgate[2][0][1][topright][\ZXfcolor][n]
        \roundgate[3][1][1][topright][\ZXfcolor][n]
        \roundgate[4][0][1][topright][\Zfcolor][n]
        \roundgate[4][2][1][topright][\Zfcolor][n]
        \draw(.5,0)--++(0,-.5);
        \sqrstate[4.5][-0.5]
        \sqrstate[4.5][0.5]
        \sqrstate[4.5][1.5]
        \sqrstate[4.5][2.5]
        \cstate[.5][0]
        \cstate[1.5][.5]
        \cstate[2.5][1.5]
        \cstate[3.5][2.5]
        \draw [decorate, decoration = {brace}]   (3.5,-.7) -- (0.5,-.7);
        \node[scale=1.5] at (2,-1.3) {$2\ell$};
    }
    \right) \notag\\ \notag\\
    A_\pm &= \left(
    \fineq[-0.8ex][0.5][1]{
        \roundgate[0][0][1][topright][\Zfcolor][n]
        \sqrstate[0.5][-0.5]
        \sqrstate[0.5][0.5]
        \cstate[-.5][.5]
        \cstate[-.5][-.5]
    }
    \right)^{2\ell}
    \pm\,
    \left(
    \fineq[-0.8ex][0.5][1]{
        \roundgate[0][0][1][topright][\Zfcolor][n]
        \sqrstate[0.5][-0.5]
        \sqrstate[0.5][0.5]
        \cstate[-.5][.5]
        \cstate[-.5][-.5]
    }
    \right)^{\ell}
\label{eq:eigenvectors}
\end{align}
such that
\begin{equation}
\begin{aligned}
    \braket*{L_\mu^{(n)}}{R_\nu^{(n)}} &= \delta_{\mu\nu} \\
    T_\ell^{(n)}\ket*{R_\pm^{(n)}} &= \frac{1}{d^{2n}}
    \fineq[-0.8ex][0.6][1]{
        \roundgate[0][0][1][topright][\Zfcolor][n]
        \cstate[-.5][.5]
        \cstate[-.5][-.5]
        \sqrstate[.5][.5]
        \sqrstate[.5][-.5]
    }\,\, \ket*{R_\pm^{(n)}} \\
    \bra*{L_\pm^{(n)}}T_\ell^{(n)} &= \bra*{L_\pm^{(n)}}\,\, \frac{1}{d^{2n}}
    \fineq[-0.8ex][0.6][1]{
        \roundgate[0][0][1][topright][\Zfcolor][n]
        \cstate[-.5][.5]
        \cstate[-.5][-.5]
        \sqrstate[.5][.5]
        \sqrstate[.5][-.5]
    } .
\end{aligned}
\end{equation}
It was proven in Ref.~\cite{foligno2024quantum} that for DU2 gates the $d^2\times d^2$ matrix with elements
\be
\begin{aligned}
    \bra*{kl}P_\Lambda\ket*{ij} \equiv
    \fineq[-0.8ex][0.6][1]{
        \roundgate[0][0][1][topright][\Zfcolor][1]
        \cstate[-.5][.5]
        \cstate[-.5][-.5]
    	\node[scale=1.3] at (1, 0.6)  {$(l,j)$};
    	\node[scale=1.3] at (1, -.6)  {$(k,i)$};
    }
\end{aligned}
\ee
is a proportional to a projector such that $P_\Lambda^2 = \Lambda P_\Lambda$ and $\tr P_\Lambda = d^2$ implying $\Lambda = d^2/n_\Lambda$ where $n_\lambda$ is the number of non-zero eigenvalues. This gives
\be
\begin{aligned}
\frac{1}{d^{2n}}
\fineq[-0.8ex][0.6][1]{
        \roundgate[0][0][1][topright][\Zfcolor][n]
        \cstate[-.5][.5]
        \cstate[-.5][-.5]
        \sqrstate[.5][.5]
        \sqrstate[.5][-.5]
} = \frac{1}{d^{2n}} \tr P_\Lambda^n = \frac{1}{n_\Lambda^{n-1}}
\end{aligned}
\ee
and thus we restate
\begin{equation}
\begin{aligned}
    T_\ell^{(n)}\ket*{R_\pm^{(n)}} &= \frac{1}{n_\Lambda^{n-1}}\,\, \ket*{R_\pm^{(n)}} \\
    \bra*{L_\pm^{(n)}}T_\ell^{(n)} &= \bra*{L_\pm^{(n)}}\,\, \frac{1}{n_\Lambda^{n-1}}.
\end{aligned}
\end{equation}
To guarantee these are leading eigenvectors one must bound the spectral radius of $T_\ell^{(n)}$. This is guaranteed by the following property (proven in Appendix \ref{app:spectralradius})
\begin{property}
\label{prop:specradius}
The spectral radius of $T_\ell^{(n)}$ has a non-trivial upper bound for $n>1$
\be
\begin{aligned}
    \rho(T_\ell^{(n)}) \leq \frac{1}{d^{2n}}
    \fineq[-0.8ex][0.6][1]{
        \roundgate[0][0][1][topright][\Zfcolor][n]
        \cstate[-.5][.5]
        \cstate[-.5][-.5]
        \sqrstate[.5][.5]
        \sqrstate[.5][-.5]
    }
    = \frac{1}{n_\Lambda^{n-1}}.
\end{aligned}
\ee
\end{property}

Therefore, we can conclude that the eigenvectors in Eq.~\eqref{eq:eigenvectors} belong to the eigenspace with eigenvalue of maximal magnitude. Of course, it is again possible that there exist other eigenvectors in this space. Such eigenvectors could produce parameter-dependent $O(t^0)$ shifts to the entanglement but will not change the overall slope/velocity. Assuming that the ones in Eq.~\eqref{eq:eigenvectors} are the only eigenvectors with maximum eigenvalue we have that as $t\rightarrow\infty$
\begin{equation}
\begin{aligned}
   T_{l/r}^{(n)\, t} &\rightarrow \frac{1}{n_\Lambda^{t(n-1)}} \left( \ketbra*{R_+^{(n)}}{L_+^{(n)}} + \ketbra*{R_-^{(n)}}{L_-^{(n)}} \right).
\end{aligned}
\end{equation}
It is easily shown that for a solvable initial state
\begin{equation}
    \braket*{L_-^{(n)}}{\rho_{l/r}^{(n)}} = 0.
\end{equation}
Therefore, as $t\rightarrow\infty$ we have
\begin{equation}
    \!\! C_{l/r}(t) \simeq n_\Lambda^{t(1-n)} \braket*{L_+^{(n)}}{\rho_{l/r}^{(n)}} \braket*{\ocircle\dots \ocircle \Box\dots\Box}{R_+^{(n)}},  
\end{equation}
and thus the entanglement velocity reads as
\begin{equation}
\label{eq:entanglement_velocity}
\begin{aligned}
    v_E \equiv \lim_{t\rightarrow\infty}\lim_{L_A\rightarrow\infty}\lim_{L\rightarrow\infty} \frac{S_A^{(n)}(t)}{4t\log(d)} = \frac{\log(n_\Lambda)}{2\log(d)}
\end{aligned}
\end{equation}
as the inner products are independent of time. Remarkably, this is the same result as the DU2 case~\cite{foligno2024quantum, rampp2024entanglement}. We have shown that, even when we separate the DU2 gates by many generic $s$ non-DU2 gates, for lo      ng enough times the entanglement shows the same growth and even the same slope/velocity. This occurs with an arbitrarily low (but finite) density of DU2 gates.

More generally, let us consider the `entanglement membrane'~\cite{jonay2018coarse, zhou2019emergent, zhou2020entanglement, sommers_2024} generated by asymptotically solvable circuits. The latter comes from regarding the tensor networks describing entanglement related quantities as partition functions. For large times, these partition functions are expected to be dominated by domain wall configurations which can be characterised by a line tension --- or membrane tensions in higher dimensions. Once identified, the line tension can be used to infer many properties of the dynamics by assuming other quantities are also dominated by such domain wall configurations. We assume that, despite the inhomogeneity of the circuit, over large scales the line tension behaves as it would in a homogeneous system. Namely, we assume the line tension to be self-averaging. Therefore, we can probe the line tension by defining the following quantity~\cite{jonay2018coarse, zhou2019emergent, zhou2020entanglement, sommers_2024}
\begin{widetext}
\begin{equation}
\begin{aligned}
Z_n(x,y,t) \!=\! \frac{1}{d^{2nL}}
\cdots
\fineq[-0.8ex][0.58][1]{
    \foreach \i in {0,...,6}{
        \foreach \j in {0,...,2}{
            \roundgate[2*\i][2*\j][1][topright][\ZXfcolor][n]
            \roundgate[2*\i+1][2*\j+1][1][topright][\ZXfcolor][n]
        }
    }
    \foreach \j in {0,...,2}{	
        \roundgate[2*1][2*\j][1][topright][\Zfcolor][n]
        \roundgate[2*3][2*\j][1][topright][\Zfcolor][n]
        \roundgate[2*6][2*\j][1][topright][\Zfcolor][n]
    }
    
    \foreach \i in {0,...,4}{
        \cstate[\i-.5][-.5]
    }
    \foreach \i in {5,...,13}{
        \sqrstate[\i-.5][-.5]
    }
    \foreach \i in {0,...,6}{
        \cstate[\i+.5][5.5]
    }
    \foreach \i in {7,...,13}{
        \sqrstate[\i+.5][5.5]
    }
    \draw [decorate, decoration = {brace}]   (3.5,-.7) -- (0.5,-.7);
        \node[scale=1.5] at (2,-1.3) {$2x$};
            \draw [decorate, decoration = {brace}]    (0.5,5.75) -- (6.5,5.75);
        \node[scale=1.5] at (3.5,6.35) {$2y$};
}
\cdots,
\end{aligned}
\end{equation}
\end{widetext}
which measures the entanglement of the time evolution operator under a particular bipartition of input and output legs. Namely, we consider bipartitions into contiguous blocks with cuts at position $x$ and $y$ for the input and output legs respectively. The factor $d^{-2nL}$ is a normalisation. 

Keeping $x$ fixed and then taking $t\rightarrow\infty$ with $y = x + vt$ we expect that a straight domain wall connecting $(x,0)$ and $(y,t)$ should emerge as the dominant configuration such that
\begin{equation}
    -\frac{1}{n-1}\log Z_n(x,y,t) \sim 2t \log d \,\, \mathcal{E}_n(v)
\end{equation}
where we have introduced the factor of $2$ to the typical definition to compensate for the half-integer time steps used in this work. As it is defined across large times and distances, the line tension should only reflect the solvability that the asymptotically solvable circuit will attain for $t\gg\ell$. One can verify that this is true by elementary application of the solvability conditions Eq.\eqref{eq:asysolvable}. As $\ell$ is kept fixed in the limit, for any non-zero $v$ it will eventually become true that $|y-x| > \ell$ such that there exists a column of DU2 gates in-between the cuts. Once this occurs, the value of $Z_n(x,y,t)$ becomes \textit{completely independent} of $\ell$. One can easily verify that  in this case the DU2 result holds such that
\begin{equation}
    \mathcal{E}_n(v) = |v| + \frac{1-|v|}{2} \frac{\log(n_\Lambda)}{\log(d)}.
\label{eq:membrane}
\end{equation}
Of course, for $v=0$ we have $x=y$ and this logic does not hold. However, Eq.~\eqref{eq:membrane} is still valid as the solvability conditions leave the network in a similar form to Eq.~\eqref{eq:ent_cone_diagram} where the value may be discerned from the leading eigenvalues of $T_\ell^{(n)}$, which we have already characterised.

The line tension gives us the expected results for the other quantities in the system. The point $\mathcal{E}_n(v=0)=v_E$ returns the entanglement velocity that we shown is exactly valid for quenches from solvable states. Furthermore, the solution to $\mathcal{E}_n(v_B)=v_B$ returns the so-called butterfly velocity $v_B=1$ (provided $n_\Lambda\neq1$), indicating that at large times $t\gg\ell$ operators spread at maximal velocity.

\subsection{Early times}
\label{sec:earlytime}

Having characterised the long time behaviour, it is natural to ask what happens at early times. This is also equivalent to considering the limit in which we send $\ell\rightarrow\infty$ to regain a homogeneous system. In particular, one might be concerned that these dynamics must also be somehow special due to the complex relation the generic $s$ gates have with the DU2 gates (Eq.~\eqref{eq:asysolvable}). In the absence of the solvability conditions there are only few statements that can be made analytically about the dynamics. In this section we investigate the model at the integrable point analytically and then the outside this point numerically. We shall see here that the early time dynamics can be qualitatively different to the long-time solvable dynamics that we have been able to characterise so far. This leads us to the conclusion that these dynamics are in no sense solvable and that in the general case only the long-time dynamics may be characterised analytically.

\subsubsection{Dynamics at the free point}

As mentioned in the introduction, when $d=2$ and the longitudinal fields are zero the time evolution operator may be reduced to free-fermionic form by performing a Jordan-Wigner transformation. Considering the homogeneous case, we can then understand both the entanglement structure and the effect of inserting distant $s_x=0$ inhomogeneities by describing the structure of the eigenmodes. The explicit diagonalisation of the propagator is provided in Appendix~\ref{app:diagonalisation}, while the key upshot is that we may use the following form
\be
\mathbb{U} = e^{i H_{\rm eff}},
\ee
with a Floquet Hamiltonian given by 
\be
\label{eq:Heff}
H_{\rm eff} = \sum_{j=1}^2 \sum_{k= -\pi}^{\pi} \varepsilon_j(k) \beta^\dag_j(k)\beta^{\phantom\dag}_j(k) + \text{const},
\ee
where $\beta_j$ are canonical fermionic operators (they anticommute for different $j$s) and the dispersion relations are given by
\be
 \varepsilon_1(k) = |k|, \,\,\,\,  \varepsilon_2(k)= 2\cos^{-1}(\cos(k/2)\sin(\pi s/2))\,.
\ee
The behaviour of the entanglement after a quench may be determined by applying the quasiparticle picture~\cite{calabrese2005evolution}. To this end we shall consider an initial state that, at the same time, is solvable according to our definition (cf.~Eqs.~\eqref{eq:solvablestate_1} and~\eqref{eq:solvablestate_2}), is Gaussian, and produces pairs of quasiparticle excitations. We shall choose the following state
\be
    \ket*{\psi} = \bigotimes_{x=1}^L \frac{1}{\sqrt{2}} (\ket*{00}_{x-1/2, x} - \ket*{11}_{x-1/2, x}).
\ee

\begin{figure}
\includegraphics[width=0.8\columnwidth]{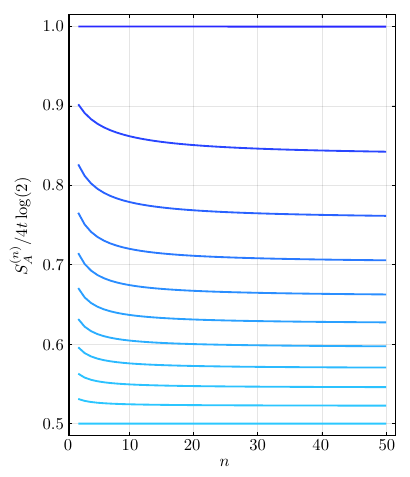}
\caption{The Rényi index dependence of the entanglement velocity in the integrable case. The different curves are calculated using different values of $s$ ranging from 1 (dark blue) to 0 (light blue) in steps of $0.1$.}
\label{fig:quasiparticle_slope}
\end{figure}

As discussed in Appendix~\ref{app:freequench}, the entanglement entropies in this state can be written as a sum of two parts for all times
\be
    S_A^{(n)}(t) = S_1^{(n)}(t) + S_2^{(n)}(t),
\ee
where $S_j^{(n)}$ is the entanglement created by the $j$-type fermions. The quasiparticle picture can be applied to determine how these quantities will evolve. For $L_A\gg t$, $L-L_A\gg t$ and $t\gg1$ the quasiparticle picture indicates that the dominant part of the entanglement will come from freely propagating entangled pairs
\bea
    S_j^{(n)} &\simeq 2t\int^{\pi}_{-\pi}\frac{dk}{2\pi} \, |v_j(k)| \, s_j^{(n)}(k), \\
    s_j^{(n)}(k) &\equiv  \frac{1}{1-n}\log[(\theta_j(k))^n + (1-\theta_j(k))^n], 
\eea
with the quasiparticle velocities defined as $v_j(k) = \varepsilon_j^\prime(k)$ and the filling fractions in the initial state
\bea
    \theta_j(k) = \bra*{\psi} \beta_j^\dag(k)\beta_j(k) \ket*{\psi}.
\eea
Using the definitions given in Appendix~\ref{app:diagonalisation} for $k>0$ one finds the following forms
\be
\begin{aligned}
    &\theta_1(k) = \frac{1}{2},&  &\theta_2(k) = \frac{1}{2}\left(1 + \frac{\cos(k/2)}{\cosh(\phi_k)}\right), \\
    &\theta_1(-k) = \frac{1}{2},&  &\theta_2(-k) = \frac{1}{2}\left(1 - \frac{\cos(k/2)}{\cosh(\phi_k)}\right),
\end{aligned}
\ee
where
\bea
    \sinh(\phi_k) = \sin(k/2) \tan(\pi s/2). 
\eea
Immediately, this allows us to see that the first contribution to the entanglement is
\be
    S_1^{(n)}(t) = 2t \log(2). 
\ee
This contribution is independent of $s$ (cf.~Appendix~\ref{app:diagonalisation}) as it corresponds the the maximum velocity modes propagating only along the $\pi/4$ bonds. This is the piece that creates to the entanglement velocity calculated using the solvability conditions in Eq.~\eqref{eq:entanglement_velocity} in the regime of large time compared to the distance between the two nearest zeros $t\gg\ell$. The remaining piece $S_2^{(n)}(t)$ cannot be expressed in terms of elementary functions of $s$: its expression in terms of special functions is long and not insightful so we do not give it here. Instead, we illustrate its properties in Fig~\ref{fig:quasiparticle_slope}.

The central conclusion to draw is that in the homogeneous case the dynamics at the free point does not appear to be similar to the DU2 circuit in any sense. Firstly, the dynamical correlations are not confined to individual lines in spacetime due to the second type of modes in the system being dispersive. One might argue that these modes are contained to good approximation within an interior light cone  $v=\pm \max_{k\in[-\pi,\pi)}|v_2(k)|$ and that could be used to reduce the tensor network in Fig.~\ref{fig:correlationcircuit}. However, due to the widening support of the correlations in time this approach will not be able to reduce the network down to a repeated contraction of a finite tensor as can be achieved for a solvable circuit. Secondly, the now trademark flat R\'enyi spectrum seen in solvable circuits is not observed in this system: whilst $S_1^{(n)}$ is independent of $n$, $S_2^{(n)}$ is not (cf.\ Fig~\ref{fig:quasiparticle_slope}). 

The long time structure will clearly change when we add $s_x=0$ points to the system. These points are only seen by the $\beta_2(k)$ modes which experience them as open boundaries (cf.\ Appendix~\ref{app:diagonalisation} and the discussion in Sec.~\ref{sec:mainres}). With two of these points positioned at $x=\pm\ell/2$ and with the edge of a subsystem at $x=0$ we expect these modes to provide a linear contribution to the entanglement until $t=\ell/2$ at which point it becomes possible for modes reflected at the boundary to cross the subsystem edge. This then causes the contribution from the $\beta_2(k)$ modes to begin to fluctuating as different quasiparticle pairs go from straddling the boundary of the subsystem to both being on either side. Focussing on a single set of modes with a given velocity $v_2(k)$ it is easy to see that the contribution to the entanglement should come as a triangle wave with period $\ell/v_2(k)$~\cite{modak_2020, capizzi_2023}. This implies the form
\bea
    S_2^{(n)}(t) &= 2\ell \int^{\pi}_{-\pi}\frac{dk}{2\pi} \, \min(\tau(k), 1-\tau(k)) \, s_2^{(n)}(k), \\
    \tau(k) &= \frac{|v_2(k)| t}{\ell} - \left\lfloor \frac{|v_2(k)| t}{\ell} \right\rfloor.
\ea
\label{eq:quasiparticle_finitesize}
\ee
From the original circuit perspective, the construction of these oscillations will occur within the other leading eigenvalues of the transfer matrix $T_\ell^{(n)}$ that cannot be identified by the solvability conditions alone.

\subsubsection{Dynamics for generic gate choice}

When the circuit is both interacting and non-solvable it becomes much more difficult to say anything substantive about the dynamics. Here we demonstrate numerically that in the homogeneous case for $d=2$ the generic $s$ case shows the same departure from solvable dynamics as the free fermionic case discussed above. The transition into solvable dynamics in the inhomogeneous case when $t\sim\ell$, however, occurs differently to Eq~\eqref{eq:quasiparticle_finitesize}. 

\begin{figure}
\includegraphics[width=1.0\columnwidth]{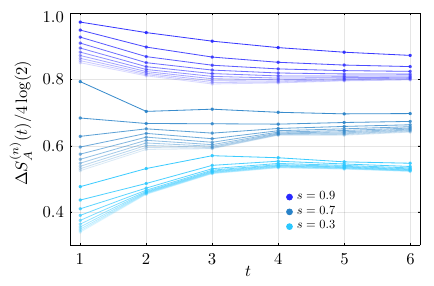}
\caption{The gradients of various entanglement entropies ($\Delta S^{(n)}_A(t) = S^{(n)}_A(t) - S^{(n)}_A(t-1)$) after a quench from a simple Bell pair state using three different homogeneous circuits with varying values of $s$. The three sets of curves that are plotted with different colours from light blue to dark blue use $s=0.3, 0.7, 0.9$ respectively and all use a uniform field $h=\pi/8$. Each set of curves contains $10$ curves that are arranged vertically and correspond to the R\'enyi index $n$ ranging from $1$ to $10$ in descending order. For $n=1$ we plot the von-Neumann entropy.}
\label{fig:quench_velocity}
\end{figure}

In the non-solvable homogeneous case we plot a selection of different entanglement gradients in Fig~\ref{fig:quench_velocity}. For the limited times we are able to access we see that for generic choices of $s$ the entanglement velocity does not appear to settle into $n$-independent values. This indicates that again the Schmidt spectrum of the wave function is becoming increasingly non-flat under these dynamics. Furthermore, we see that the effect of increasing $s$ is to monotonically increase the asymptotic value of the gradient. It is very tempting to imagine that the entanglement velocity in homogeneous circuits for $s>0$ is always greater than that of the $s=0$ circuit and that this might be a consequence of the solvability conditions Eq.~\eqref{eq:asysolvable}. However, we have no argument for this.

\begin{figure}
\includegraphics[width=1.0\columnwidth]{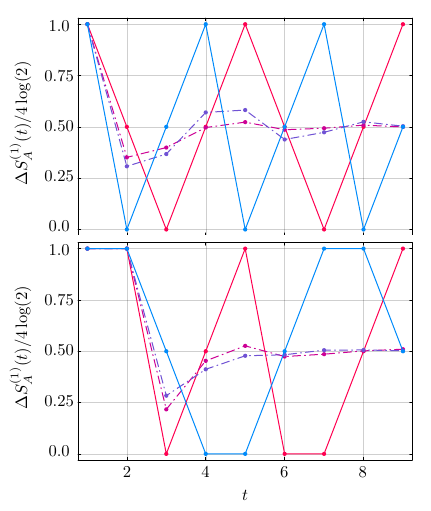}
\caption{Gradient of the von-Neumann entanglement entropy after a quench from a Bell pair state with $s_x=0$ points positioned at every $2\ell=4$ bonds in the \textit{upper panel} (as pictured in Eq.~\eqref{eq:entanglement_cone}) and every $2\ell=8$ bonds in the \textit{lower panel}. The different curves are calculated for quenches with $s=1.0$ but varying longitudinal fields $h=0, \pi/12, \pi/6, \pi/4$ from red to blue. Solid lines indicate those that will not converge due to the presence of conservation laws.}
\label{fig:quench_inhom_velocity}
\end{figure}

\begin{figure}
\includegraphics[width=1.0\columnwidth]{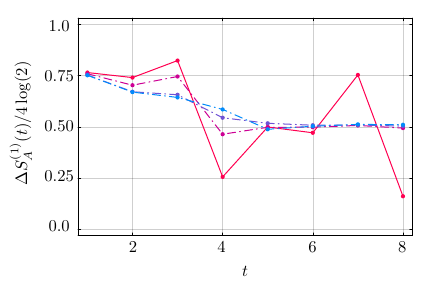}
\caption{Gradient of the von-Neumann entanglement entropy after a quench from a Bell pair state with $s_x=0$ points positioned at every $2\ell=8$ bonds (same as the \textit{lower panel} of Fig~\ref{fig:quench_inhom_velocity}) but now with $s=2/3$ for all other gates. The different curves are calculated for quenches with $s=2/3$ but varying longitudinal fields $h=0, \pi/12, \pi/6, \pi/4$ from red to blue. Solid lines indicate those that will not converge due to the presence of conservation laws.}
\label{fig:quench_inhom_velocity2}
\end{figure}

To investigate the transition from the non-solvable into the solvable dynamics we consider the gradient of the von-Neumann entropy in the inhomogeneous case, shown in Fig~\ref{fig:quench_inhom_velocity} and Fig~\ref{fig:quench_inhom_velocity2}. In Fig~\ref{fig:quench_inhom_velocity} we have the typical gates as $s=1.0$ (DU) and then position the $s=0$ (DU2) gates every $\ell=4,8$ sites with each of the subsystem boundaries positioned halfway between two such points. For $2t-1<\ell/2$ we see that, as expected, the entanglement grows at the maximal velocity, as it does for a homogeneous DU circuit. Once $2t-1\geq\ell/2$ the DU2 gates become relevant and the entanglement velocity begins to shift. However, depending on the parametrisation of the gate this occurs differently. For the non-ergodic gates identified in Sec~\ref{sec:correlations} we see that the velocity does not converge and oscillates around its mean value of $1/2$. For the ergodic gates, the oscillations appear heavily damped leading to a rapid convergence on the DU2 velocity of $1/2$. The periodicity of the oscillations in the integrable case (red line) is owed to the linear dispersion of the trapped fermions that comes from the choice of $s=1$ inside the subsystem. The oscillations at the non-integrable Clifford yet non-ergodic point $s=1, h=\pi/4$ (blue line) are also periodic yet have a different period to the integrable case. In Fig~\ref{fig:quench_inhom_velocity2} we consider the same setup with $\ell=8$ but change the generic gates to use $s=2/3$. As can be seen in Fig~\ref{fig:subleading_numerics} we will no longer see non-ergodic behaviour at $h=\pi/4$. This is in line with the behaviour of the correlation functions discussed in Sec.~\ref{sec:correlations} where we saw that the gap remains open for $s\neq1$. The immediate consequence of this is that we only have one line which does not converge in Fig~\ref{fig:quench_inhom_velocity2} (red line) that corresponds to the integrable system. In this case the oscillations are not periodic as the dispersion of the trapped fermions is non-linear and thus produces a range of velocities at which information can spread. This distinction can be understood through Eq.~\eqref{eq:quasiparticle_finitesize}.

We now return to considering the homogeneous circuit with no DU2 gates inserted. Whilst breaking the flatness of the R\'enyi spectrum is a good indication of a lack of solvability, we can confirm the latter directly by probing the entanglement structure of the fixed points discussed in Sec~\ref{sec:mainres}. This is a convenient choice as it does not depend on the initial state and allows us to directly test the bulk properties of the circuits. Considering for definiteness the left fixed point we have 
\bea
    \bra*{L_t} = 
    \fineq[-0.8ex][0.5][1]{
        \foreach \i in {0,...,3}{
            \pgfmathparse{3-\i}
            \foreach \j in {0,...,\pgfmathresult}{
                \roundgate[\i+\j][-\i+\j][1][topright][\ZXfcolor][1]
            }
            \cstate[-0.5+\i][0.5+\i]
            \cstate[-0.5+\i][-0.5-\i]
        }
        \draw [decorate, decoration = {brace}]   (-.75,.75)--++(3,3);
        \node[scale=2] at (-1.25+1.5,1.25+1.5)  {$t$};    
        \draw [decorate, decoration = {brace}]   (3.75, 3.5)--++(0,-4);
        \draw [decorate, decoration = {brace}]   (3.75, 3.5-5)--++(0,-2);
        \node[scale=2] at (4.5,1.5)  {$A$};    
        \node[scale=2] at (4.5,2-4.5)  {$\bar{A}$};
    }
\ea\,.
\label{eq:fixed_point_nonsolv}
\ee
To assess one possible way for this tensor, viewed as a state with $2t$ qudits of local dimension $d^2$, to be computationally simple we can ask whether it can be approximated by a low-entangled MPS form as $t$ grows large. We may define the density matrix corresponding to this state
\be
    \rho = \frac{\ketbra*{L_t}}{\braket*{L_t}}, 
\ee
and then construct a reduced density matrix $\rho_A = \tr_{\bar{A}} \rho$ across some bipartition like the one pictured in Eq.~\eqref{eq:fixed_point_nonsolv}. The entanglement growth of this reduced density matrix growing linearly with $t$ implies that an efficient MPS representation of this state is not possible and is therefore a strong indicator that this circuit is not solvable~\cite{banuls2009matrix, lerose2020influence, foligno2023temporal}.

\begin{figure}
\includegraphics[width=1.0\columnwidth]{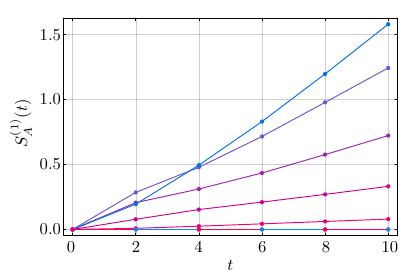}
\caption{The von-Neumann entanglement between two halves of the fixed point (cf. Eq.~\eqref{eq:fixed_point_nonsolv}). The curves are calculated with circuits all using $h=\pi/8$ and from red to blue with $s=0, 1/6, \dots, 5/6, 1$. The alternating flat red and blue line on the bottom of the graph represents the overlapping $s=0$ DU2 and $s=1$ DU circuits that have no entanglement across the cut.}
\label{fig:fixepoint_entanglement}
\end{figure}
\begin{figure}
\includegraphics[width=1.0\columnwidth]{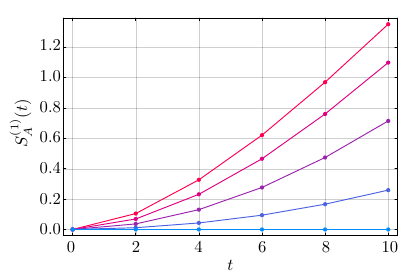}
\caption{The von-Neumann entanglement of the fixed point across the bipartition pictured in Eq.~\eqref{eq:fixed_point_nonsolv}. The curves are calculated with circuits all using $h=\pi/8$ and from red to blue with $s=0.9,0.925,0.95,0.975,1$.}
\label{fig:fixepoint_entanglement2}
\end{figure}

The entanglement growth of the fixed point is displayed in Fig~\ref{fig:fixepoint_entanglement}. For the moderate values of $s$ shown in this figure we see that the slope of the line appears to increase with $s$ but that when approaching the DU point $s=1$ the slope only reaches its asymptotic value after a delay. The mechanism by which the limit $s\rightarrow1$ produces the flat entanglement line seen for DU gates is unclear. In particular, it may be the case that 
\be
    \lim_{s\rightarrow1}\lim_{t\rightarrow\infty} \frac{S_A^{(1)}(t)}{t} \neq \lim_{t\rightarrow\infty} \lim_{s\rightarrow1}\frac{S_A^{(1)}(t)}{t}.  
\ee
Our numerical observations suggest this possibility: as shown in Fig~\ref{fig:fixepoint_entanglement2}, the delay for the curves to reach their asymptotic value appears to grow as the gates get closer to the DU point, suggesting that if the $s\rightarrow1$ limit is taken first the curve will become flat but that if the $t\rightarrow\infty$ limit is taken first we will always reach a slope that will not vanish when $s\rightarrow1$. However, it is unclear whether the asymptotic slope  of these curves will be monotonic in $s$ or whether at some critical $s=s_c$ the asymptotic slope begins to decrease again. Whilst an interesting question this uncertainty is not core to our claim here. We can clearly state that the entanglement of the fixed point does not stay bounded for any values of $s$ except the already known solvable points $s=0$ and $s=1$.

\section{Discussion}
\label{sec:conclusions}

In this paper we introduced a family interacting quantum circuits, which we dubbed `asymptotically solvable', where the dynamics for timescales larger than a tuneable threshold can be accessed exactly, while they appear generic for shorter timescales.

Asymptotically solvable circuits involve an unavoidable degree of spatial inhomogeneity: their distinctive properties require the presence of `special' ergodic gates applied at distances that are not scaling with the system size. The distance between the two special gates enclosing a local subsystem determines the timescale after which its dynamics become exactly accessible, while for shorter timescales it sets an effective system size allowing for efficient numerical evaluation. For periodic arrangements of the inhomogeneities, i.e., in the presence of translational invariance, these circuits can be shown to exactly flow to the DU2 family~\cite{yu2024hierarchical} --- a recently introduced generalisation of dual-unitary circuits~\cite{bertini2019exact} --- under an appropriate spatiotemporal renormalisation group procedure. Asymptotically solvable circuits, however, show a much richer pattern of spatiotemporal correlations compared to both dual-unitary circuits and DU2: the correlations of these circuits are not reduced to a few lines of the causal light cone but are non-trivial within a full area of the light cone whose size is controlled by the distance between the nearest special gates. At the same time, they also admit correlations moving at the speed of light showing that effect of the special gates is not just to cut the system into disconnected parts but rather to `filter' the correlations allowed to propagate to large scales.  

Our work prompts many fascinating questions for future research. A particularly compelling one concerns the possibility of attaining large-scale solvability without ever requiring special gates. Our analysis suggests that this is not possible for strictly homogeneous circuits, however, it might be possible by considering circuits where the gates' parameters have an appropriate spatial modulation despite each gate being generic. In this case, solvable gates might emerge as an asymptotic fixed point of an appropriate spatiotemporal renormalisation group flow. This question becomes particularly relevant when thinking about embedding non-trivial conservation laws in asymptotically solvable circuits. The structure of correlations in the simplest case discussed here only allows for conservation laws with strictly conserved densities or that are moving ballistically at the speed of light. Removing the special gates should allow to study cases where the conserved charges display diffusive or even super-diffusive~\cite{krajnik2022absence, feldmeier2022emergent, denardis2023nonlinear, gopalakrishnan2024distinct, krajnik2024dynamical, rosenberg2024dynamics} transport.    

An exciting suggestion of this work is that even in the absence of solvable impurity gates one might make statements about the entanglement dynamics (i.e. non-trivially bounding from above and below the entanglement velocities) of a non-solvable circuit that is merely related to a solvable circuit through our graphical identities. We were able to make these arguments analytically in the integrable case through the use of the quasiparticle picture but observed numerically that that the results appear to extend to the non-integrable circuit as well. This may suggest that such a statement can be made by use of the graphical conditions alone. We leave these questions to further work.

Another interesting direction concerns the characterisation of spectral properties of asymptotically solvable circuits. For a spatially periodic distribution of the defects our renormalisation group argument suggests that the latter should reduce to those of DU2 circuits for energy scales below a certain threshold set by the distance of two neighbouring special gates. The correlations should, however, depart from the latter over longer energy scales, suggesting that the distance between special gates plays the role of a Thouless energy~\cite{thouless1977maximum}. The situation becomes less obvious in the case of an aperiodic special gate distribution. Other metrics associated with quantum chaos, such as out of time order correlators (OTOCs) and operator space entanglement are also an potential prospect as the renormalisation argument suggests we should be able to demonstrate the emergence of the solvable DU2 behaviour in these quantities as well, particularly the OTOC.

\begin{acknowledgments}
We thank Pavel Kos for valuable comments on the draft and helpful discussions. We acknowledge financial support from the Royal Society through the University Research Fellowship No.\ 201101. 
\end{acknowledgments}

\section*{Data Availability}
The data used to support the findings of this paper are publicly available \cite{supporting_data}.

\appendix

\section{Equivalent representations of the qubit evolution operator}
\label{app:gateformulations}

The evolution operator for the circuit takes the brickwork form
\bea
\mathbb{U} = &\mathbb{U}_o\mathbb{U}_e,\\
\eea
with
\bea
\mathbb{U}_e =\bigotimes_{x=0}^{L-1} U_{x,x+1/2}, \quad
\mathbb{U}_o =\bigotimes_{x=1}^{L} U_{x-1/2,x}\, .
\eea
We use
\be
\label{eq:asysolvable_app} 
\begin{aligned}
&U_{x,x+1/2} = (w_x\otimes w_{x+1/2}) e^{i\frac{\pi}{4} (Z\otimes Z + s_x Y\otimes Y)},\\
&w_x = e^{ih_x Z} e^{i\frac{\pi}{4} X}\!\!.
\end{aligned}
\ee
This may be rewritten as 
\begin{align}
    U_{x,x+1/2} = (e^{ih_x Z}\otimes & e^{ih_{x+1/2} Z}) e^{i\frac{\pi}{4} s_x Z\otimes Z} \\ 
    \cross &(e^{i\pif{4} X}\otimes e^{i\pif{4} X}) e^{i\frac{\pi}{4} Z\otimes Z}\notag 
\end{align}
In total then 
\bea
    \mathbb{U} = \mathbb{W}^\dag e^{i H_o } e^{iH_K} e^{iH_e} e^{iH_K} \mathbb{W}
\eea
where we define
\begin{align}
    \mathbb{W} &= e^{i\pif{4}\sum_{x=1}^L Z_x Z_{x+1/2}} \\
    H_K &= \pif{4} \sum_{j=1}^{2L} X_{j/2}, \qquad
    H_F = \sum_{j=1}^{2L} h_{j/2} X_{j/2} \notag\\
    H_e &= \sum_{x=1}^L [\pif{4}Z_{x-1/2}Z_{x} + \pif{4} s_x Z_{x}Z_{x+1/2}] + H_F, \notag\\
    H_o &= \sum_{x=1}^L [\pif{4} s_{x-1/2} Z_{x-1/2}Z_{x} + \pif{4} Z_{x}Z_{x+1/2}] + H_F.\notag
\end{align}
This can be rewritten as 
\bea
    &\mathbb{U} = \mathbb{W}^\dag  e^{-iH_K} \,\mathbb{U}_{KI}\,\, e^{iH_K} \mathbb{W} \\
    &\mathbb{U}_{KI} = e^{iH_K} e^{iH_o} e^{iH_K} e^{iH_e}
\eea
such that the evolution over many steps is equivalent to that of the (modulated) kicked Ising $\mathbb{U}_{KI}$ up to a similarity transformation that can be written as a tensor product over pairs of sites. This transformation can be easily accounted for and does not significantly affect the quantities considered in this paper so we neglect it.

As mentioned in the main text, the circuit produced using the gates in Eq.~\eqref{eq:asysolvable_app} can also be written as a circuit using a slightly different form of gate shown in Eq.~\eqref{eq:asysolvablelarged} that is more easily generalised to higher dimensions. We show this here. First note that we can write
\bea
    e^{i\frac{\pi}{4} Z\otimes Z} &= e^{i\pif{4}} (e^{i\pif{4}Z}\otimes e^{i\pif{4}Z})\,\, U_C \\
    U_C &= \left(\ketbra*{1}{1}\otimes\mathbbm{1} + \ketbra*{2}{2}\otimes Z \right) \\
    &= \left(\mathbbm{1} \otimes \ketbra*{1}{1} + Z \otimes \ketbra*{2}{2} \right)
\eea
such that we may rearrange
\bea
\!\!\!U_{x,x+1/2} = (w_x e^{i\pif{4}Z}\otimes w_{x+1/2}e^{i\pif{4}Z})\,\, e^{i\frac{\pi}{4} s_x X\otimes X} \,\, U_C
\eea
where we have dropped the irrelevant phase factor. Finally we can identify the single site gates as
\be
    w_x e^{i\pif{4}Z} = e^{ih_x Z} e^{i\pif{4}X} e^{i\pif{4}Z} = i \, e^{i(h_x-\pif{4}) Z} \, H_2
\ee
where $H_2 = (Z + X)/\sqrt{2}$ is the $2\cross 2$ Hadamard gate. Therefore, up to an irrelevant phase factor, the gate is of the form presented in Eq.~\eqref{eq:asysolvablelarged} in the main text. The only adjustment is that the longitudinal fields are shifted $h\rightarrow h-\pif{4}$.

\section{Explicit representation generalised Pauli matrices}
\label{sec:clockalgebra}

The explicit representation for clock, shift and generalised Hadamard matrices reads as 
\be
\label{eq:clockandshift}
\begin{aligned}
& Z_d = \begin{pmatrix}
1 & 0 & 0 & \cdots & 0\\
0 & \omega & 0 & \cdots & 0\\
0 & 0 & \omega^2 & \cdots & 0\\
\vdots & \vdots & \vdots & \ddots & \vdots\\
0 & 0 & 0 &\cdots & \omega^{d-1}
\end{pmatrix},\\
& X_d = 
\begin{pmatrix}
0 & 1 & 0 &\cdots & 0\\
0 & 0 & 1 & \cdots & 0\\
0 & 0 & 0 &\cdots & 0\\
\vdots & \vdots & \vdots & \ddots & \vdots\\
1 & 0 & 0 &\cdots & 0
\end{pmatrix},
\end{aligned}
\ee
and 
\be
H_d = \frac{1}{\sqrt{d}}
\begin{pmatrix}
1 & 1 & 1 &\cdots & 1\\
1 & \omega & \omega^2 & \cdots & \omega^{d-1}\\
1 & \omega^2 & \omega^4 &\cdots & \omega^{2(d-1)}\\
\vdots & \vdots & \vdots & \ddots & \vdots\\
1 & \omega^{d-1} & \omega^{2(d-1)} &\cdots & \omega^{(d-1)^2}
\end{pmatrix}\,,
\ee
where $\omega$ denotes a primitive $d$-th root of unity 
\be
\omega = e^{i \frac{2\pi}{d}}. 
\ee
These matrices fulfil 
\be
\begin{aligned}
&X_d^d = Z_d^d = \mathbbm{1},  &  &X_d Z_d =\omega Z_d X_d,\\
&H_d Z H^\dag_d = X_d\,.  
\end{aligned}
\ee

\section{Proof of Eq.~\eqref{eq:inhom_condition} for $d>2$}
\label{sec:proofgend}

In this appendix we show that the identification in Eq.~\eqref{eq:identification} with $U_s$ given in Eq.~\eqref{eq:asysolvablelarged} gives gates fulfilling Eq.~\eqref{eq:inhom_condition} for all $d\geq 2$. The proof follows directly the one in Sec.~\ref{sec:proofd2}. 

First, setting 
\begin{align}
\fineq[-0.8ex][0.75][1]{
\roundgate[0][0][1][topright][white][$Z$]}&=\sum_{j=1}^d  \ketbra{j}{j}\otimes Z_d^j,\\
 \fineq[-0.8ex][0.75][1]{
\roundgate[0][0][1][bottomright][grey2][$Z$]}&= \sum_{j=1}^d  \ketbra{j}{j}\otimes Z_d^{-j}\notag\\
& = \sum_{j=1}^d  Z_d^{-j}\otimes \ketbra{j}{j},
\end{align}
we immediately find  
\be
\label{eq:Zidentityhigherd}
\fineq[-0.8ex][0.75][1]{
\roundgate[0][0][1][topright][white][$Z$]
\roundgate[-1][0][1][topright][grey2][$Z$]
} = \sum_{j=0}^{d-1} \fineq[-0.8ex][0.7][1]{
    \draw(-.25, -.5)--++(0,1.0);
    \draw(.25,  -.5)--++(0,1.0);
    \cstate[-0.25][0][][]
     \cstate[0.25][0][][]
    \node[scale=1.3] at (-.85,0)  {$Z_d^{-j}$};
    \node[scale=1.3] at (.65,0)  {$Z_d^j$};
} = \sum_{j=0}^{d-1} \fineq[-0.8ex][0.7][1]{
    \draw(-.5, -.25)--++(1.0, 0);
    \draw(-.5,  .25)--++(1.0, 0);
       \cstate[0][0.25][][]
     \cstate[0][-0.25][][]
    \node[scale=1.3] at (0,-.65)  {$Z_d^{-j}$};
    \node[scale=1.3] at (0,.85)  {$Z_d^{j}$};
},
\ee
where the second equality can again be verified by observing that the matrix elements coincide. Analogously we have 
\be
\label{eq:ZXidentityhigherd}
\fineq[-3ex][0.75][1]{
\roundgate[0][0][1][topright][white][$Z$]
\roundgate[-1][0][1][topright][grey2][$Z$]
    \cstate[-0.5][.5][][]
    \node[scale=1.3] at (-.5,.95)  {$X_d^j$};
} = 0\,, \,\,\,
\fineq[1ex][0.75][1]{
\roundgate[0][0][1][topright][white][$Z$]
\roundgate[-1][0][1][topright][grey2][$Z$]
    \cstate[-0.5][-.5][][]
    \node[scale=1.3] at (-.5,-.95)  {$X_d^j$};
} = 0\,,\quad j=1,\ldots,d-1. 
\ee
We also note 
\be
\left(\frac{1}{d} \sum_{j=1}^{d} X_d^j \otimes X_d^j\right)^2 = \frac{1}{d} \sum_{j=1}^{d} X_d^j \otimes X_d^j, 
\ee
which implies 
\be
e^{\frac{i\pi s}{2d} \sum_{j=1}^{d} X_d^j \otimes X_d^j} = 1 + (i^s -1) \frac{1}{d} \sum_{j=1}^{d} X_d^j \otimes X_d^j.
\ee
Therefore, defining some graphical notation for the gates without the single-site unitaries 
\bea
\fineq[-0.8ex][0.75][1]{
\roundgate[0][0][1][topright][red5]}&= e^{\frac{i\pi s}{2d} \sum_{j=1}^{d} X_d^j \otimes X_d^j}\left( \ketbra{j}{j}\otimes Z_d^j \right), \\
\fineq[-0.8ex][0.75][1]{
\roundgate[0][0][1][bottomright][green2]}&= \left[^{\frac{i\pi s}{2d} \sum_{j=1}^{d} X_d^j \otimes X_d^j}\left( \ketbra{j}{j}\otimes Z_d^j \right)\right]^\dag,
\eea
we find 
\be
\begin{aligned}
\fineq[-2.2ex][0.75][1]{
    \roundgate[0][0][1][topright][red5]
    \roundgate[-1][0][1][topleft][green2]
} =& |A|^2 \fineq[-.8ex][0.75][1]{
\roundgate[-1][0][1][topright][grey2][$Z$]
\roundgate[0][0][1][topright][white][$Z$]
}\\
&+ |B|^2 \sum_{j=1}^{d-1} \fineq[-2.6ex][0.75][1]{
\roundgate[-1][0][1][topright][grey2][$Z$]
\roundgate[0][0][1][topright][white][$Z$]
\cstate[.4][0.4][][]
\cstate[-1.4][0.4][][]
\draw[thick](.5, .5)--++(.1, .1);
\draw[thick](-1.5, .5)--++(-.1, .1);

\node[scale=1.3] at (-1.5,.95)  {$X_d^j$};
\node[scale=1.3] at (.5,.95)  {$X_d^j$};
}, 
\end{aligned}
\ee
where we used Eq.~\eqref{eq:ZXidentityhigherd} and set
\be
A= (1+\frac{i^s-1}{d}), \qquad  B = \frac{i^s-1}{d}. 
\ee
Using Eq.~\eqref{eq:Zidentityhigherd} and transposing the leftmost $X^j_d$ to account for the direction of matrix multiplication for spatial propagation (left to right in the diagrams) we have
\be
\begin{aligned}
\fineq[-2.2ex][0.75][1]{
    \roundgate[0][0][1][topright][red5]
    \roundgate[-1][0][1][topleft][green2]
} =& (|A|^2+d|B|^2) \fineq[-0.8ex][0.7][1]{
    \draw(-.5, -.25)--++(1.0, 0);
    \draw(-.5,  .25)--++(1.0, 0);}+ |A|^2 \sum_{j=1}^{d-1} \fineq[-0.8ex][0.7][1]{
    \draw(-.5, -.25)--++(1.0, 0);
    \draw(-.5,  .25)--++(1.0, 0);
       \cstate[0][0.25][][]
     \cstate[0][-0.25][][]
    \node[scale=1.3] at (0,-.65)  {$Z_d^{-j}$};
    \node[scale=1.3] at (0,.85)  {$Z_d^{j}$};}\notag\\
    &  + |B|^2 \sum_{j,k=1}^{d-1} \fineq[-0.8ex][0.7][1]{
    \draw(-.5, -.25)--++(1.0, 0);
    \draw(-.5,  .25)--++(1.0, 0);
       \cstate[0][0.25][][]
     \cstate[0][-0.25][][]
    \node[scale=1.3] at (0,-.65)  {$ Z_d^{-j} $};
    \node[scale=1.3] at (0,.85)  {$X^{-k} Z_d^{j} X^k$};}.
\end{aligned}
\ee
Recalling the algebra of $Z_d$ and $X_d$ we find
\be
\sum_{j,k=1}^{d-1} \fineq[-0.8ex][0.7][1]{
    \draw(-.5, -.25)--++(1.0, 0);
    \draw(-.5,  .25)--++(1.0, 0);
       \cstate[0][0.25][][]
     \cstate[0][-0.25][][]
    \node[scale=1.3] at (0,-.65)  {$ Z_d^{-j} $};
    \node[scale=1.3] at (0,.85)  {$X_d^{-k} Z_d^{j} X_d^k$};}= -\sum_{j=1}^{d-1} \fineq[-0.8ex][0.7][1]{
    \draw(-.5, -.25)--++(1.0, 0);
    \draw(-.5,  .25)--++(1.0, 0);
       \cstate[0][0.25][][]
     \cstate[0][-0.25][][]
    \node[scale=1.3] at (0,-.65)  {$ Z_d^{-j} $};
    \node[scale=1.3] at (0,.85)  {$ Z_d^{j}$};}.
\ee
Therefore, we finally obtain 
\be
\fineq[-0.8ex][0.75][1]{
    \roundgate[0][0][1][topright][red5]
    \roundgate[-1][0][1][topleft][green2]
} = \fineq[-0.8ex][0.7][1]{
    \draw(-.5, -.25)--++(1.0, 0);
    \draw(-.5,  .25)--++(1.0, 0);}+ \left[1-\frac{4}{d} \sin(\frac{\pi}{4}s)^2\right] \sum_{j=1}^{d-1} \fineq[-0.8ex][0.7][1]{
    \draw(-.5, -.25)--++(1.0, 0);
    \draw(-.5,  .25)--++(1.0, 0);
       \cstate[0][0.25][][]
     \cstate[0][-0.25][][]
    \node[scale=1.3] at (0,-.65)  {$Z_d^{-j}$};
    \node[scale=1.3] at (0,.85)  {$Z_d^{j}$};}.
\ee
In the final step we used 
\be
|A|^2+d |B|^2 = 1, \quad |A|^2-|B|^2 = 1- \frac{4}{d} \sin(\frac{\pi}{4}s)^2. 
\ee
Multiplying the top left and right legs respectively by $w_+^\dag$ and $w_+$ and folding the green gate below the orange the second relation above is represented as
\be
\label{eq:properties_paulihigherd}
\fineq[-0.8ex][0.7][1]{
   \roundgate[0][0][1][topright][\ZXfcolor][1]
    \cstate[-0.5][-0.5]
    \cstate[-0.5][0.5]
}
=
\fineq[-0.8ex][0.7][1]{
    \draw(-.5, -.5)--++(1.0, 0);
    \draw(-.5,  .5)--++(1.0, 0);
    \cstate[-0.5][-0.5]
    \cstate[-0.5][0.5]
}+ 
 \left[1-\frac{4}{d} \sin(\frac{\pi}{4}s)^2\right] \sum_{j=1}^{d-1}
\fineq[-0.8ex][0.7][1]{
    \draw(-.5, -.5)--++(1.0, 0);
    \draw(-.5,  .5)--++(1.0, 0);
	\node[scale=1.3] at (-1.1,-.5)  {$Z_d^{-j}$};
	\node[scale=1.3] at (-1.1,.5)  {$\widetilde{X}^{j}_d$};
    \cstate[-0.5][-0.5][][black]
    \cstate[-0.5][0.5][][black]
},
\ee
where we introduced 
\be
\widetilde{X}_d = e^{- i \sum_{j=1}^{d-1} h_{j,\pm} Z^j_q} {X}_d e^{i \sum_{j=1}^{d-1} h_{j,\pm} Z^j_q}\,. 
\ee
Note that for $d>4$ the family in Eq.~\eqref{eq:asysolvablelarged} does not contain a dual unitary point as the coefficient of the second term in Eq.~\eqref{eq:properties_paulihigherd} does not vanish for any real $s$. 

Proceeding analogously we find 
\begin{align}
& \fineq[-0.8ex][0.7][1]{
   \roundgate[0][0][1][topright][\ZXfcolor][1]
    \cstate[-0.5][-0.5][][]
    \cstate[0.5][0.5][][]
    \cstate[-0.5][0.5]
    \node[scale=1.3] at (-1.6,-.5)  {$Z_d^{n} X_d^m$};
    \node[scale=1.3] at (1.1,.5)  {$Z_d^{k}$};
}
=
\fineq[-0.8ex][0.7][1]{
    \roundgate[0][0][1][topright][\Zfcolor][1]
    \cstate[-0.5][-0.5][][]
    \cstate[0.5][0.5][][]
    \cstate[-0.5][0.5]
    \node[scale=1.3] at (-1.6,-.5)  {$Z_d^{n} X_d^m$};
    \node[scale=1.3] at (1.1,.5)  {$Z_d^{k}$};
} 
= 0
 \label{eq:properties_paulihigherd3}
\end{align}
where the first equality can be shown by commuting the $e^{\frac{i\pi s}{2d} \sum_{j=1}^{d} X_d^j\otimes X_d^j}$ through the Hadamard gates at the cost of rotating the $X_d\rightarrow Z_d^{-1}$ and then simply commuting through the further diagonal gates, finally allowing it to cancel due to unitarity. The second equality is shown by noting that as a simple consequence of Eq.~\eqref{eq:ZXidentityhigherd}
\begin{align}
& 
\fineq[-0.8ex][0.7][1]{
    \roundgate[0][0][1][topright][\Zfcolor][1]
    \cstate[-0.5][-0.5][][]
    \cstate[-0.5][0.5]
    \node[scale=1.3] at (-1.6,-.5)  {$Z_d^{n} X_d^m$};
}
= 0
\end{align}
Therefore, we see that any off diagonal
operator on the bottom left corner produces off-diagonal operators in the top right corners. Furthermore, when those off-diagonal operators reach the $s=0$ gate at the end of the chain their amplitudes vanish. We can therefore conclude the proof as in Sec.~\ref{sec:proofd2}.

\section{Conditions for non-ergodicity}
\label{app:nonergodicity}
{
In this appendix we detail the proof of Property~\ref{prop:ergodic}. Much of this proof follows from a similar one in Ref.~\cite{Kos_2021}. To do this we shall assume the existence of a second uni-modular eigenvalue $|\lambda_1|=1$ and work backwards to infer what conditions must be satisfied for its existence. We assume there exists a normalised eigenvector $\ket*{v}$ such that
\begin{equation}
    T_\ell \ket*{v} = e^{i\theta}\ket*{v},
    \quad\quad
    \braket{\ocircle\dots\ocircle}{v} = 0.
\end{equation}
where in this Appendix we will always be using \textit{normalised vectorised operators} such that
\be
\begin{aligned}
\ket*{\ocircle}&=\frac{1}{\sqrt{d}}\sum_{i=1}^d \ket*{i}\otimes\ket*{i}, \\
\ket*{\sigma^{(m)}}&=\frac{1}{\sqrt{d}}\sum_{i=1}^d (\sigma^{(m)})_{ij} \ket*{i}\otimes\ket*{j}
\end{aligned}
\ee
and we will use the shorthand for the vectorised Pauli operators $\{\ket*{X},\ket*{Y},\ket*{Z}\} = \{\ket*{\sigma^{(1)}},\ket*{\sigma^{(2)}},\ket*{\sigma^{(3)}}\}$.
We will now show that $\bra*{v}$ must also be a left eigenvector of $T_\ell$ with the same eigenvalue
\be
\bra*{v}T_\ell = \bra*{v}e^{i\theta},
\ee
as a consequence of the uni-modularity eigenvalue. We will focus on the two hermitian operators $T_\ell^\dag T_\ell$ and $T_\ell T_\ell^\dag$, both of whom are non-expansive. Trivially, we have $\bra*{v}T_\ell^\dag T_\ell\ket*{v}=1$, but we also must have 
\be
\begin{aligned}
    \bra*{v}T_\ell T_\ell^\dag\ket*{v} &= \sum_\alpha |\bra*{v} T_\ell\ket*{\alpha}|^2 \geq 1, \\
    \bra*{v}T_\ell T_\ell^\dag\ket*{v} &= \sum_{i} \sigma_i(T_\ell)^2 \cdot|\braket{i}{v}|^2 \leq 1, \\
    \implies \bra*{v}T_\ell T_\ell^\dag\ket*{v} &= 1.
\end{aligned}
\ee
where in the first inequality we expanded in an orthonormal basis $\{\ket{\alpha}\}$ containing $\ket{v}$, while in the second we expanded in the eigenbasis of $T_\ell T_\ell^\dag$ and denoted by $\sigma_i(T_\ell)^2$ its eigenvalues (which are the singular values of $T_\ell$ squared). Note that the first inequality relies on the fact that the eigenvalue is uni-modular and if we lower its magnitude the inequalities will have finite overlap and we will not be able to make the final statement. 

Now, we can turn the second statement into an equality
\be
\begin{aligned}
    \bra*{v}T_\ell T_\ell^\dag\ket*{v} &= \sum_{i} \sigma_i(T_\ell)^2 \cdot|\braket{i}{v}|^2 = 1, \\
    \implies \braket{i}{v} &= 0 \quad\text{if}\quad \sigma_i(T_\ell) \neq 1.
\end{aligned}
\ee
This means $\ket*{v}$ is living in the uni-modular eigenspace, so
\begin{equation}
    T_\ell T_\ell^\dag\ket*{v} = T_\ell^\dag T_\ell\ket*{v} = \ket*{v},
\label{eq:normaltransfer_eigen}
\end{equation}
where we the logic follows equally for $T_\ell^\dag T_\ell\ket*{v}$. Now
\be
\begin{aligned}
    T_\ell^\dag T_\ell\ket*{v} &= \ket*{v} \\
    T_\ell^\dag \ket*{v} &= e^{-i\theta}\ket*{v} \\
    \bra*{v} T_\ell &= \bra*{v}e^{i\theta}.
\end{aligned}
\ee
The benefit of this setup is that the relations in Eq.~\eqref{eq:normaltransfer_eigen} are very pleasant to work with. We have
\be
\begin{aligned}
T_\ell^\dag T_\ell &= \frac{1}{4^2}
\fineq[-0.8ex][0.55][1]{
    \roundgate[0][3][1][bottomright][\Zfccolor][1]
    \roundgate[1][2][1][bottomright][\ZXfccolor][1]
    \roundgate[2][3][1][bottomright][\ZXfccolor][1]
    \roundgate[3][2][1][bottomright][\ZXfccolor][1]
    \roundgate[4][3][1][bottomright][\Zfccolor][1]
    \cstate[-0.5][2.5]
    \cstate[-0.5][3.5]
    \cstate[4.5][2.5]
    \cstate[4.5][3.5]
    \roundgate[0][0][1][topright][\Zfcolor][1]
    \roundgate[1][1][1][topright][\ZXfcolor][1]
    \roundgate[2][0][1][topright][\ZXfcolor][1]
    \roundgate[3][1][1][topright][\ZXfcolor][1]
    \roundgate[4][0][1][topright][\Zfcolor][1]
    \cstate[-0.5][-0.5]
    \cstate[-0.5][0.5]
    \cstate[4.5][-0.5]
    \cstate[4.5][0.5]
}\\
& = \frac{1}{4^2}
\fineq[-0.8ex][0.55][1]{
    \roundgate[0][2.5][1][bottomright][\Zfccolor][1]
    \roundgate[4][2.5][1][bottomright][\Zfccolor][1]
    \cstate[-0.5][2]
    \cstate[-0.5][3]
    \cstate[4.5][2]
    \cstate[4.5][3]
    \roundgate[0][.5][1][topright][\Zfcolor][1]
    \roundgate[4][.5][1][topright][\Zfcolor][1]
    \cstate[-0.5][0]
    \cstate[-0.5][1]
    \cstate[4.5][0]
    \cstate[4.5][1]
    \draw(.5, 1)--++(0, 1.0);
    \draw(1.5, 0)--++(0, 3.0);
    \draw(2.5, 0)--++(0, 3.0);
    \draw(3.5, 1)--++(0, 1.0);
}
\end{aligned}
\ee
and
\be
\begin{aligned}
T_\ell T_\ell^\dag &= \frac{1}{4^2}
\fineq[-0.8ex][0.55][1]{
    \roundgate[0][2.5][1][topright][\Zfcolor][1]
    \roundgate[1][3.5][1][topright][\ZXfcolor][1]
    \roundgate[2][2.5][1][topright][\ZXfcolor][1]
    \roundgate[3][3.5][1][topright][\ZXfcolor][1]
    \roundgate[4][2.5][1][topright][\Zfcolor][1]
    \cstate[-0.5][2]
    \cstate[-0.5][3]
    \cstate[4.5][2]
    \cstate[4.5][3]
    \roundgate[0][0.5][1][bottomright][\Zfccolor][1]
    \roundgate[1][-0.5][1][bottomright][\ZXfccolor][1]
    \roundgate[2][0.5][1][bottomright][\ZXfccolor][1]
    \roundgate[3][-0.5][1][bottomright][\ZXfccolor][1]
    \roundgate[4][0.5][1][bottomright][\Zfccolor][1]
    \cstate[-0.5][0]
    \cstate[-0.5][1]
    \cstate[4.5][0]
    \cstate[4.5][1]
    \draw(0.5, 1)--++(0, 1.0);
    \draw(1.5, 1)--++(0, 1.0);
    \draw(2.5, 1)--++(0, 1.0);
    \draw(3.5, 1)--++(0, 1.0);
}\\
&= \frac{1}{4^2}
\fineq[-0.8ex][0.55][1]{
    \roundgate[0][2.5][1][topright][\Zfcolor][1]
    \roundgate[1][3.5][1][topright][\ZXfcolor][1]
    \roundgate[3][3.5][1][topright][\ZXfcolor][1]
    \roundgate[4][2.5][1][topright][\Zfcolor][1]
    \cstate[-0.5][2]
    \cstate[-0.5][3]
    \cstate[4.5][2]
    \cstate[4.5][3]
    \roundgate[0][0.5][1][bottomright][\Zfccolor][1]
    \roundgate[1][-0.5][1][bottomright][\ZXfccolor][1]
    \roundgate[3][-0.5][1][bottomright][\ZXfccolor][1]
    \roundgate[4][0.5][1][bottomright][\Zfccolor][1]
    \cstate[-0.5][0]
    \cstate[-0.5][1]
    \cstate[4.5][0]
    \cstate[4.5][1]
    \draw(0.5, 1)--++(0, 1.0);
    \draw(1.5, 0)--++(0, 3.0);
    \draw(2.5, 0)--++(0, 3.0);
    \draw(3.5, 1)--++(0, 1.0);
}
\end{aligned}
\ee
where we are denoting the conjugate gates in red. We see that hermitian operators with a large enough support will always split into two distinct pieces separated by identities. We name these pieces
\be
\begin{aligned}
l_1 = \frac{1}{4}
\fineq[-0.8ex][0.6][1]{
    \roundgate[0][2.5][1][bottomright][\Zfccolor][1]
    \cstate[-0.5][2]
    \cstate[-0.5][3]
    \roundgate[0][.5][1][topright][\Zfcolor][1]
    \cstate[-0.5][0]
    \cstate[-0.5][1]
    \draw(.5, 1)--++(0, 1.0);
}\,\,,
\quad&\quad
r_1 = \frac{1}{4}
\fineq[-0.8ex][0.6][1]{
    \roundgate[4][2.5][1][bottomright][\Zfccolor][1]
    \cstate[4.5][2]
    \cstate[4.5][3]
    \roundgate[4][.5][1][topright][\Zfcolor][1]
    \cstate[4.5][0]
    \cstate[4.5][1]
    \draw(3.5, 1)--++(0, 1.0);
}\,\,, \\
l_2 = \frac{1}{4}
\fineq[-0.8ex][0.6][1]{
    \roundgate[0][2.5][1][topright][\Zfcolor][1]
    \roundgate[1][3.5][1][topright][\ZXfcolor][1]
    \cstate[-0.5][2]
    \cstate[-0.5][3]
    \roundgate[0][0.5][1][bottomright][\Zfccolor][1]
    \roundgate[1][-0.5][1][bottomright][\ZXfccolor][1]
    \cstate[-0.5][0]
    \cstate[-0.5][1]
    \draw(0.5, 1)--++(0, 1.0);
    \draw(1.5, 0)--++(0, 3.0);
}\,\,,
\quad&\quad
r_2 = \frac{1}{4}
\fineq[-0.8ex][0.6][1]{
    \roundgate[3][3.5][1][topright][\ZXfcolor][1]
    \roundgate[4][2.5][1][topright][\Zfcolor][1]
    \cstate[4.5][2]
    \cstate[4.5][3]
    \roundgate[3][-0.5][1][bottomright][\ZXfccolor][1]
    \roundgate[4][0.5][1][bottomright][\Zfccolor][1]
    \cstate[4.5][0]
    \cstate[4.5][1]
    \draw(2.5, 0)--++(0, 3.0);
    \draw(3.5, 1)--++(0, 1.0);
}\,.
\end{aligned}
\ee
For general gates, these channels are Hermitian, non-expansive and positive semi-definite so must have real eigenvalues $\lambda\in[0, 1]$. We want to show that the pairs of channels on each edge share unit eigenvectors such that we can construct $\ket*{v}$ that is both an eigenvector of $T^\dag_\ell T_\ell$ and $T_\ell T^\dag_\ell$. We now argue that we can consider
\be
\begin{aligned}
 &(l_1\otimes\mathbbm{1}) l_2 \ket*{a} = \ket*{a}, \quad (\mathbbm{1}\otimes r_1) r_2 \ket*{b} = \ket*{b}, \\ &\braket{\ocircle\ocircle}{a} = \braket{\ocircle\ocircle}{b} = 0,
\end{aligned}
\ee
as necessary and sufficient conditions for the existence of a shared second unit eigenvector of $l_1\otimes\mathbbm{1}$ ($\mathbbm{1}\otimes r_1$) and $l_2$ ($r_2$). This is because the only way that the product of two hermitian non-expansive positive semi-definite operators (i.e. $\lambda\in[0, 1]$) can have a uni-modular eigenvector is if the vector is a uni-modular eigenvector of both operators separately.

It is quite easy to determine if there are any extra 1s in the spectra of these operators. We consider
\begin{equation}
    \!\!\!n_l \!=\! \lim_{k\rightarrow\infty} \tr \left[(l_1\otimes\mathbbm{1}) l_2 \right]^k\!\!\!,
    \quad
    n_r \!=\! \lim_{k\rightarrow\infty} \tr \left[(\mathbbm{1}\otimes r_1) r_2 \right]^k\!\!\!.
\label{eq:num_unimodular_leftright}
\end{equation}
where $\mathbbm{1}$ is the identity on both the forward and backward copies. Despite the fact that these operators are not Hermitian, they are non-expansive and we argued above that they cannot have any uni-modular eigenvalues that are not $1$. Therefore, Eq.~\eqref{eq:num_unimodular_leftright} will converge to integers. 
We can calculate the values of $n_l$ and $n_r$. We can rearrange the terms in the products such that
\begin{equation}
    n_l = \lim_{k\rightarrow\infty} \tr \left[D_l D_l^\dag \right]^k,
    \quad
    n_r = \lim_{k\rightarrow\infty} \tr \left[D_r D_r^\dag \right]^k,
\end{equation}
where
\be
\begin{aligned}
D_l = \frac{1}{4}
\fineq[-0.8ex][0.6][1]{
    \roundgate[0][-1][1][topright][\Zfcolor][1]
    \roundgate[1][0][1][topright][\ZXfcolor][1]
    \roundgate[0][1][1][topright][\Zfcolor][1]
    \cstate[-0.5][-1.5]
    \cstate[-0.5][-.5]
    \cstate[-0.5][1.5]
    \cstate[-0.5][.5]
},
\quad\quad
D_r = \frac{1}{4}
\fineq[-0.8ex][0.6][1]{
    \roundgate[1][-1][1][topright][\Zfcolor][1]
   \roundgate[0][0][1][topright][\ZXfcolor][1]
    \roundgate[1][1][1][topright][\Zfcolor][1]
    \cstate[1.5][-1.5]
    \cstate[1.5][-.5]
    \cstate[1.5][1.5]
    \cstate[1.5][.5]
}.
\end{aligned}
\label{eq:string_merge}
\ee
The question is now whether or not there is a second uni-modular \textit{singular value} of $D_l$ and/or $D_r$. This is a very physical question: we are asking if the non-trivial operator strings produced by spatial evolution from the infinite temperature boundary can combine into larger strings without their amplitudes decaying.

We now focus on the left edge as the analysis is identical for both edges. By writing out the expansions of $D_l$ we can check the number of singular values. Before doing this, let us fix notation. We will take the subsystem to be positioned such that the leftmost spin the transfer matrix acts on is positioned at $x=1/2$ such that $s_1=s_{1+\ell}=0$.

To expand $D_l$ it is helpful to state the operator expansions of the following objects
\be
\begin{aligned}
\fineq[-0.8ex][0.7][1]{
   \roundgate[0][0][1][topright][\ZXfcolor][1]
    \cstate[-0.5][-0.5][][]
    \cstate[-0.5][0.5]
	\node[scale=1.3] at (-1.1,-.5)  {$X$};
}
&= 
\sin(\frac{\pi s}{2})
\fineq[-0.8ex][0.7][1]{
    \draw(-.5, -.5)--++(1.0, 0);
    \draw(-.5,  .5)--++(1.0, 0);
	\node[scale=1.3] at (-1.1,.5)  {$\widetilde{X}$};
    \cstate[-0.5][-0.5]
    \cstate[-0.5][0.5][][black]
},
\\
\fineq[-0.8ex][0.7][1]{
   \roundgate[0][0][1][topright][\ZXfcolor][1]
    \cstate[-0.5][-0.5][][]
    \cstate[-0.5][0.5]
	\node[scale=1.3] at (-1.1,-.5)  {$Y$};
}
&= 
\sin(\frac{\pi s}{2})
\fineq[-0.8ex][0.7][1]{
    \draw(-.5, -.5)--++(1.0, 0);
    \draw(-.5,  .5)--++(1.0, 0);
	\node[scale=1.3] at (-1.1,-.5)  {$X$};
	\node[scale=1.3] at (-1.1,.5)  {$\widetilde{Y}$};
    \cstate[-0.5][-0.5][][]
    \cstate[-0.5][0.5][][black]
},
\end{aligned}
\ee
simple substitution therefore allows us to write
\begin{align}
\fineq[-0.8ex][0.7][1]{
   \roundgate[0][0][1][topright][\ZXfcolor][1]
    \cstate[-0.5][-0.5][][]
    \cstate[-0.5][0.5]
	\node[scale=1.3] at (-1.1,-.5)  {$\widetilde{Y}$};
}
=& 
\sin(\frac{\pi s}{2}) \cos(2h)
\fineq[-0.8ex][0.7][1]{
    \draw(-.5, -.5)--++(1.0, 0);
    \draw(-.5,  .5)--++(1.0, 0);
	\node[scale=1.3] at (-1.1,-.5)  {$X$};
	\node[scale=1.3] at (-1.1,.5)  {$\widetilde{Y}$};
    \cstate[-0.5][-0.5][][]
    \cstate[-0.5][0.5][][black]
}\notag\\
+&\sin(\frac{\pi s}{2}) \sin(2h)
\fineq[-0.8ex][0.7][1]{
    \draw(-.5, -.5)--++(1.0, 0);
    \draw(-.5,  .5)--++(1.0, 0);
	\node[scale=1.3] at (-1.1,.5)  {$\widetilde{X}$};
    \cstate[-0.5][-0.5]
    \cstate[-0.5][0.5][][black]
}, \label{eq:properties1}\\
\fineq[-0.8ex][0.7][1]{
   \roundgate[0][0][1][topright][\ZXfcolor][1]
    \cstate[-0.5][-0.5][][]
    \cstate[-0.5][0.5]
	\node[scale=1.3] at (-1,-.5)  {$\widetilde{X}$};
}
=& 
\sin(\frac{\pi s}{2}) \cos(2h)
\fineq[-0.8ex][0.7][1]{
    \draw(-.5, -.5)--++(1.0, 0);
    \draw(-.5,  .5)--++(1.0, 0);
	\node[scale=1.3] at (-1.1,.5)  {$\widetilde{X}$};
    \cstate[-0.5][-0.5]
    \cstate[-0.5][0.5][][black]
}\notag\\
-&\sin(\frac{\pi s}{2}) \sin(2h)
\fineq[-0.8ex][0.7][1]{
    \draw(-.5, -.5)--++(1.0, 0);
    \draw(-.5,  .5)--++(1.0, 0);
	\node[scale=1.3] at (-1.1,-.5)  {$X$};
	\node[scale=1.3] at (-1.1,.5)  {$\widetilde{Y}$};
    \cstate[-0.5][-0.5][][]
    \cstate[-0.5][0.5][][black]
},\label{eq:properties2}
\end{align}
where $\tilde X$ and $\tilde Y$ are defined in Eqs.~\eqref{eq:tildeops} and any explicitly written longitudinal field $h$ corresponds to that used in the definition of the $\tilde X, \tilde Y$ on the LHS of the equation.
Finally, we will need
\begin{align}
&\hspace{-.35cm}\fineq[-0.8ex][0.6][1]{
   \roundgate[0][0][1][topright][\ZXfcolor][1]
    \cstate[-0.5][-0.5]
    \cstate[-0.5][0.5][][]
	\node[scale=1.3] at (-1.1,.5)  {$Z$};}
=
\sin(\frac{\pi s}{2}) 
\fineq[-0.8ex][0.6][1]{
    \draw(-.5, -.5)--++(1.0, 0);
    \draw(-.5,  .5)--++(1.0, 0);
	\node[scale=1.3] at (-1.1,-.5)  {$Z$};
	\node[scale=1.3] at (-1.1,.5)  {$\widetilde{X}$};
    \cstate[-0.5][-0.5][][black]
    \cstate[-0.5][0.5][][black]
} \label{eq:properties_additional} \\
&\hspace{-.35cm}\fineq[-0.8ex][0.6][1]{
   \roundgate[0][0][1][topright][\ZXfcolor][1]
    \cstate[-0.5][-0.5][][]
    \cstate[-0.5][0.5][][]
	\node[scale=1.3] at (-1.1,.5)  {$Z$};
	\node[scale=1.3] at (-1.1,-.5)  {$\widetilde{Y}$};
}
=
-\cos(2h)\left(
\fineq[-0.8ex][0.6][1]{
    \draw(-.5, -.5)--++(1.0, 0);
    \draw(-.5,  .5)--++(1.0, 0);
	\node[scale=1.3] at (-1.1,-.5)  {$Y$};
	\node[scale=1.3] at (-1.1,.5)  {$Z$};
    \cstate[-0.5][-0.5][][black]
    \cstate[-0.5][0.5][][]
}
+\cos(\frac{\pi s}{2}) 
\fineq[-0.8ex][0.6][1]{
    \draw(-.5, -.5)--++(1.0, 0);
    \draw(-.5,  .5)--++(1.0, 0);
	\node[scale=1.3] at (-1.1,.5)  {$\wt{X}$};
	\node[scale=1.3] at (-1.1,-.5)  {$X$};
    \cstate[-0.5][-0.5][][]
    \cstate[-0.5][0.5][][black]
}
\right) \notag\\
&\qquad\qquad +\sin(2h)\left(
\fineq[-0.8ex][0.6][1]{
    \draw(-.5, -.5)--++(1.0, 0);
    \draw(-.5,  .5)--++(1.0, 0);
	\node[scale=1.3] at (-1.1,-.5)  {$Z$};
    \cstate[-0.5][-0.5][][black]
    \cstate[-0.5][0.5]
}
+\cos(\frac{\pi s}{2}) 
\fineq[-0.8ex][0.6][1]{
    \draw(-.5, -.5)--++(1.0, 0);
    \draw(-.5,  .5)--++(1.0, 0);
	\node[scale=1.3] at (-1.1,.5)  {$\wt{Y}$};
    \cstate[-0.5][-0.5]
    \cstate[-0.5][0.5][][black]
}
\right)\notag
\end{align}
Defining 
\begin{equation}
    J_x = \frac{\pi s_{x}}{4}
\end{equation}
for brevity and using the relations above one finds
\begin{align}
    D_l =& \kb{\oc\oc}{\oc\oc} 
    + \left( \cos2J_{\frac{1}{2}} \ket*{\oc\tilde Y} + \sin2J_{\frac{1}{2}} \ket*{\tilde Y \tilde X} \right) \bra*{\oc Z} \notag \\
    &+ \sin2h_{\frac{1}{2}} \left(\sin2J_{\frac{1}{2}} \ket*{\oc\tilde X} + \cos2J_{\frac{1}{2}} \ket*{\tilde Y\tilde Y} \right) \bra*{Z\oc} \notag\\
    &+ \cos2h_{\frac{1}{2}} \left(\sin2J_{\frac{1}{2}} \ket*{\oc\tilde Y} - \cos2J_{\frac{1}{2}} \ket*{\tilde Y\tilde X} \right) \bra*{ZX} \notag\\
    &+ \sin2h_{\frac{1}{2}} \kb{\tilde Y \oc}{ZZ} - \cos2h_{\frac{1}{2}} \kb{\tilde Y Z}{ZY}
\end{align}
where $h_{\frac{1}{2}}$ is the longitudinal field applied to the leftmost spin acted on by $T_\ell$. This form verifies immediately that $D_l$ always has \textit{at least two unit singular values} $D_l^\dag D_l \ket*{\oc\oc} = \ket*{\oc\oc}$ and $D_l^\dag D_l \ket*{\oc Z} = \ket*{\oc Z}$, with more appearing at the special points $h_{\frac{1}{2}}=0, \pi/4$. As we discuss in the main text, one can verify numerically that there do indeed exist points with all $h_x=\pif{4}$ where $T_\ell$ possesses non-trivial conserved quantities. Therefore, we will not be able to rule out other unimodular eigenvalues in this case. We can however address the other regions.

\subsection{Unit eigenvectors for $ 0<|h_{\frac{1}{2}}|<\pi/4$}
We verified above that in this region $n_l=2$ and thus there exists a second mutual unit eigenvector of $l_1\otimes\mathbbm{1}$ and $l_2$. The two unit eigenvectors can be checked to be
\be
\begin{aligned}
\ket*{a_0} &= \ket*{\oc\oc}, \\
\ket*{a_1} &= (U_{\frac{1}{2}, 1} \otimes U_{\frac{1}{2}, 1}^*) \ket*{\oc Z} \\
&= \cos2J_{\frac{1}{2}}\ket*{\oc\tilde Y} + \sin2J_{\frac{1}{2}}\ket*{Z\tilde X}, 
\end{aligned}
\ee
where the latter is trivially an eigenstate $l_2$ and can be seen to be an eigenstate of $l_1\otimes\mathbbm{1}_1$ through $D_l^\dag D_l \ket*{\oc Z} = \ket*{\oc Z}$. Therefore, any unit eigenvector of $T_\ell$ must take the form
\be
\begin{aligned}
\ket*{v_{2\ell}} = \sum_{i=0}^1 \ket*{a_i}\otimes\ket*{\psi_i},
\quad\quad \sum_{i=0}^1 \braket*{\psi_i} = 1.
\end{aligned}
\ee
We now prove that $\braket*{\psi_1}=0$ within this region of the parameter space. To do this, we run a self-consistency check: assuming $\ket*{v_{2\ell}}$ to be an eigenstate we must have 
\be
\begin{aligned}
    \| (P_{Z \oc}\otimes\mathbbm{1}_{2\ell-2}) T_\ell \ket*{v_L} \| = 0,
\end{aligned}
\ee
where $P_{\alpha\beta} = \ketbra*{\alpha\beta}{\alpha\beta}$ and $\mathbbm{1}_{x}$ is the identity on $x$ doubled qubits. Using the operator expansions in Eqs.~\eqref{eq:properties2} and~\eqref{eq:properties_additional} and the fact that $\ket{\psi_i}$ must be eigenvectors of $r_1$ we have
\be
\label{eq:appA_psi1_zero}
\begin{aligned}
    \bra*{\psi_1}\left(\mathcal{M}_{\wt{X}Z}\otimes\mathbbm{1}_{2\ell-3}\right)\ket*{\psi_1} = 0,
\end{aligned}
\ee
where
\be
\begin{aligned}
    \mathcal{M}_{\alpha\beta} = \frac{1}{4}
\fineq[-0.8ex][0.6][1]{
    \roundgate[0][2.5][1][bottomright][\ZXfccolor][1]
    \cstate[-0.5][2][][]
    \cstate[-0.5][3][][]
    \roundgate[0][.5][1][topright][\ZXfcolor][1]
    \cstate[-0.5][0][][]
    \cstate[-0.5][1][][]
    \draw(.5, 1)--++(0, 1.0);
    
	\node[scale=1.3] at (-1.1,2)  {$\beta$};
	\node[scale=1.3] at (-1.1,3)  {$\alpha$};
	\node[scale=1.3] at (-1.1,0)  {$\alpha$};
	\node[scale=1.3] at (-1.1,1)  {$\beta$};
}\,\,
\end{aligned},
\ee
with the gates pictured here being folded variants of $U_{1,3/2}$. Outside of $s_1=1$, this channel is positive definite and diagonal in the Pauli basis
\be
\begin{aligned}
\mathcal{M}_{\wt{X}Z} =& \quad
\cos^22h_1\left( \cos^22J_{1} \kb{\ocircle}{\ocircle} + \kb{Z}{Z} \right) \\
&+ \sin^22h_1\left( \cos^22J_{1} \kb{X}{X} + \kb{Y}{Y} \right).
\end{aligned} 
\ee
Therefore Eq.~\eqref{eq:appA_psi1_zero} implies that for $0<|h_1|<\pi/4$ and  $s_1\neq1$
\begin{equation}
    \braket*{\psi_1}{\psi_1} = 0.
\end{equation}
To check the point where $s_1=1$ we look at
\be
\begin{aligned}
    \| (P_{ZZ}\otimes\mathbbm{1}_{2\ell-2}) T_\ell \ket*{v_\ell} \| &= 0 \\
    \implies     
    \bra*{\psi_1}\left(\mathcal{M}_{\wt{X}Y}\otimes\mathbbm{1}_{2\ell-3}\right)\ket*{\psi_1} &= 0.
\end{aligned}
\ee
Evaluated \textit{specifically at} $s_1=1$ we have that $\mathcal{M}_{\wt{X}Y}$ is also positive definite and diagonal in the Pauli basis
\begin{align}
&\mathcal{M}_{\wt{X}Y} = \sin^22h_1\left(\sin^22h_1\kb{\ocircle}{\ocircle} + \cos^22h_1\kb{X}{X}\right) \notag\\
&+ \cos^22h_1\left(\cos^22h_1\kb{Z}{Z} + \sin^22h_1\kb{Y}{Y}\right).
\end{align} 
Therefore, we find that also at  $0<|h_1|<\pi/4$ and $s_1=1$ we must have $\braket*{\psi_1}{\psi_1}=0$.

In conclusion: whether $s_1 = 1$ or not, if $0<h_{\frac{1}{2}}<\pif{4}$ and $0<h_{1}<\pif{4}$ we must have $\braket*{\psi_1}{\psi_1}=0$ and therefore
\be
\begin{aligned}
\ket*{v_L} = \ket*{\ocircle\ocircle}\otimes\ket*{\psi_0},
\quad\quad \braket*{\psi_0} = 1.
\end{aligned}
\ee
In these cases we must have
\be
\begin{aligned}
\ket*{v_{2\ell}} = \ket*{\oc\oc}\otimes\ket*{v_{2\ell-2}}
\end{aligned}
\ee
and likewise on the right edge. We now see that $\ket*{v_{2\ell-2}}$ must be a unimodular eigenvector of 
\be
\begin{aligned}
(\bra*{\oc\oc}\otimes\mathbbm{1}_{2\ell-2}) T_\ell (\ket*{\oc\oc}\otimes\mathbbm{1}_{2\ell-2}).
\end{aligned}
\label{eq:projected_transfer_app}
\ee
At this point if $\ell$ is large enough to require it we may repeat the analysis presented so far but for the smaller channel defined by Eq.~\eqref{eq:projected_transfer_app}. In doing this, one ends up asking the same questions about the similarly defined channels $l_1'\otimes\mathbbm{1}$ and $l_2'$ that act on the sites $3/2, 2$. In this case, it is immediately obvious that the equivalent edge channel
\be
\begin{aligned}
l_1' = \frac{1}{4}
\fineq[-0.8ex][0.6][1]{
    \roundgate[0][2.5][1][bottomright][\ZXfccolor][1]
    \cstate[-0.5][2]
    \cstate[-0.5][3]
    \roundgate[0][.5][1][topright][\ZXfcolor][1]
    \cstate[-0.5][0]
    \cstate[-0.5][1]
    \draw(.5, 1)--++(0, 1.0);
}
= \kb{\oc}{\oc} + \cos^22J_1 \kb{Z}{Z}
\end{aligned}
\ee
does not have any non-trivial unit eigenvectors since by definition $0 < s_1 \leq 1$. Therefore, we can conclude that $\ket*{v_{2\ell-2}}$ is trivial on it's edge site, giving
\be
\begin{aligned}
\ket*{v_{2\ell}} = \ket*{\oc\oc}\otimes\ket*{\oc}\otimes\ket*{v_{2\ell-3}}.
\end{aligned}
\ee
The next site can be probed by projecting $l_2'$ like so
\be
\begin{aligned}
(\bra*{\oc}\otimes\mathbbm{1}_1)\, l_2'\, (\ket*{\oc}\otimes\mathbbm{1}_1) = 
\frac{1}{8}
\fineq[-0.8ex][0.6][1]{
    \roundgate[0][2.5][1][topright][\ZXfcolor][1]
    \roundgate[1][3.5][1][topright][\ZXfcolor][1]
    \cstate[-0.5][2]
    \cstate[-0.5][3]
    \cstate[0.5][4]
    \roundgate[0][0.5][1][bottomright][\ZXfccolor][1]
    \roundgate[1][-0.5][1][bottomright][\ZXfccolor][1]
    \cstate[0.5][-1]
    \cstate[-0.5][0]
    \cstate[-0.5][1]
    \draw(0.5, 1)--++(0, 1.0);
    \draw(1.5, 0)--++(0, 3.0);
}\,\,,
\end{aligned}
\ee
where the constant has changed due to the normalisation of the $\ket*{\oc}$ states. The expansion of this channel gives
\begin{align}
&= \kb{\oc}{\oc} + \cos^22J_1\sin^22h_1 \sin^22J_{\frac{3}{2}} \kb{\tilde X}{\tilde X} \\
+&(\cos^22J_{\frac{3}{2}} + \sin^22J_{\frac{3}{2}} \cos^22J_1\cos^22h_1) \kb{\tilde Y}{\tilde Y} \notag
\end{align} 
Again we can see that since $s_1>0$ by definition, this channel does not have any non-trivial unimodular eigenvectors. This leads us to conclude that the support of the eigenvector on next site must also be trivial.
\be
\begin{aligned}
\ket*{v_{2\ell}} = \ket*{\oc\oc}\otimes\ket*{\oc\oc}\otimes\ket*{v_{2\ell-4}}.
\end{aligned}
\ee
Since $s_x>0$ everywhere inside the subsystem this process will continue until we arrive at the other edge of the subsystem and are left with the channel
\be
\!\!\!(\!\bra*{\oc\oc}\otimes\!\dots\!\otimes\!\bra*{\oc\oc}\otimes\mathbbm{1}_2) T_\ell (\ket*{\oc\oc}\otimes\!\dots\!\otimes\ket*{\oc\oc}\otimes\mathbbm{1}_2\!)
\ee
which is graphically written as
\be
\begin{aligned}
\frac{1}{4}
\fineq[-0.8ex][0.6][1]{
    \roundgate[0][-1][1][topright][\ZXfcolor][1]
    \roundgate[1][0][1][topright][\ZXfcolor][1]
    \roundgate[2][-1][1][topright][\Zfcolor][1]
    \cstate[-0.5][-1.5]
    \cstate[-0.5][-.5]
    \cstate[2.5][-.5]
    \cstate[2.5][-1.5]
}.
\end{aligned}
\ee
This channel may be dealt with by hand upon which one finds that the only non-zero eigenvalues are $\{1,\,\, \cos2J_{\ell-1}\cos2h_{\ell-1/2}\cos2h_{\ell}\}$ where the unit eigenvalue of course corresponds to the $\ket*{\oc\oc}$. Once again, we are left to conclude that on the last two remaining sites the eigenvector must be trivial. We find therefore that all uni-modular eigenvectors of $T_\ell$ must be non-trivial within two sites of the left edge of the channel if they are to be orthogonal to the completely trivial infinite temperature state. This logic may be carried out identically for the right edge.

In total, we have found that a necessary condition for the existence of a non-trivial uni-modular eigenvector is that it must have non-trivial support at the edges. When investigating the channels responsible for the edge behaviour we found that for the left edge to be non trivial we must have $h_{\frac{1}{2}}, h_1 \in \{0, \pif{4}\}$ and symmetrically that for the right edge to be non trivial we must have $h_{\ell-\frac{1}{2}}, h_\ell \in \{0, \pif{4}\}$.

}

{
\section{Bounding the Spectral Radius}
\label{app:spectralradius}

In this appendix we prove that the eigenvectors in Eq.~\eqref{eq:eigenvectors} correspond to the eigenvalues with maximal possible magnitude by bounding the spectral radius of $T_\ell^{(n)}$ to 
\be
\frac{1}{d^{2n}}\fineq[-0.8ex][0.7][1]{
    \roundgate[0][0][1][topright][\Zfcolor][n]
    \sqrstate[0.5][-0.5]
    \sqrstate[0.5][0.5]
    \cstate[-.5][.5]
    \cstate[-.5][-.5]
}
=
n_\Lambda^{1-n}.
\ee
The spectral radius of a matrix may be calculated in the following way
\begin{equation}
    \rho(T_\ell^{(n)}) = \lim_{t\rightarrow\infty} \| (T_\ell^{(n)})^t\|_2^{\frac{1}{t}}
\end{equation}
where we are defining the $p$-norms as 
\be
\|A\|_{2p} \equiv [\tr(AA^\dag)^p]^{\frac{1}{2p}}.
\ee
Now, we consider the 2-norm
\be
\begin{aligned}
(d^{2nt} \| (T_\ell^{(n)})^t\|_2)^2 &= \tr \left(
\fineq[-0.8ex][0.5][1]{
    \transferT[0][5.3][4][n][\Zfccolor][\ZXfccolor]
	\node[scale=2.5,rotate=90] at (2.0,4.2)  {$\cdots$};
    \transferT[0][2][4][n][\Zfccolor][\ZXfccolor]
    \transfer[0][0][4][n][\Zfcolor][\ZXfcolor]
	\node[scale=2.5,rotate=90] at (2.0,-1.2)  {$\cdots$};
    \transfer[0][-3.3][4][n][\Zfcolor][\ZXfcolor]


}
\right)
\end{aligned}, 
\ee
where there are $t$ applications of the un-normalised transfer matrix followed by $t$ applications of it's conjugate (in red) that are ultimately joined together at the top and bottom by the trace. This can be reformulated as the norm of an operator acting in space. To do this we unfold the contractions on the left and right edges and fold in the contractions on the top and bottom to get the following
\be
\begin{aligned}
=
\tr\left(
\fineq[-0.8ex][0.5][1]{
    \transfer[0][0][4][1][\Zfcolor][\ZXfcolor][0]
    \transfer[0][-2][4][topright][\Zfcolor][\ZXfcolor][0]
    \transfer[0][-4][4][topright][\Zfcolor][\ZXfcolor][0]
    \transfer[0][-6][4][topright][\Zfcolor][\ZXfcolor][0]
    \transfer[0][-8][4][topright][\Zfcolor][\ZXfcolor][0]

    \transfer[5][0][4][topleft][\Zfccolor][\ZXfccolor][0]
    \transfer[5][-2][4][topleft][\Zfccolor][\ZXfccolor][0]
    \transfer[5][-4][4][topleft][\Zfccolor][\ZXfccolor][0]
    \transfer[5][-6][4][topleft][\Zfccolor][\ZXfccolor][0]
    \transfer[5][-8][4][topleft][\Zfccolor][\ZXfccolor][0]

    \cstate[.5][1.5]
    \cstate[1.5][1.5]
    \cstate[2.5][1.5]
    \cstate[3.5][1.5]
    \cstate[5.5][1.5]
    \cstate[6.5][1.5]
    \cstate[7.5][1.5]
    \cstate[8.5][1.5]

    \cstate[0.5][-8.5]
    \cstate[1.5][-8.5]
    \cstate[2.5][-8.5]
    \cstate[3.5][-8.5]
    \cstate[5.5][-8.5]
    \cstate[6.5][-8.5]
    \cstate[7.5][-8.5]
    \cstate[8.5][-8.5]
}
\right)^n \\
\end{aligned}
\ee
where now the multiplication and trace functions from left to right and we have pictured a specific $t=5$ to serve as an example. Now we can recognise this as meeting the definition of the operator norm given above


\be
\begin{aligned}
=
\left( \left\|
\fineq[-0.8ex][0.5][1]{
    \transfer[0][0][4][topright][\Zfcolor][\ZXfcolor][0]
    \transfer[0][-2][4][topright][\Zfcolor][\ZXfcolor][0]
    \transfer[0][-4][4][topright][\Zfcolor][\ZXfcolor][0]
    \transfer[0][-6][4][topright][\Zfcolor][\ZXfcolor][0]
    \transfer[0][-8][4][topright][\Zfcolor][\ZXfcolor][0]
    \cstate[.5][1.5]
    \cstate[1.5][1.5]
    \cstate[2.5][1.5]
    \cstate[3.5][1.5]
    \cstate[.5][-8.5]
    \cstate[1.5][-8.5]
    \cstate[2.5][-8.5]
    \cstate[3.5][-8.5]
}
\right\|_{2n}\right)^{2n}
\end{aligned}
\ee
This quantity is no easier to calculate than those given before. The difficulty here comes from the bullet states applied to the top and bottom of the diagram that block the application of the solvability conditions. However, we now show that they do not meaningfully contribute to the large $t$ asymptotic value. We can use operator norm inequalities to cut them out of the diagram in order to isolate all $t$-dependence into an operator whose norm we may calculate. We will apply H\"older's inequality for operator norms
\be
    \|AB\|_r \leq \|A\|_p\|B\|_q, 
\ee
which is valid for all $p,q,r\in[1,\infty]$ such that 
\be
\frac{1}{p} + \frac{1}{q} = \frac{1}{r}\,, 
\ee
in order to state that
\be
\begin{aligned}
\hspace{-0.4cm}
\left\|
\fineq[-0.8ex][0.45][1]{
    \transfer[0][0][4][topright][\Zfcolor][\ZXfcolor][0]
    \transfer[0][-2][4][topright][\Zfcolor][\ZXfcolor][0]
    \transfer[0][-4][4][topright][\Zfcolor][\ZXfcolor][0]
    \transfer[0][-6][4][topright][\Zfcolor][\ZXfcolor][0]
    \transfer[0][-8][4][topright][\Zfcolor][\ZXfcolor][0]
    \cstate[.5][1.5]
    \cstate[1.5][1.5]
    \cstate[2.5][1.5]
    \cstate[3.5][1.5]
    \cstate[.5][-8.5]
    \cstate[1.5][-8.5]
    \cstate[2.5][-8.5]
    \cstate[3.5][-8.5]
}
\right\|
\hspace{-0.12cm}
\leq
\hspace{-0.12cm}
\left\|
\fineq[-0.8ex][0.45][1]{
\foreach \i in {0,...,4}{
    \roundgate[0][-2*\i][1][topright][\Zfcolor][1]
}
\foreach \i in {1,...,4}{
    \roundgate[1][1-2*\i][1][topright][\ZXfcolor][1]
}
\foreach \i in {1,...,3}{
    \roundgate[2][-2*\i][1][topright][\ZXfcolor][1]
}
\roundgate[3][-3][1][topright][\ZXfcolor][1]
\roundgate[3][-5][1][topright][\ZXfcolor][1]
\roundgate[4][-4][1][topright][\Zfcolor][1]

\draw[white] (0.5, 1.5) -- (0.51, 1.55);
\draw[white] (0.5, -8.5) -- (0.51, -8.55);

\draw[thick] (0.5, 0.5) -- (4.5, .5);
\draw[thick] (1.5, -0.5) -- (4.5, -.5);
\draw[thick] (2.5, -1.5) -- (4.5, -1.5);
\draw[thick] (3.5, -2.5) -- (4.5, -2.5);

\draw[thick] (0.5, -8.5) -- (4.5, -8.5);
\draw[thick] (1.5, -7.5) -- (4.5, -7.5);
\draw[thick] (2.5, -6.5) -- (4.5, -6.5);
\draw[thick] (3.5, -5.5) -- (4.5, -5.5);

}
\right\|
\left\|
\fineq[-0.8ex][0.45][1]{


\roundgate[1][1][1][topright][\ZXfcolor][1]
\roundgate[2][0][1][topright][\ZXfcolor][1]
\roundgate[3][-1][1][topright][\ZXfcolor][1]
\roundgate[4][-2][1][topright][\Zfcolor][1]
\roundgate[3][1][1][topright][\ZXfcolor][1]
\roundgate[4][0][1][topright][\Zfcolor][1]
\draw[thick] (3.52, -2.5) -- (.5, -2.5);
\draw[thick] (2.52, -1.5) -- (.5, -1.5);
\draw[thick] (1.52, -.5) -- (.5, -.5);

\draw[thick] (4.5, -3.5) -- (.5, -3.5);
\draw[thick] (4.5, -4.5) -- (.5, -4.5);

\roundgate[2][-8][1][topright][\ZXfcolor][1]
\roundgate[3][-7][1][topright][\ZXfcolor][1]
\roundgate[4][-6][1][topright][\Zfcolor][1]
\roundgate[4][-8][1][topright][\Zfcolor][1]
\draw[thick] (3.52, -5.5) -- (.5, -5.5);
\draw[thick] (2.52, -6.5) -- (.5, -6.5);
\draw[thick] (1.52, -7.5) -- (.5, -7.5);

\cstate[.5][1.5]
\cstate[1.5][1.5]
\cstate[2.5][1.5]
\cstate[3.5][1.5]

\draw[thick] (1, -8.5) -- (.5, -8.5);
\cstate[1][-8.5]
\cstate[1.5][-8.5]
\cstate[2.5][-8.5]
\cstate[3.5][-8.5]
}
\right\|_{\infty}\hspace{-0.5cm}
\end{aligned}
\ee
where the two norms without subscripts are of the same type. The method for making the split at larger $t$ is to simply pad the middle of the operator under the infinity norm with more and more identity tensors/straight lines such that no more gates must be included in it for the inequality to hold. This means that the piece we have cut off the original diagram is constant with respect to time
\be
\begin{aligned}
\left\|
\fineq[-0.8ex][0.45][1]{

\roundgate[1][1.5][1][topright][\ZXfcolor][1]
\roundgate[2][0.5][1][topright][\ZXfcolor][1]
\roundgate[3][-.5][1][topright][\ZXfcolor][1]
\roundgate[4][-1.5][1][topright][\Zfcolor][1]
\roundgate[3][1.5][1][topright][\ZXfcolor][1]
\roundgate[4][0.5][1][topright][\Zfcolor][1]
\draw[thick] (3.52, -2.) -- (.5, -2);
\draw[thick] (2.52, -1.) -- (.5, -1);
\draw[thick] (1.52, -.0) -- (.5, -.0);

\draw[thick] (4.5, -3) -- (.5, -3);

\node[scale=2.8,rotate=90] at (2.6,-3.7)  {$\cdots$};

\draw[thick] (4.5, -4.5) -- (.5, -4.5);

\roundgate[2][-8][1][topright][\ZXfcolor][1]
\roundgate[3][-7][1][topright][\ZXfcolor][1]
\roundgate[4][-6][1][topright][\Zfcolor][1]
\roundgate[4][-8][1][topright][\Zfcolor][1]
\draw[thick] (3.52, -5.5) -- (.5, -5.5);
\draw[thick] (2.52, -6.5) -- (.5, -6.5);
\draw[thick] (1.52, -7.5) -- (.5, -7.5);

\cstate[.5][2]
\cstate[1.5][2]
\cstate[2.5][2]
\cstate[3.5][2]

\draw[thick] (1, -8.5) -- (.5, -8.5);
\cstate[1][-8.5]
\cstate[1.5][-8.5]
\cstate[2.5][-8.5]
\cstate[3.5][-8.5]
}
\right\|_{\infty}
=
\left\|
\fineq[-0.8ex][0.45][1]{
\roundgate[1][1][1][topright][\ZXfcolor][1]
\roundgate[2][0][1][topright][\ZXfcolor][1]
\roundgate[3][-1][1][topright][\ZXfcolor][1]
\roundgate[4][-2][1][topright][\Zfcolor][1]
\roundgate[3][1][1][topright][\ZXfcolor][1]
\roundgate[4][0][1][topright][\Zfcolor][1]
\draw[thick] (3.52, -2.5) -- (.5, -2.5);
\draw[thick] (2.52, -1.5) -- (.5, -1.5);
\draw[thick] (1.52, -.5) -- (.5, -.5);

\roundgate[2][-6][1][topright][\ZXfcolor][1]
\roundgate[3][-5][1][topright][\ZXfcolor][1]
\roundgate[4][-4][1][topright][\Zfcolor][1]
\roundgate[4][-6][1][topright][\Zfcolor][1]
\draw[thick] (3.52, -3.5) -- (.5, -3.5);
\draw[thick] (2.52, -4.5) -- (.5, -4.5);
\draw[thick] (1.52, -5.5) -- (.5, -5.5);

\cstate[.5][1.5]
\cstate[1.5][1.5]
\cstate[2.5][1.5]
\cstate[3.5][1.5]

\draw[thick] (1, -6.5) -- (.5, -6.5);
\cstate[1][-6.5]
\cstate[1.5][-6.5]
\cstate[2.5][-6.5]
\cstate[3.5][-6.5]
}
\right\|_{\infty}
= \text{const}
\end{aligned}
\ee
as $\|A\otimes\mathbbm{1}\|_\infty=\|A\|_\infty$. The remaining norm can be expressed as
\be
\begin{aligned}
\left\|
\fineq[-0.8ex][0.45][1]{
\foreach \i in {0,...,4}{
    \roundgate[0][-2*\i][1][topright][\Zfcolor][1]
}
\foreach \i in {1,...,4}{
    \roundgate[1][1-2*\i][1][topright][\ZXfcolor][1]
}
\foreach \i in {1,...,3}{
    \roundgate[2][-2*\i][1][topright][\ZXfcolor][1]
}
\roundgate[3][-3][1][topright][\ZXfcolor][1]
\roundgate[3][-5][1][topright][\ZXfcolor][1]
\roundgate[4][-4][1][topright][\Zfcolor][1]

\draw[white] (0.5, 0.5) -- (0.51, 0.65);
\draw[white] (0.5, -8.5) -- (0.51, -8.55);

\draw[thick] (0.5, 0.5) -- (4.5, .5);
\draw[thick] (1.5, -0.5) -- (4.5, -.5);
\draw[thick] (2.5, -1.5) -- (4.5, -1.5);
\draw[thick] (3.5, -2.5) -- (4.5, -2.5);

\draw[thick] (0.5, -8.5) -- (4.5, -8.5);
\draw[thick] (1.5, -7.5) -- (4.5, -7.5);
\draw[thick] (2.5, -6.5) -- (4.5, -6.5);
\draw[thick] (3.5, -5.5) -- (4.5, -5.5);

}
\right\|_{2n}
=
\left(
\fineq[-0.8ex][0.45][1]{
\foreach \i in {0,...,4}{
    \roundgate[0][-2*\i][1][topright][\Zfcolor][n]
}
\foreach \i in {1,...,4}{
    \roundgate[1][1-2*\i][1][topright][\ZXfcolor][n]
}
\foreach \i in {1,...,3}{
    \roundgate[2][-2*\i][1][topright][\ZXfcolor][n]
}
\roundgate[3][-3][1][topright][\ZXfcolor][n]
\roundgate[3][-5][1][topright][\ZXfcolor][n]
\roundgate[4][-4][1][topright][\Zfcolor][n]

\draw[white] (0.5, 0.5) -- (0.51, 0.65);
\draw[white] (0.5, -8.5) -- (0.51, -8.55);

\foreach \i in {0,...,9}{
    \cstate[-.5][0.5-\i]
}
\foreach \i in {0,...,4}{
    \sqrstate[0.5+\i][0.5-\i]
    \sqrstate[0.5+\i][-8.5+\i]
}
}
\right)^{\frac{1}{n}}
\end{aligned}
\ee
where we remind the reader that the operator within the norm on the l.h.s.\ has two copies which results in the power of $1/n$ rather than the expected $1/2n$ on the r.h.s.. This contraction can now be performed easily by applying the solvability conditions
\be
\begin{aligned}
\left\|
\fineq[-0.8ex][0.45][1]{
\foreach \i in {0,...,4}{
    \roundgate[0][-2*\i][1][topright][\Zfcolor][1]
}
\foreach \i in {1,...,4}{
    \roundgate[1][1-2*\i][1][topright][\ZXfcolor][1]
}
\foreach \i in {1,...,3}{
    \roundgate[2][-2*\i][1][topright][\ZXfcolor][1]
}
\roundgate[3][-3][1][topright][\ZXfcolor][1]
\roundgate[3][-5][1][topright][\ZXfcolor][1]
\roundgate[4][-4][1][topright][\Zfcolor][1]

\draw[white] (0.5, 0.5) -- (0.51, 0.65);
\draw[white] (0.5, -8.5) -- (0.51, -8.55);

\draw[thick] (0.5, 0.5) -- (4.5, .5);
\draw[thick] (1.5, -0.5) -- (4.5, -.5);
\draw[thick] (2.5, -1.5) -- (4.5, -1.5);
\draw[thick] (3.5, -2.5) -- (4.5, -2.5);

\draw[thick] (0.5, -8.5) -- (4.5, -8.5);
\draw[thick] (1.5, -7.5) -- (4.5, -7.5);
\draw[thick] (2.5, -6.5) -- (4.5, -6.5);
\draw[thick] (3.5, -5.5) -- (4.5, -5.5);

}
\right\|_{2n}
=
\left(
\fineq[-0.8ex][0.6][1]{
\roundgate[0][0][1][topright][\Zfcolor][n]
\cstate[-.5][0.5]
\cstate[-.5][-0.5]
\sqrstate[.5][0.5]
\sqrstate[.5][-0.5]

\draw[white] (0, 0.5) -- (0, 0.75);
\draw[white] (0, -0.5) -- (0, -0.75);
}
\right)^{\frac{t}{n}},
\end{aligned}
\ee
which is a result that generalises to any larger times.
We can now write that
\bea
(d^{2nt} \| (T_\ell^{(n)})^t\|_2)^2 \leq 
\left(
\fineq[-0.8ex][0.6][1]{
\roundgate[0][0][1][topright][\Zfcolor][n]
\cstate[-.5][0.5]
\cstate[-.5][-0.5]
\sqrstate[.5][0.5]
\sqrstate[.5][-0.5]

\draw[white] (0, 0.5) -- (0, 0.75);
\draw[white] (0, -0.5) -- (0, -0.75);
}
\right)^{2t}
(\text{const})^{2n}.
\eea
Taking the large time limit of this bound, we find that the constant drops out as desired
\begin{equation}
\begin{aligned}
\rho(T_\ell^{(n)}) \leq \frac{1}{d^{2n}} \left(
\fineq[-0.8ex][0.6][1]{
    \roundgate[0][0][1][topright][\Zfcolor][n]
    \cstate[-.5][.5]
    \cstate[-.5][-.5]
    \sqrstate[.5][.5]
    \sqrstate[.5][-.5]
}
\right)
= \frac{1}{n_\Lambda^{n-1}}
\end{aligned}
\end{equation}
This upper bound is indeed the eigenvalue corresponding to the eigenvectors in  Eq.~\eqref{eq:eigenvectors}. Therefore, we have proven that these eigenvectors are indeed two of the leading eigenvectors.

\section{Diagonalization of the propagator at the free point}
\label{app:diagonalisation}

In this appendix we detail the diagonalisation of the propagator for $d=2$,  $h_x=0$ and $s_x=s$. We work with the kicked Ising representation of the propagator (cf.~Appendix~\ref{app:gateformulations}), which can be expressed as 
\bea
    \mathbb{U} &= e^{i\pif{4}\sum_{j=1}^{2L}X_{j/2}} e^{iH_{o}} e^{i\pif{4}\sum_{j=1}^{2L}X_{j/2}} e^{iH_{e}}, \\
    H_e &= \sum_{x=1}^L [\pif{4}Z_{x-1/2}Z_{x} + f Z_{x}Z_{x+1/2}], \\
    H_o &= \sum_{x=1}^L [f Z_{x-1/2}Z_{x} + \pif{4} Z_{x}Z_{x+1/2}],
\ea
\ee
where we set $f = {\pi}s/4$. A favourable rearrangement of this expression is the following
\begin{align}
&\!\!\mathbb{U} = {\rm T}\exp(i \int_0^1 {\rm d}\tau H(\tau)) =\,U_{1}\, U_{2}\prod_{n=1}^{2L} X_{n/2}, \notag\\
&   \!\!U_{1} = \exp[i\pif{4}\sum_{m=1}^{L} Y_{m}Y_{m+\frac{1}{2}}] \!\exp[i\pif{4}\sum_{m=1}^{L} Z_{m-\frac{1}{2}}Z_{m}]\!, \notag\\
&    \!\!U_{2} = \exp[if\sum_{m=1}^{L} Y_{m-\frac{1}{2}}Y_{m}] \!\exp[if\sum_{m=1}^{L} Z_{m}Z_{m+\frac{1}{2}}]\!.
\label{eq:appC_floquet_ferm}
\end{align}
It is easily seen that
\be
\begin{aligned}
 [\prod_{n=1}^{2L} X_{n/2}, U_1] = [\prod_{n=1}^{2L} X_{n/2}, U_2] = [U_1, U_2] = 0\,. 
\label{eq:appC_commutations}
\end{aligned}
\ee
Therefore, we may diagonalise them all simultaneously. One can immediately recognise that the two propagators $U_1$ and $U_2$ will have the same spectra as homogeneous Kicked Ising models. This is because they both take the form
\be
    U_\mu = \exp(i\sum_{m=1}^{L} t_{m-\frac{1}{2}}) \exp(i\sum_{m=1}^{L} t_{m}),
\ee   
where $\{t_j\}$ fulfil  
\bea    
    &t_j^2 = \text{const}, \\
    &\{t_m,t_{m+\frac{1}{2}}\}=0, \\ 
    &[t_m,t_n] = 0, \qquad |m-n| > \frac{1}{2}.
\eea
This algebra immediately implies that the model can be written as a product of quadratic forms of Majorana fermions~\cite{fendley_2019, chapman_2020}. 
In particular, due to the fine-tuned angle $\pi/4$ the operator $U_1$ is actually equivalent to the self-dual Kicked Ising model which is dual unitary, whilst $U_2$ is generically away from this critical point. We therefore can expect the modes of $U_1$ to all travel at maximal velocity $|v|=1$ and the modes of $U_2$ to travel at a range of smaller velocities.

To piece this out we introduce Majorana fermions with the following Jordan-Wigner transformation
\bea
    a_{m-\frac{1}{2}} &= \left( \prod_{n=1}^{m-1} X_{n/2} \right) Y_{m/2}, \\
    a_{m} &= \left( \prod_{n=1}^{m-1} X_{n/2} \right) Z_{m/2},
\eea
that are Hermitian and satisfy the Clifford algebra
\be
    a_m = a_m^\dag, \qquad \{a_m, a_n\} = 2\delta_{mn}.
\ee
We must take care with the boundary conditions for these operators. Using the fact that $L$ is always even we work in the basis where
\be
\mathcal P = \prod_{n=1}^{2L} X_{n/2},
\ee
is diagonal and equal to $+1$ or $-1$ which we refer to as the even and odd sectors respectively. In each sector we have a different boundary condition: periodic $(a_{m+2L}=a_{m})$ in the odd sector and anti-periodic $(a_{m+2L}=-a_{m})$ in the even sector.

\subsection{Odd Sector}
Here we take $\mathcal P = -1$ and work with periodic boundary conditions. With these definitions, we may write
\begin{align}
    &\!\!\!U_{1} \!= \!\exp[i\pif{4}\sum_{m=1}^{L} \!\!a_{2m}a_{2m+\frac{1}{2}}\!]\!\exp[{i \pif{4}\sum_{m=1}^{L} \!\!a_{2m}a_{2m-\frac{3}{2}}}\!]\!, \hspace{-1cm}\\
    &\!\!\!U_{2} \!=\! \exp[i f\sum_{m=1}^{L} \!\!a_{2m-1}a_{2m-\frac{1}{2}}\!] \exp[i f \sum_{m=1}^{L} \!\!a_{2m+1}a_{2m-\frac{1}{2}}\!]\!.  \notag
\end{align}
We shall adopt a notation that makes the unit cell more clear, for $m = 1,2,\dots,L$
\bea
    \gamma_{m-\frac{1}{2}}^{(1)} = a_{2m-\frac{3}{2}}, &\qquad  \gamma_{m}^{(1)} = a_{2m}, \\
    \gamma_{m-\frac{1}{2}}^{(2)} = a_{2m-1}, &\qquad  \gamma_{m}^{(2)} = a_{2m-\frac{1}{2}}, \\
\eea
under which
\begin{align}
    &\!\!\!\!U_{1} \!= \exp[-i \pif{4}\sum_{m=1}^{L} \gamma_{m+\frac{1}{2}}^{(1)} \gamma_{m}^{(1)}] \!\!\exp[i\pif{4}\sum_{m=1}^{L} \gamma_{m}^{(1)}\gamma_{m-\frac{1}{2}}^{(1)}]\!, \notag\\
    &\!\!\!\!U_{2} \!= \exp[-i f\!\!\sum_{m=1}^{L} \gamma_{m}^{(2)} \gamma_{m-\frac{1}{2}}^{(2)}] \!\!\exp[i f \!\!\sum_{m=1}^{L} \gamma_{m+\frac{1}{2}}^{(2)} \gamma_{m}^{(2)}]\!. \!\!
\end{align}
Here we draw attention to the fact that not only do $U_1$ and $U_2$ commute, they also act on disjoint Hilbert spaces. 

An important identity for interpreting these forms is 
\be
    [\gamma_a\gamma_b,\gamma_c] = 2( \gamma_a \delta_{bc} - \gamma_b \delta_{ac}),
\ee
where one should read equations without upper indices as equations with all operators of the same upper index. We shall diagonalise these operators in their adjoint representation. To that end it is convenient to introduce the vectorised notation
\be
    \ket*{x} = i^{2x} \gamma_x
\label{eq::appC_vectorised}
\ee

\subsubsection{Diagonalising $U_1$}
We first define the Liouvillian $\mathcal{L}_O = [O, \,\cdot\,]$ and introduce $A,B$ such that
\be
    U_1 = \exp(i\pif{4} B) \exp(i\pif{4} A),
\ee
Then, we have that for $m = 1,2,\dots,L$
\bea
    \mathcal{L}_A \ket*{m}  &= -2 \ket*{m - 1/2}, \\
    \mathcal{L}_A \ket*{m-1/2} &= -2 \ket*{m}, \\
    \mathcal{L}_B \ket*{m}  &= 2 \ket*{m + 1/2}, \\
    \mathcal{L}_B \ket*{m+1/2} &= 2 \ket*{m}. \\
\eea
This gives
\bea
    e^{i\pif{4}\mathcal{L}_A} \ket*{m}  &= -i \ket*{m - 1/2}, \\
    e^{i\pif{4}\mathcal{L}_A} \ket*{m-1/2} &= -i \ket*{m}, \\
    e^{i\pif{4}\mathcal{L}_B} \ket*{m}  &= i \ket*{m + 1/2}, \\
    e^{i\pif{4}\mathcal{L}_B} \ket*{m-1/2}  &= i \ket*{m-1}. \\
\eea
Putting these together it is trivial to construct eigenstates
\bea
    \ket*{k,-} &= \frac{1}{\sqrt{L}}\sum_{m=1}^L e^{-ikm} \ket*{m}, \\
    \ket*{k,+} &= \frac{1}{\sqrt{L}}\sum_{m=1}^L e^{-ikm} \ket*{m-1/2},
\eea
where $k\in R$ for the Ramond sector momenta 
\be
R = \{ \frac{2\pi n}{L} : n = -\frac{L}{2}, \dots, \frac{L}{2}-1\},
\ee
that are chosen to give periodic boundary conditions. Returning to the operator formalism, we can construct the eigenmodes easily but we must be careful to not generate any redundancies. It is important to note that the use of the $i^{2m}$ factor in Eq.~\eqref{eq::appC_vectorised} is equivalent to shifting the momenta of these modes by $\pi$. We will make definitions explicitly using $k\in R : k>0$
\bea
    \beta_1^\dag(k) &= \frac{1}{\sqrt{2L}}\sum_{m=1}^L e^{-ikm} \gamma_{m-\frac{1}{2}}^{(1)}, \\
    \beta_1(-k) &= \frac{i}{\sqrt{2L}}\sum_{m=1}^L e^{-ikm}  \gamma_m^{(1)}.
\eea
that are related to $\ket*{k-\pi,+}$ and $\ket*{k-\pi,-}$ respectively. For the $k=0$ and $k=-\pi$ operators we define
\bea
    \beta_1^\dag(0) &= \frac{1}{2\sqrt{L}}\sum_{m=1}^L (\gamma_{m-\frac{1}{2}}^{(1)} + i\gamma_{m}^{(1)} ), \\
    \beta_1^\dag(-\pi) &= \frac{1}{2\sqrt{L}}\sum_{m=1}^L (-1)^m (\gamma_{m-\frac{1}{2}}^{(1)} + i\gamma_{m}^{(1)} ). \\
\eea
These operators are well defined as canonical fermions
\be
    \{\beta_1(k), \beta_1^\dag(k')\} = \delta_{k,k'},
\ee
with all other anti-commutators vanishing. Their evolution is particularly simple 
\be
    U_1 \, \beta_1^\dag(k) \, U_1^\dag = -e^{i|k|} \,\beta_1^\dag(k).
\ee
It is helpful to keep this $-1$ separate as it represents the creation operator connecting the odd and even sectors.

\subsubsection{Diagonalising $U_2$}
We again define introduce $A,B$ such that
\be
    U_2 = \exp(if B) \exp(if A)
\ee
Then, we have that for $m = 1,2,\dots,L$
\bea
    \mathcal{L}_A \ket*{m}  &= -2 \ket*{m + 1/2} \\
    \mathcal{L}_A \ket*{m+1/2} &= -2 \ket*{m} \\
    \mathcal{L}_B \ket*{m}  &= 2 \ket*{m - 1/2} \\
    \mathcal{L}_B \ket*{m-1/2} &= 2 \ket*{m} \\
\eea
Which gives
\begin{align}
    &e^{if\mathcal{L}_A} \ket*{m}  = \cos(2f)\ket*{m} - i\sin(2f)\ket*{m+1/2}  \notag\\
    &e^{if\mathcal{L}_A} \ket*{m-1/2} = \cos(2f)\ket*{m-1/2} - i\sin(2f)\ket*{m-1} \notag\\
    &e^{if\mathcal{L}_B} \ket*{m}  = \cos(2f)\ket*{m} + i\sin(2f)\ket*{m - 1/2} \\
    &e^{if\mathcal{L}_B} \ket*{m-1/2}  = \cos(2f)\ket*{m-1/2} + i\sin(2f)\ket*{m}\!.\notag
\end{align}
This process is symmetric under translations of $\Delta m=1$ (2 sites), so we assume the eigenvector takes the form
\be
    \hspace{-.25cm}\ket{k,\pm} = \frac{1}{\sqrt{L}}\sum_{m=1}^L e^{-ikm} \left[a_k^{(\pm)} \ket*{m\!-\!1/2}\! +\! b_k^{(\pm)}\ket*{m} \right]\!.
\ee
There there are two vectors such that
\bea
    e^{if\mathcal{L}_B}e^{if\mathcal{L}_A}\ket{k,\pm} &= e^{\pm i\epsilon(k)}\ket{k,\pm}, \\
    \epsilon(k) &= 2\sin^{-1}(|\sin(k/2)\sin(2f)|),
\eea
with
\bea
    a_k^{(\pm)} &= \frac{e^{\mp\varphi_k/2}}{\sqrt{2\cosh(\varphi_k)}}, \\
    b_k^{(\pm)} &= \pm ie^{-ik/2} \frac{e^{\pm\varphi_k/2}}{\sqrt{2\cosh(\varphi_k)}}, \\
    \sinh(\varphi_k) &= \cos(k/2)\tan(2f).
\eea
Undoing the vectorisation as we did above we can construct the appropriate modes for $k\in R : k\geq0$ that are related to $\ket*{k-\pi,\pm}$
\begin{align}
    &\!\!\!\!\beta_2^\dag(k) \!=\!\! 
    \sum_{m=1}^L \!e^{-ikm}  
    \frac{ (e^{-\phi_k/2}\gamma_{m-\frac{1}{2}}^{(2)} \!-\! i e^{(\phi_k-ik)/2}\gamma_{m}^{(2)} )} {2\sqrt{L \cosh{\phi_k}}}\!,\\
    &\!\!\!\!\beta_2(-k) \!=\!\! \sum_{m=1}^L \!\!e^{-ikm}  
    \frac{ (e^{\phi_k/2}\gamma_{m-\frac{1}{2}}^{(2)} \!+\! i e^{-(\phi_k+ik)/2}\gamma_{m}^{(2)} )} {2\sqrt{L \cosh{\phi_k}}}\!,
\end{align}
where
\be
    \sinh(\phi_k) = \sinh(\varphi_{k-\pi}) = \sin(k/2)\tan(2f),
\ee
and 
\bea
    U_2 \, \beta_2^\dag(k) \, U_2^\dag &= -e^{i\varepsilon(k)}\, \beta_2^\dag(k) \\
    \varepsilon(k) = \epsilon(k-\pi)-\pi &= \cos^{-1}(\cos(k/2)\sin(2f)).
\eea
Again we have taken care to make sure there is a $-1$ factor out the front.

\subsection{Odd Sector}
Here we take $\mathcal P = 1$ and work with anti-periodic boundary conditions $(a_{m+2L}=-a_{m})$. With these definitions, we may again write
\begin{align}
    &\!\!\!U_{1} \!= \!\exp[i\pif{4}\sum_{m=1}^{L} \!\!a_{2m}a_{2m-\frac{3}{2}}\!]\!\exp[{i \pif{4}\sum_{m=1}^{L} \!\!a_{2m}a_{2m+\frac{1}{2}}}\!] \hspace{-1cm}\\
    &\!\!\!U_{2} \!=\! \exp[i f\sum_{m=1}^{L} \!\!a_{2m-1}a_{2m-\frac{1}{2}}\!] \exp[i f \sum_{m=1}^{L} \!\!a_{2m+1}a_{2m-\frac{1}{2}}\!]\!.  \notag
\end{align}
which is identical to the previous form with only the boundary conditions changing. The process here is now identical except for the fact that we must use a different set of momenta that are compatible with the anti-periodic boundaries. We refer to these as the Neveu-Schwarz sector momenta 
\be
{\rm NS} = \left\{ \frac{2\pi (n+1/2)}{L} : n = -\frac{L}{2}, \dots, \frac{L}{2}-1\right\}.
\ee 
The notable feature here is no $k=0$ mode. Therefore, we may use the definitions of the fermion operators presented in the previous section for the momenta non-zero momenta in the Ramond sector.

\section{Quench at the Free Point}
\label{app:freequench}

Here we provide details on the structure of the free fermionic quench considered in the main text. The structure of the evolution operator is described in Appendix~\ref{app:diagonalisation}. The initial state considered is the following state in the computational basis for even $L$
\bea
    \ket*{\psi} &= \bigotimes_{x=1}^L \frac{1}{\sqrt{2}} \sum_{i,j=0}^1 (Z)_{ij} \ket*{ij}_{x-1/2,x}\\
    &= \prod_{m=1}^L \frac{1}{\sqrt{2}}(c^\dag_{m-\frac{1}{2}} + c^\dag_{m} ) \ket{0}
\ea
\label{eq:appD_state}
\ee
where $c_{j/2}^\dag = \frac{1}{2} (a_{j-\frac{1}{2}} + i a_{j})$ and $\ket*{0}$ the state annihilated by all $c_j$. This state is an eigenstate of $\prod_{j=1}^{2L} X_{j/2}$ with eigenvalue $1$ so we must use antiperiodic boundary conditions $(a_{m+2L}=-a_{m})$ and thus momenta from the Neveu-Schwarz sector. The correlations in this state satisfy Wick's theorem and have the property that
\be
    \bra*{\psi} \gamma_m^{(1)}\gamma_n^{(2)} \ket*{\psi} = 0
\ee
This implies that the initial density matrix may be written as a quadratic form of the Majorana modes and that, in particular, it splits into the following tensor product
\be
    \rho = \rho_1\otimes\rho_2,
\ee
where $\rho_j$ involves only Majoranas of type $\gamma_m^{(j)}$. The evolution of this density matrix is then given as
\be
    \rho(t) = \rho_1(t) \otimes \rho_2(t), \qquad \rho_j(t) = U_j\, \rho_j(0)\, (U_j^\dag)^t.
\ee
This means that the entanglement entropies are expressed as a sum of two contributions depending respectively only on the $j=1$ and $j=2$ degrees of freedom, i.e., 
\be
    S^{(n)}(t) = S^{(n)}_1(t) + S^{(n)}_2(t).
\ee
At this point we have identified the fact that the quench problem separates into two parts that are both homogeneous and free-fermionic. This allows us to apply the quasiparticle picture in its most familiar form.

}

\bibliography{multisite_v3}

\end{document}